%% file: bare_jrnl.tex
\documentclass[journal]{IEEEtran}
\ifCLASSINFOpdf
\else
\fi
\newcommand{\change}[1]{{\color{blue} {#1}}}

\newcommand{\CHENG}{}
\renewcommand{\change}{}

\newcommand{\CHENGB}{}
\newcommand{\revise}[1]{{ {#1}}}
\newcommand{\revision}[1]{{{#1}}}

\newcommand{\chengr}{}
\newcommand{\jn}{}

\newcommand{\bluecolor}[1]{{ {#1}}}

\def\extend{1}

\usepackage{hyperref}

\usepackage{array}
\usepackage{svg}
\usepackage{subcaption}
\usepackage{tikz}
\usepackage{pgfplots}
\usepackage{verbatim}
\usepackage{amsmath,bm}

\usepackage{tkz-euclide}
\usepackage{marginnote}

\usepackage{diagbox}
\usepackage{xcolor}

\usepackage[a-2b]{pdfx}

\usepackage[ruled,linesnumbered, vlined]{algorithm2e}

\newtheorem{problem}{Problem}
\newtheorem{definition}{Definition}
\newtheorem{lemma}{Lemma}
\newtheorem{proof}{Proof}
\newtheorem{example}{Example}

\usetikzlibrary{patterns}

\usepackage[numbers]{natbib} 
\usepackage{footnote}
\usepackage{amssymb}

\begin{document}
\title{BMTree: Designing, Learning, and Updating Piecewise Space-Filling Curves for Multi-Dimensional Data Indexing}

\author{
Jiangneng~Li, %~\IEEEmembership{Member,~IEEE,}
Yuang~Liu, 
Zheng~Wang, 
Gao~Cong,  
Cheng~Long, 
Walid G. Aref,~\IEEEmembership{Fellow,~IEEE,}
Han Mao~Kiah, 
and~Bin~Cui,~\IEEEmembership{Fellow,~IEEE}
% Michael~Shell,~\IEEEmembership{Member,~IEEE,}
% John~Doe,~\IEEEmembership{Fellow,~OSA,}
% and~Jane~Doe,~\IEEEmembership{Life~Fellow,~IEEE}% <-this % stops a space
\thanks{
J. Li, Y. Liu, Z. Wang, G. Cong, C. Long, and H.M. Kiah are with Nanyang Technological University, Singapore, 639798.
E-mail: \{jiangnen002@e., s230084@e., zheng011@e., gaocong@, c.long@, HMKiah@\}ntu.edu.sg;
W. G. Aref is with Purdue University, West Lafayette, IN, USA, 47907. E-mail: aref@purdue.edu;
B. Cui is with School of Computer Science, Peking University, Beijing, China, 100871. E-mail: bin.cui@pku.edu.cn.
}% <-this % stops a space
% \thanks{J. Doe and J. Doe are with Anonymous University.}% <-this % stops a space
% 
% 
% \thanks{Manuscript received April 19, 2005; revised August 26, 2015.}
}

\maketitle

\begin{abstract}
% To index multi-dimensional data,  s
Space-filling curves (SFC, for short)  have been widely applied to index multi-dimensional data, which first 
% used to
maps the data to one dimension, and then a one-dimensional indexing method, e.g., the B-tree indexes the mapped data.
 Existing SFCs
 adopt a single mapping scheme for the whole {\CHENGB data space}. However, 
a single mapping scheme {\CHENGB often does not} perform well on all the 
data space. 
In this paper, we propose a new type of SFC {\CHENGB called  piecewise SFCs that adopts} different mapping schemes for different data subspaces. 
{\CHENGB Specifically, we propose a data structure termed the Bit Merging tree (BMTree) that can generate data subspaces and their SFCs simultaneously, and achieve desirable properties of the SFC for the whole data space. Furthermore,}
% to represent data subspaces and their selected SFCs. 
we develop a reinforcement learning-based solution to build the BMTree, aiming to achieve excellent query performance. \bluecolor{To update the BMTree efficiently when the distributions of data and/or queries change,  we develop a new mechanism that achieves fast detection of distribution shifts in data and queries, and enables {\em partial} retraining of the BMTree. The retraining mechanism achieves  performance enhancement efficiently since it avoids retraining the BMTree from scratch.}
Extensive experiments show 
the effectiveness and efficiency of the BMTree with the proposed learning-based methods.
% that our proposed method outperforms {\CHENGB existing} SFCs in terms of query performance.
\end{abstract}


% \fi
\IEEEpeerreviewmaketitle

\input{sections/introduction}

\input{sections/preliminary2}

\input{sections/method4}

\input{sections/method_update}

\input{sections/analysis}

\iffalse
\input{sections/discussion}
\fi

\input{sections/evaluation2}

% \input{sections/result}

\input{sections/result2}

\input{sections/related_work2}

\input{sections/conclusion}

% or


% \appendices

% % you can choose not to have a title for an appendix
% % if you want by leaving the argument blank
% \section{}

\ifCLASSOPTIONcaptionsoff
  \newpage
\fi



%\bibliographystyle{IEEEtran}

\bibliographystyle{IEEETrans}
\bibliography{ref}


% \end{thebibliography}


% \iffalse
% \end{IEEEbiography}


% % insert where needed to balance the two columns on the last page with
% %\newpage

% \fi


%\vfill


\end{document}

%% file: sections/introduction.tex
\section{Introduction}\label{sec:intro}
% PostGis~\citep{postgis}
%Organizing and managing data records are always fundamental operations in a database system. As for data records with a explicit primary key, data are naturally sorted by the key value. However, considering data records with multidimensional features (e.g., spatial-temporal data), it is non-trivial to design an ordering which may tremendously influence downstream tasks such as query and retrieval. Apart from other mapping functions, 

\IEEEPARstart{A} space-filling curve (SFC, for short) 
%corresponds to a mapping function $T: \mathbf{x} \mapsto v$ that 
is a way to map a multi-dimensional data point $\mathbf{x}$ to a one-dimensional value, say $v$ that can be represented by a mapping function $T: \mathbf{x} \mapsto v$.
SFC mappings been widely used for multi-dimensional indexing.  The idea is to first map multi-dimensional data points to one-dimensional values using the SFC mapping function,
% (which are one-dimensional)
and then use one-dimensional indexing methods, e.g., a conventional B-Tree~\cite{bayer1997universal} or any of the recent learned indexes~\cite{kraska2018case, DBLP:conf/sigmod/DingMYWDLZCGKLK20, ding2020tsunami,DBLP:journals/pvldb/MarcusKRSMK0K20}, to index the mapped values.
This has been explored both in the literature~\cite{faloutsos1988gray, faloutsos1989fractals, lee2010z, yiu2008b, zhou2020data,wang2019learned, qi2020effectively} and 
%in industry (
by various database systems,
%such as 
e.g., PostgreSQL~\citep{postgis}, Amazon DynamoDB~\citep{dynamodb}, Apache HBase~\citep{nishimura2011md}, and many other systems.

\begin{figure}%[ht]
% \vspace*{-2mm}
	\centering	\includegraphics[scale=0.38]{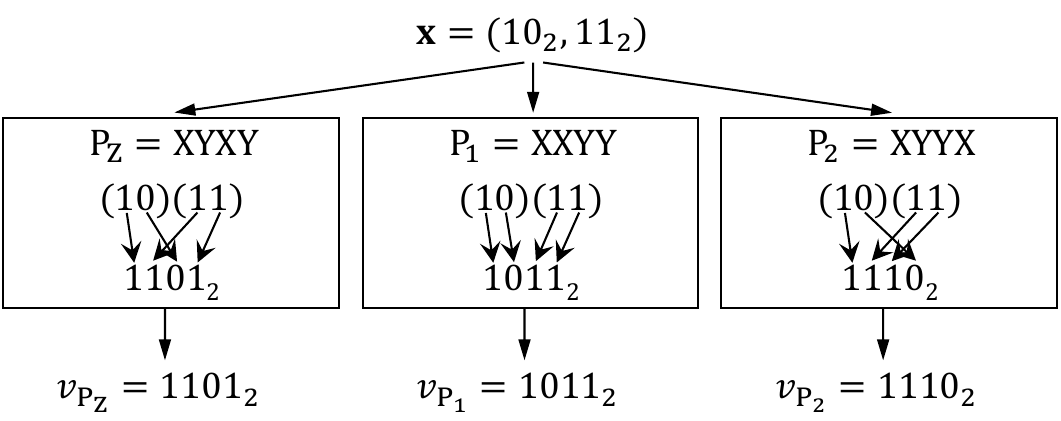}
	\vspace*{-1mm}
	\caption{The example illustrates mapping a point using various Bit Merging Patterns (BMP), where each mapping corresponds to a difference SFC. The 
    illustrated BMP mappings are as follows: $\texttt{P}_Z$ ({\tt XYXY}) is the BMP of the Z-curve, $\tt XXYY$ is the BMP of the saw-tooth curve (a column-wise scan~\cite{jagadish1990linear}), and $\tt XYYX$ is the third BMP.
    % \walid{Is the above correct? I believe XXYY is the saw-tooth SFC not the C-curve. Please check.}
% 	\hanmao{For the BMPs, I would prefer if you write $\tt XYXY$, $\tt XXYY$, $\tt XYYX$ instead.}
	}
	
	\label{fig:different_bit_patterns}
	\vspace*{-2mm}
\end{figure}

There are extensive studies on designing SFCs, e.g., the Z-curve~\citep{orenstem1984class, orenstein1986spatial, orenstein1989redundancy}, the C-curve~\citep{jagadish1990linear}, and the Hilbert curve~\cite{jagadish1990linear, mokbel2003analysis, moon2001analysis, kipf2020adaptive}. 
\revision{
% \marginpar{\#1 W4D1}
For example, the Z-curve adopts a  \emph{bit interleaving} mapping 
%schema 
scheme~\citep{samet2006foundations} that first converts the dimensions of input data to \emph{bit strings}. Refer to  Figure~\ref{fig:different_bit_patterns} for illustration, where  data point ${\bf x}=(2,3)$ is converted into its two corresponding binary strings (one string per coordinate) with 2 bits for each dimension: $({\tt 10}_2, {\tt 11}_2)$. Then, the bit interleaving merges bits alternatively from the different bit strings to form one SFC value (in Figure~\ref{fig:different_bit_patterns}, the bit interleaving %corresponds to 
adopts the {\tt XYXY} merging scheme that merges the {\chengr bit strings} {\tt XX} and {\tt YY} to the SFC value {\tt XYXY}, e.g., mapping Point ${\bf x}$ to ${\tt 1101}_2$).}

However, one common problem is that each  SFC has its own fixed mapping scheme/function that cannot be adjusted to fit  different {\chengr datasets}. 
%$f$ with which an input $\mathbf{x}$ will have a predetermined $v$ and not adjustable based on the database. For example, the Z-curve strictly order data points in a zig-zag way, as shown in Figure \ref{fig:popular_sfcs}(b).
The choice of one SFC for a dataset  significantly affects  query performance, and no single SFC can dominate the performance for all datasets and all query workloads (as shown in Fig.~\ref{fig:motivation_non_universal}). 
% \walid{Can we have a reference here to support the above statement?}
%
%There are also studies on  analysing and selecting the optimal SFC~\cite{moon2001analysis, xu2014optimality} for multidimensional data. 
%
%However, these studies do not design a new SFC by taking into account of the data  and query workload properties. 
%
\revision{%To this end, 
To tailor a new SFC to fit  the data  and query workload properties,
QUILTS~\citep{nishimura2017quilts} 
%proposes the first solution by 
extends 
% Z-curve's mapping scheme called
bit interleaving
% ~\citep{samet2006foundations}, 
% which converts the dimensions of input data to bit strings and alternatively merges bits from different bit strings to form an SFC value, 
%There are multiple 
by considering other ways of merging bit strings. For example, instead of merging bits following {\tt XYXY}, we can merge bits by following {\tt XXYY} or {\tt XYYX} to generate different SFC values at different regions of the multi-dimensional data set. 
%described in Figure~\ref{fig:different_bit_patterns}).
% \walid{Figure 1 does not explain how the QUILTS paper works, so it is confusing to refer to it here.}
Each pattern of merging bits is termed a \emph{bit merging pattern} (BMP, for short), where each BMP can describe a different SFC (as will be explained in Section~\ref{sec:preliminary} in greater detailed).}
%
% which defines
% a sequential process of generating the merged string for a multi-dimensional data point, i.e., deciding for each bit in the merged string, which dimension and which bit to take from a bit string of the dimension. 
% %
% For example, the BMP for Z-curve is $\tt XYXY$ for a 2 dimensional data point with bit strings $\tt XX$ and $\tt YY$.
% , which is a string deciding how to merge bits.  
QUILTS evaluates all the candidate SFCs described by BMPs based on a given workload and data, and selects the optimal one using heuristic methods.
QUILTS proposes to use multiple SFCs at the same time to index one data set so that the resulting mixed SFC is  query-aware and is skew-tolerant  for a given query pattern. However, the resulting SFC is static and hence does not change if the data distribution or the query workload changes over time. 
QUILTS makes the first attempt to utilize data and query workload properties to  select an optimal SFC. 
However, like other SFCs, {\chengr QUILTS} applies a single BMP for the entire data space (i.e., QUILTS applies one BMP to compute the SFC values of all data points). 
%\walid{This above sentence is incorrect. Quilt uses multiple BMPs that corresponds to different SFCs. For example, assume that x = $x_1x_2x_3x_4$, y = $y_1y_2y_3y_4$. QUILTS can map the first 2 bits $x_1x_2$ according to a z-curve BMS and the second $x_3x_4$ according to the C-curve according to the underlying query and data distributions. However, once this mapping is done, it is static and cannot change, and hence a justification for the research in this paper. Please rephrase accordingly.} 
Optimal SFCs may differ for different data subspaces. A detailed example of this situation is given in Section~\ref{sec:limitation} that illustrates this problem.
% Similar to all the other SFCs, QUILTs applies an SFC with a single BMP for the whole data space.
% %, which may not perform optimally. 
% However, the optimal SFCs for different data subspaces may be different. 
% For example, 
% in Figure~\ref{fig:compare_bmtree_quilts}, Z-curve works optimally for the window queries represented by blue \revision{($2 \times 2$)} rectangles  while C-curve is optimal for the yellow \revision{($1 \times 4$)} rectangles.
\iffalse
For example, in Figure~\ref{fig:compare_bmtree_quilts}, the Z-curve works best for queries with \revision{$2 \times 2$} blue rectangles, while C-curve is optimal for the query with  a \revision{$1 \times 4$} yellow rectangle. 
No single SFC can achieve the best performance for both types of queries.
\fi
% on different subspaces. 
% (2) The heuristic rules do not fit the realistic application. When there is more than one query type, heuristics loses the effectiveness and may need to conduct a brute force algorithm comparing all SFCs in the candidate SFC set, which is time-consuming.
%
Another issue of {\chengr QUILTS} is that it does not provide an effective way of generating and evaluating candidate SFCs. The heuristic rules used by QUILTS are designed for very specific types of window queries (e.g., with a fixed area) and do not fit  general query processing scenarios, where the workload includes more than one query type (with different areas or aspect ratios).
% of a query rectangle). 
For example,  a heuristic rule used by QUILTS assumes that grid cells intersecting with a query should be continuous in  SFC order, which 
%could never be met with 
may not hold for queries with different aspect ratios (this is   elaborated on in greater detail in Section \ref{sec:limitation}). Further, QUILTS does not consider the scenario where the distributions  of data and/or queries are changed, which results in sub-optimal query performance if the SFC is not updated.
% (details are given in Motivation 3 of Section \ref{sec:limitation}). 

% {\bf Gao you can think if you can make a summary of Motivation 2 and put here. if not,  we could also refer to motivation 2}

%A brute force algorithm may need executing when these rules are broken, which could be time-consuming.

%These limitations motivate us to develop a new SFC design to solve concerned problems.
\bluecolor{
To address the limitations of an SFC with a single BMP, i.e., a single mapping scheme, our idea is to design different BMPs for different subspaces based on the data and query workload features, aiming to optimize  query performance.
%\walid{I believe this is the wrong motivation statement because QUILTS already do that. I believe the correct motivation is to focus on the dynamic nature of query workloads and once the query workload changes, the BMTree can change the SFC used accordingly. This should be a relatively easy fix in the writing of this introduction.}
The resulting SFCs 
used in the BMTree would comprise multiple BMPs, each corresponding to a subspace that we refer to by  
%
%For limitation one, we introduce a new SFC design machinery that adopts different BMPs for different subspaces (since different subspaces may have different underlying data and query distributions) and then combines the functions for different subspaces into one for the whole space (by ensuring that the functions for different subspaces are adopted in a systematic way). 
%
a \emph{piecewise} SFC. 
We focus on three aspects of the piecewise SFC construction framework: 
(1) Designing the piecewise SFC: We propose a binary tree structure to help construct and design a piecewise SFC. (2) Learning the piecewise SFC. We design a data-driven learning-based method that can automatically construct the piecewise SFC according to the query workload to optimize query performance. (3) Updating piecewise SFC: We develop a mechanism aiming to efficiently update the piecewise SFC w.r.t. the updated query and data scenario.
}

% \walid{One important dimension that you need to elaborate on: 1) According to my recollection, Quilts only use two SFCs, the z-curve and the c-curve, and do not consider any other SFCs. Also, the selection of the mix of Z- and C-curve bits is static, i.e., once chosen it does not change even if the query workload changes. These can be strong selling points for the BMTree.}
% To design piecewise SFCs, we address the following three challenges. 
%since different BMPs may be adopted for different subspaces. 
%
%, and it would potentially have better performance than a normal SFC. 
\iffalse
Figure \ref{fig:compare_bmtree_quilts} illustrates an 
%intuitive 
example of piecewise SFCs,
%. In the piecewise SFC, 
where we choose Z-curve for the left half subspace and C-curve for the right half subspace, thus achieving the optimal performance for both blue and yellow query rectangles (which needs 12 cell scans). 
\fi

\iffalse
\begin{figure}[ht]
	% \vspace{-3mm}
	\centering
	\small
    \includegraphics[width=0.9\linewidth]{VLDB_Learned_SFC/figures/intro_quilts_piecewise.pdf}
	% \vspace*{-4mm}
	\caption{Comparison between QUILTS and a piecewise SFC example, where the green %shade highlights the grids that 
	cells are scanned (14 for 
	%the Z-curve 
	QUILTS and 12 for the piecewise SFC) for queries in the blue and yellow dashed rectangles.
	%are two query types.
	}\label{fig:compare_bmtree_quilts}
	% \vspace*{-3mm}
\end{figure}
\fi

%
\subsection{Designing Piecewise SFCs}
% 
% It is a new and open problem of 
How to design effective BMPs for different subspaces, while guaranteeing desirable properties of the overall mapping function to indexing data, is non-trivial.
%
%A BMP represents 
%
%Recall that the bit merging process is 
%
%to merge multiple bit strings (each for one dimension of a multi-dimensional data point $\mathbf{x}$) into one (representing the mapped value $v$). It can be regarded as 
%
%a sequential process of sequentially deciding for each bit in the merged string, which bit to take from a bit string for one dimension of a multi-dimensional data point $\mathbf{x}$. 
%
To achieve that, we propose 
%a novel idea of 
to 
seamlessly integrate  subspace partitioning and BMP generation.  
We develop a new structure termed the  \emph{Bit Merging Tree} (BMTree, for short)
%, a binary tree structure, 
to recursively generate both 
the 
subspaces and the corresponding BMPs.
% (see in Section~\ref{sec:why_bmtree}). 
%Its root node corresponds to all the data points, and  the first bit from a selected dimension is assigned to the root node to represent the first bit of BMPs; The bit of the selected dimension is used to generate two data subspaces, one for data points with value 0 on the selected bit and the other for data points with value 1 on the bit; Each subspace corresponds to a node and we choose the next dimension and assign its first unused bit to the node; The bit is the next bit of BMPs for the data at the subspace, and will be used to further partition the subspace. The procedure continues until all bits of BMPs (or the merged bit strings) have been specified. In this way, we achieve the goal of generating different  BMPs for different subspaces. 
% 
In the BMTree, (1)~Each node represents a  bit from the binary string of the selected dimension, 
% for BMPs, 
and its bit value ($0$ or $1$) 
% plays the roles of 
partitions the data at the node into two child nodes, and (2)~Each leaf node represents a subspace, and the sequence of bit string from the root to the leaf node represents the BMP for the subspace. 
% 
%Specifically,  we choose a dimension  and a bit from the selected dimension sequentially, and the selected bit (and dimension) plays two roles: 1)~it becomes a bit of the BMPs (the merged bit strings) for the current data points, and 2)~the bit of the dimension is used to generate two data subspaces, one for data points with value 0 on the selected bit and the other for data points with value 1; for each subspace, we continue this process recursively to choose the next bit until all bits of the merged bit strings have been specified. In this way, we achieve the goal of generating different  SFCs for different subspaces.
 %
% We propose the \emph{Bit Merging Tree} (BMTree), a binary tree structure, to implement this idea, where (1) each node represents a selected bit from one dimension for BMPs, and meanwhile its value (0 or 1) plays the roles of determining one of the two child nodes 
 %subspaces 
% the data point falls in and (2) each leaf node represents a subspace, and the sequence of bit string from the root to the leaf node represents the BMP for the subspace. 

% following different ways of merging the remaining bits (i.e., using different BMPs). 
% 
% 
% 
% Second, the piecewise SFC design makes it challenging to guarantee two desirable properties of the overall mapping function: 

Further, we prove 
that
the Piecewise SFC modeled by the BMTree maintains two desirable properties: Monotonicity~\citep{lee2007approaching} and Injection. 
Monotonicity is a desirable property for designing window query algorithms,
% Intuitively, monotonicity will 
which guarantees that the SFC values of data points
% falling 
in a query rectangle fall in the SFC value range formed by two boundary points of the query rectangle. 
Combining different SFCs {\CHENG from different subspaces to obtain a final SFC for the whole space} may lead to the risk of breaking the monotonicity property. Similarly, it may also lead to 
%the 
an
injection violation, i.e., that the mapping function may not return a unique mapped value for each input. 
We construct the BMTree in a principled way 
%s.t. 
such that
the two properties are guaranteed.
% Details can be found in Section~\ref{sec:why_bmtree}. 

\subsection{Learning Piecewise SFCs}

To address the limitation of heuristic algorithms in the SFC design, 
%when building the BMTree, 
we propose to model 
% the process of 
building the BMTree
%, i.e., designing and splicing BMPs, 
as 
a Markov decision process (MDP, for short)~\citep{puterman2014markov}, aiming to develop data-driven solutions 
%to 
for
designing suitable BMPs for different subspaces. Specifically, we define the states, the actions, and the rewards signals of the MDP framework
%based on the BMTree building procedure, in which optimizing the rewards is well aligned 
to build the BMTree such that  
% the piecewise SFCs modeled by the
the generated
BMTree can optimize 
%the 
query performance. 
 We leverage  reinforcement learning and  Monte Carlo Tree Search (MCTS, for short)~\citep{browne2012survey}, to learn a performance-aware policy and avoid local optimal 
 settings. 
 To improve 
 %the 
 performance, we design a greedy action selection 
 %algorithm for the MCTS algorithm.
 algorithm for  MCTS.
 %when the action space size is enormous. 
Moreover, to improve 
%the 
training efficiency, we define a metric termed \emph{ScanRange} as a proxy of the query  performance (e.g., I/O cost 
%and 
or
query latency), and apply ScanRange for the 
%reward computation.
computation of rewards.

\bluecolor{
\subsection{Updating Piecewise SFCs}

In situations where the distributions of data and queries change~\cite{lu2018learning}, the previously learned module faces an issue of 
having 
sub-optimal performance. 
Fully retraining a BMTree poses efficiency challenges due to the BMTree training cost, and the need to update all SFC values of data points maintained in the index. To address this 
issue, 
we propose a novel mechanism aligned with the BMTree structure that enables partial retraining, 
and hence
reducing the overall cost.
% Fully retraining a BMTree not only brings a training efficiency issue, but the following updating of SFC values for index maintenance is also costly. Instead of training a new BMTree from scratch when a distribution shift happens,
% we develop a novel mechanism that follows the design of the BMTree structure and allows partial retraining of the BMTree to facilitate the retraining cost. 
First, we introduce a distribution shift score to quantify 
the 
shift degree, and  decide if retraining is necessary.
% and decide if a retraining should happen.
Then, we 
% follow the design of BMTree 
% \walid{What do you mean by: we follow the design of BMTree? this is unclear. Please rephrase.}
% and 
develop an optimization potential score 
to identify which nodes of BMTree, when optimized, can significantly enhance query performance.
% measuring optimizing which node can achieve notable enhancement of query performance.
% ,  deciding which portion of the BMTree should be retrained.
We partially delete the nodes of the BMTree  
that need 
to be retrained, and develop an adapted training reinforcement learning environment (with the states, actions, and rewards adapted for partial retraining) and regenerate the BMTree 
%w.r.t. 
with respect to
the updated data and query 
%workload. 
workloads.
% First, we introduce a method that efficiently detects the a portion of BMTree to be retrained to maximize the performance improvement. We measure the distribution shifts of data and query based on the structure of BMTree (See in Section~\ref{sec:retrain_detect}).  We then adapt the introduced BMTree RL training environment which is then applied to regenerate the BMTree w.r.t. the updated distribution.
}

The main contributions of this paper are as follows:

\noindent(1) 
We propose the idea of piecewise SFCs
% for designing SFCs, 
that allows to 
design different BMPs for different subspaces by considering the data and query workload properties to deal with non-uniformly distributed data and query workloads. %situation.
%, aiming to achieve better query performance. 
% To the best of our knowledge, the idea is new in the literature. 

\noindent(2) To design piecewise SFCs, we 
%propose 
introduce
the 
% Bit Merging Tree (BMTree)
BMTree
to partition the
% whole 
data space into subspaces, and generate {\chengr a BMP} for each subspace.  We prove that the piecewise SFC represented by a BMTree satisfies two properties, namely injection and monotonicity.
% , which are important for designing query processing algorithms.

%We study the SFC Design problem and design a piecewise SFC architecture to achieve potentially better performance. We model the whole mapping function of the piecewise SFC as a binary tree named BMTree. We prove BMTree maintains two properties, namely injection, and monotonicity.

\noindent(3) To build a BMTree, we develop an 
% reinforcement learning 
RL-based solution by modeling BMP design as a MDP, and design an MCTS-based BMTree construction algorithm.
% To better explore child nodes during rollouts, we design a greedy algorithm to guide the action selection. We also develop the ScanRange metric to measure the window query performance to speed up the learning procedure. 
% To achieve good performance, w
% We design a greedy based action selection algorithm to guide MCTS. 
%the RL when the action space size become enormous. 
We develop 
%a fast computing 
the  ScanRange metric to efficiently measure the window query performance on an SFC. As a result, the ScanRange metric speeds up the learning procedure.
% \walid{In the above sentence, it is unclear: Does the window query performance speeds up the learning procedure, or does the ScanRange metric speeds up the learning procedure? Please rephrase this sentence to make this clear.}
% reward computing during the learning procedure.
% To the best of our knowledge, this is among the first  learning based approaches for designing an SFC.

%\noindent(3) We introduce a metric named ScanRange reflecting the window query performance. Optimizing ScanRange consistently optimizes the query performance of different index structures.

\bluecolor{\noindent (4) To efficiently update a BMTree, we develop a mechanism that allows partially retraining 
of 
the BMTree when data and/or query distributions shift, and 
%achieves 
enhances
the query performance with reasonable retraining costs. 
% including the retraining sub-space detection and retraining procedure.
}

\noindent(5) We integrate our learned SFCs into the B$^+$-Tree index inside PostgreSQL and 
% , and  compare with baseline SFCs on the querying performance under PostgreSQL. 
% We also apply our learned SFCs to a 
%recent 
inside the learned spatial index RSMI~\cite{qi2020effectively}. 
Experimental results under both settings consistently show 
that the 
BMTree outperforms the baselines in terms of query performance. \bluecolor{Further, the partial retraining mechanism achieves notable performance enhancement that is competitive to full retraining while achieving 
%an 
over  $2\times$ speedup compared to full retraining.}
% in the experimental study. 

\bluecolor{Compared to the previously published paper~\cite{li2023towards},  this paper introduces over $35\%$ new content. We extend the BMTree by incorporating a novel reconstruction mechanism that enables it to quickly adapt itself to distribution shifts and achieves better query performance, which is not supported by other SFC methods.
% \walid{Do you mean by the above the incremental training? If yes, state so as this is significant. Also, detail more the experiments below, otherwise, the reviewers may argue that the contributions may not be enough.}
We also include additional experiments to evaluate the proposed mechanism under different shift settings, including data shift, query shift, and their combination. }

%% file: sections/preliminary2.tex
\section{ PROBLEM STATEMENT \& PRELIMINARIES }
\label{sec:preliminary}

% We first define the problem of SFC design and then discuss previous work for this problem.

%\subsection{Notations}
%\label{sec:notation}

\subsection{Problem Definition}\label{sec:problem}

Let $\mathcal{D}$ be a database, where each data point $\mathbf{x} \in \mathcal{D}$ has $n$ dimensions, denoted by $\mathbf{x} = (d_1, d_2, \ldots, d_n)$.
% Each dimension $x_i$ ($1 \le i \le n$) can be converted to a bit sequence as follows. 
For ease of presentation, we  consider only 2-dimensional data points $\mathbf{x} = (x, y)$, 
% Note that the proposed solution 
that can be easily extended to $n$ dimensions. 
% Data 
$\mathbf{x}$ can be converted to bit strings as: $\mathbf{x} = \left((x_1x_2 \ldots x_m)_2, (y_1y_2 \ldots y_m)_2\right)$.
%
%
%  \begin{equation}
%  	\label{eq:binary_encode}
%  	\mathbf{x} = \left((x_1x_2 \ldots x_m)_2, (y_1y_2 \ldots y_m)_2\right)
% %  	x_i = (b^i_1b^i_2 \ldots b^i_m)_2
% %  	, \; m = \left\lceil log_2(x_i) \right\rceil,
%  \end{equation}
where each $x_i$, $y_j$ ($1 \le i,j \le m$) 
%denotes a binary number 
%
are 0 or 1
(i.e., $x_i, y_j \in \{0, 1\}$) and $m$ is the 
%predecided 
length of the bit string that is dependent on 
%carnality 
the cardinality 
of the dimensions 
%$i$
$x$ and $y$. Take $\mathbf{x} = (4, 5)$ for example, it can be converted in base 2 to $\mathbf{x} = (100_2, 101_2)$. 
% \commentI{[How to handle fractional values in $\mathbf{x}$, how to write this in prior works?]} 
%In a generalized situation, 
In previous studies on SFC based multidimensional indexes, e.g.,~\citep{bayer1997universal,samet2006foundations,skopal2006new}, values of data points are typically mapped to fine-grained grid cells for discretization.
% Note that  $x_i$ can be transformed into an integer if it is not originally. Data space can be partitioned into small cubes and the cube's index can be used to represent $x_i$ as an integer \change{(So called Z-address in~\citep{bayer1997universal})}.
%
%The core problem of 
SFC maps %an $n$-dimensional 
$\mathbf{x}$ into a scalar value $v$ (called SFC value) with a mapping function 
% $T$ (i.e.,
$C(\mathbf{x}) \rightarrow v$.
% ). 
%
An SFC value $v$ can be 
%regarded 
used as the key value of data $\mathbf{x}$ 
%, and 
%
%the SFC value  can be used 
to determine the order of $\mathbf{x}$ in $\mathcal{D}$.
%
%Therefore, the SFC values can be applied to construct multidimensional data access structures on top of a one-dimensional index such as B+ tree.

%We study the Space-Filling Curve (SFC) problem, which is defined as follows.

\begin{problem}[SFC Design]
Given a database $\mathcal{D}$ and a query workload $Q$, we aim to develop a mapping function $T$ that maps each data point $\mathbf{x} \in \mathcal{D}$ into an SFC value $v$, such that 
 with an index structure (e.g., a B$^+$-Tree) built on the SFC values of that data points in $\mathcal{D}$, the query performance (e.g., 
 %in terms of 
 I/O cost and querying time) on $Q$ is optimized.
%

%is able to produce excellent query performance (e.g., I/O cost and querying time) on $Q$.
\label{prob:sfc}
\end{problem}

\iffalse
Apart from database $\mathcal{D}$, the SFC problem takes as input (1)~a query workload $Q$, and (2)~an index structure. 
% We cannot find a real-life query workload in the literature, and by following the previous work
\revision{
% \marginpar{\#1 D3}
We {\chengr follow the previous work~\cite{nishimura2017quilts,qi2020effectively} and generate window query workloads}
% We generate window queries 
with three different distributions, including uniform (\texttt{UNI}), Gaussian (\texttt{GAU}), and skew (\texttt{SKE}) distributions.}
% , to form query workloads. 
%For the latter, 
We adopt the B$^+$-Tree and RSMI~\cite{rsmi2020code} that represent 
%non-learning 
classic and learning-based index structures, respectively. %More details can be found in Section~\ref{sec:setup}.
\fi

\subsection{Preliminaries on SFC}
\label{sec:prior}

% We first present two desirable properties for a mapping function $T$, called  \emph{injection} and \emph{monotonicity}. Then, we present the curve design methods in the Z-curve and Quilts, which also satisfy the two properties.
We present two desired properties for a mapping function $T$, namely \emph{Injection} and \emph{Monotonicity}. Then, we describe the curve design methods in the Z-curve and Quilts that also satisfy these properties.

% \smallskip
\noindent\textbf{Injection\footnote{This property is defined on discretized input. No injection is guaranteed in continuous space since no bijection mapping exists between {$\mathbb{R}$} and {$\mathbb{R}^n$}~\cite{peano1890courbe}.
}.}
% 
% 
%
%Consider  $\mathbf{x}$ as bit sequence encoding, 
An SFC design is expected to satisfy the 
%criterion 
%property named injection, which
injection property that
guarantees a unique mapping from $\mathbf{x}$ to $v$. This is to ensure that each SFC value $v$ can be used as a key value of $\mathbf{x}$ for ordering and indexing data.
%and indexing.  
%querying.
% A mapping function $T(\cdot)$ is injective if a unique value $v$ is mapped to 
%by 
% exactly one data point $\mathbf{x}$. A SFC mapping function needs to guarantee to be injective.
%to be an SFC mapping function. 
It is defined as follows.
\begin{definition}[Injection] Given a function $C: \mathbf{x} \rightarrow v$, $C$ is injective if $\mathbf{x}$ maps to a unique value $v$, s.t. $\forall \mathbf{x_1} \not= \mathbf{\mathbf{x_2}}, C(\mathbf{x_1} ) \not= C(\mathbf{x_2} )$.
\end{definition}
\revise{\noindent The injection property is desirable for an index to narrow the search space for better query performance. Consider an extreme situation where all data points map to the same value. Then, an index based on the SFC values cannot narrow the search space for a query.}

% \smallskip
\noindent\textbf{Monotonicity.} 
%, and guarantees the query results exactly when an index structure is built based on these SFC values.
The monotonicity~\citep{lee2007approaching} is defined as follows.

\begin{definition}[Monotonicity] Given two $n$-dimensional data points (denoted as $\mathbf{x}^{\prime}$ and $\mathbf{x}^{\prime\prime}$), whose SFC values are denoted by $C(\mathbf{x}^{\prime})$ and $C(\mathbf{x}^{\prime\prime})$. When a mapping function $C$ holds monotonicity, if $d^{\prime}_i \ge d^{\prime\prime}_i$ is satisfied for $\forall i \in [1, n]$, it always has $C(\mathbf{x}^{\prime}) \ge C(\mathbf{x}^{\prime\prime})$. 
\end{definition}

Maintaining monotonicity is a desirable  
%important 
%criterion 
property
for mapping data points to SFC values 
as explained below. 
%with special benefits for query algorithm design.
%
% Figure~\ref{fig:monotonicity} illustrates an example of 2-dimensional window query. 
Assuming the origin of the space is at the lower left, given a 2-dimensional window query represented by its minimum (bottom-left corner) and maximum (top-right corner) points (i.e., $\mathbf{q}_{min} = (x_{min},y_{min})$, $\mathbf{q}_{max} = (x_{max},y_{max})$). Let $\mathcal{P} = \{(x, y) \; | \; x_{min} \le x \le x_{max}, y_{min} \le y \le y_{max} \}$ denotes the query results bounded by the query window.
% Let $\mathbf{p}=(x, y) \in P$ denote the query results bounded by the query window (i.e., $x_{min} \le x \le x_{max}$ and $y_{min} \le y \le y_{max}$)
%
%With monotonicity holds, 
If the monotonicity property holds,
the result points in $\mathcal{P}$ are within the range bounded by 
%be found based on searching with \mathbf{q}_{min}$
the SFC values of $\mathbf{q}_{min}$ and $\mathbf{q}_{max}$. The reason is that for any data point $\mathbf{p} \in \mathcal{P}$, whose SFC value $C(\mathbf{p})$ 
always holds that
% is still within the SFC values of the two window points, i.e.,
$C(\mathbf{q}_{min}) \le C(\mathbf{p}) \le C(\mathbf{q}_{max})$. 
The property is desirable since it enables us to design simple and efficient algorithms for processing a window query by checking data points whose SFC values are within the bounded range only;
Otherwise, the algorithm does not work. 
For example, the Hilbert curve and its variants~\cite{jagadish1990linear, mokbel2003analysis, moon2001analysis} do not satisfy the monotonicity property, %in their mappings, 
which makes it hard to identify the scanning range for a window query in the space of their SFC values, and requires maintaining additional structure to design more complicated algorithms~\citep{lawder2001querying}.
% which makes it complicated to design algorithms for identifying the scanning range in the space of their SFC values, \change{and require additional structure and corresponding algorithm in order to conduct window query~\citep{lawder2001querying}.}

%It strongly eases implementing a window query algorithm.

% \smallskip
\noindent\textbf{Computing SFC values in the Z-curve~\cite{orenstem1984class, orenstein1986spatial, orenstein1989redundancy} and in QUILTS~\cite{nishimura2017quilts}.} 
% A mapping function is to map a data point into a scalar value $v$.
%
Both Z-curve and 
%its extension 
QUILTS 
%on the other hand, 
guarantee the injection and monotonicity properties. Figure~\ref{fig:different_bit_patterns} 
%provides an example to show
examplifies
 how the Z-curve and QUILTS map a data point $\mathbf{x}$ to a scalar SFC value $v$. 
 % They map $\mathbf{x}$ in its bit string encoding to $v$, which is a merged sequence %regarded as 
 % of binary numbers.
%
% and thus the query results based on their SFC values are generally approximate. 
The curve design in the Z-curve and QUILTS are presented as follows.

The SFC value of $\mathbf{x}$ in the Z-curve is %generated 
computed
via bit interleaving, which generates a binary number consisting of bits (0 or 1) filled alternatively from each dimension's bit string.
% , namely the interleaving sequence. 
The Z-curve value of a 2-dimensional data point $\mathbf{x}$ is computed by Function $C_z$:
\begin{equation}\label{eq:z_value}
C_z(\mathbf{x}) = (x_1y_1x_2y_2\ldots x_my_m)_2
% v_z = (b^1_1b^2_1 \ldots b^n_1 \; b^1_2b^2_2 \ldots b^n_2 \; \ldots \; b^1_{m}b^2_{m} \ldots b^n_{m})_2.
\end{equation}
% The main idea of designing a mapping function in Z-curve is to alternatively scan each dimension in $\mathbf{x}$ and select the current highest bit of each dimension, to form a long binary number $v_z$. 
%computed by Z-curve. 
It assumes that all dimensions have the same bit-string length, and the zero-padding technique
% ~\cite{albawi2017understanding}
is usually applied to fit the length equally by padding zeros at the head of each bit string.
%
% An SFC value $v_q$ of 
%

QUILTS generalizes the bit interleaving pattern of the Z-curve to more general \emph{bit merging pattern}, each of which represents a way of merging bits. 
% \change{It releases the requirement of equal bit length.}
%We take two-dimensional data for example,  
For example, for two-dimensional data,
QUILTS defines a bit merging pattern as follows.

% m=3
% p=yxyyxx

% A bit merging pattern is a bitstring P of length 2m over the alphabet {x,y} such that it contains exactly m x's and m y's.

% Given any bitstring P = (p_1,p_2,..., p_2m), we define the SFC T_P((x1..xm),(y1..ym))=(z1,z2,...z_2m) as follows.
% Since P contains exactly m x's, let I={i_1,...,i_m} be the set of indices that such that p_{i_j}=x. Then we set z_{i_j}=x_j

% T_p(x1x2x3,y1y2y3)=(y1x1y2y3x2x3)

\begin{definition}[Bit Merging Pattern]\label{def:bit_merging_pattern}
% \change{
% A bit merging pattern (BMP) is a string $P$ of length $2m$ over the alphabet $\{\texttt{X}, \texttt{Y}\}$ s.t. it contains exactly $m$ \texttt{X}'s and $m$ \texttt{Y}'s. Given a $P = p_1 p_2 \ldots p_{2m}$, we define the SFC with $P$
% % , $T_P(\mathbf{x}) = (b_1 b_2 \ldots b_{2m})_2$ 
% \begin{equation}\label{eq:bit_merging_value}
% T_P(\mathbf{x}) = (b_1 b_2 \ldots b_{2m})_2
% \end{equation}
% as follows. Since $P$ contains exactly $m$ $\texttt{X}$'s, let $I = [i_1, \ldots, i_m]$ be the list of ascending ordered indices that $p_{i_j} = \texttt{X}$. Then we set $b_{i_j} = x_j$. We assign $y_j$ in the same way to $(b_1 b_2 \ldots b_{2m})_2$ and finish value computing. For example, given a $P = \texttt{XXYY}$, the value of data record $\mathbf{x}$ computed by $T_P$ is $T_P(\mathbf{x}) = (x_1x_2y_1y_2)_2$. Notice both $x$ and $y$ are subsequences of $T_P(\mathbf{x})$.
% %
%  }
%  \end{definition}
%  \hanmao{
%  Please write as follows.
% 
 A bit merging pattern (BMP) is a string $\tt P$ of length $2m$ over the alphabet $\{\texttt{X}, \texttt{Y}\}$ s.t. it contains exactly $m$ \texttt{X}'s and $m$ \texttt{Y}'s. Given a BMP point ${\tt P} = p_1 p_2 \ldots p_{2m}$, the SFC described by $\tt P$ is defined as follows.
 We set 
\begin{equation}\label{eq:bit_merging_value}
C_{\tt P}(\mathbf{x}) = (b_1 b_2 \ldots b_{2m})_2
\end{equation} according to the following rule:
% \begin{itemize}
    % \item 
    % $\bullet$
    (1) 
    Since $\tt P$ contains exactly $m$ $\texttt{X}$'s, we let $I = \{i_1, \ldots, i_m\}$ be the list of ordered indices such that $p_{i_\ell} = \texttt{X}$. Then, we set $b_{i_\ell} = x_\ell$ for $1\le \ell\le m$. 
    % \item 
    (2)
    Similarly, for the value of $y$, we consider $J = \{j_1, \ldots, j_m\}$ where $p_{j_\ell} = \texttt{Y}$, and assign $b_{j_\ell}$ the bit value of $y_\ell$.
% \end{itemize}
For example, given the BMP point ${\tt P} = \texttt{XXYY}$, the value of data point $\mathbf{x}$ computed by $C_{\tt P}$ is $C_{\tt P}(\mathbf{x}) = (x_1x_2y_1y_2)_2$. Notice that both $x$ and $y$ are subsequences of $T_{\tt P}(\mathbf{x})$.
 
\end{definition}
% ordering of bits from the dimensions (each corresponding to a \change{bit sequence}) of the input $\mathbf{x}$. By applying the pattern on  $\mathbf{x}$, a merged bit sequence $v$ will be generated as the SFC value of $\mathbf{x}$ under the pattern:
% \begin{equation}\label{eq:bit_merging_value}
% v_q = (b_1 b_2 \ldots b_M)_2, \; M=\sum^{n}_{i=1}m_i,
% \end{equation}
% where $m_i$ denotes the bit sequence length of the $i$-th dimension $d_i$, and $b_t \in \{0, 1\}$ ($1 \le t \le M$). 
%
% For example, the Z-curve corresponds to a bit merging pattern, with which we fill $x_1$ to $b_1$ and $y_1$ to $b_2$ in $v_q$, respectively, and fill other bits of $v_q$ in a similar way.
%
% To satisfy the monotonicity property, 
% \change{As discussed in \citep{nishimura2017quilts},} 
% A bit merging pattern needs satisfying the following condition: 
% for any two bits $b^i_{j^{\prime}}$ and $b^i_{j^{\prime\prime}}$ from the same dimension $x_i$, if $b^i_{j'}$ is placed before $b^i_{j^{\prime\prime}}$ in $x_i$, then $b^i_{j^{\prime}}$ should also be placed before $b^i_{j^{\prime\prime}}$ in $v_q$. 
% in each dimension, denoted by $x$, for any two bits $x_i$ and $x_j$, if $x_i$ is placed before $x_j$ in $x$ ($i < j$), then $x_i$ should also be filled in $v_q$ before $x_j$.   This condition also means $x$ and $y$ should be subsequences of $v_q$.

SFCs represented with different BMPs form an SFCs set. QUILTS considers this set and selects the optimal SFC evaluated on a given query workload as the output curve. We prove the monotonicity of SFCs with BMPs, which 
%is used to 
guarantees the monotonicity property of our method in Section \ref{sec:theoretical}. The detailed proof is given in~\cite{li2023towards}.
% {\color{blue}
\begin{lemma}[Monotonicity of SFCs with BMPs]
\label{lem:monotonicity_bit_merging_pattern}
An SFC with a BMP achieves the monotonicity property. 

\iffalse
\ifnum\extend=1
\begin{proof}
% 
% \bluecolor{The detailed proof can be found in \cite{li2023towards}.}

\iffalse
Given a BMP ${\tt P}$, we prove the monotonicity property in $2$-dimension space which is easy to extend to $n$.
Given $\mathbf{x} = (x, y) $ and $\mathbf{x}^{\prime} = (x^{\prime}, y^{\prime})$ satisfying $x \geq x^{\prime} \; \& \; y \geq y^{\prime}$, $T_{\tt P}(\mathbf{x}) = b$ and $T_{\tt P}(\mathbf{x}^{\prime}) = b^{\prime}$. (1) If $\mathbf{x} = \mathbf{x}^{\prime}$, then 
% the SFC with $P$ computes equal value 
$T_{\tt P}(\mathbf{x}) = T_{\tt P}(\mathbf{x}^{\prime})$; (2) if $\mathbf{x} \not= \mathbf{x}^{\prime}$, assume $x > x^{\prime}$. There is a smallest indice $s(x,x^{\prime})$ such that $x_{s(x,x^{\prime})} = 1 > x^{\prime}_{s(x,x^{\prime})} = 0$ while $x_i =  x^{\prime}_i$ when $0<i<s(x,x^{\prime})$. (i) If $y = y^{\prime}$, assume ${\tt P}$ put $x_{s(x,x^{\prime})}$ to bit $b_t$ of $T_P$, we have $b_t = 1 > b^{\prime}_t = 0$ while $b_j = b^{\prime}_j$ for all $0<j<t$, $T_{{\tt P}}(\mathbf{x}) > T_{{\tt P}}(\mathbf{x}^{\prime})$. (ii) If $y > y^{\prime}$, we pick the one from $x_{s(x,x^{\prime})}$ and $y_{s(y,y^{\prime})}$ which maps to the smaller index in $T_{\tt {\tt P}}$, the condition in (i) still satisfies and $T_{{\tt P}}(\mathbf{x}) > T_{{\tt P}}(\mathbf{x}^{\prime})$. Same proof applied when $x = x^{\prime} \; \& \; y > y^{\prime}$. With (1) and (2), lemma proved.
\fi
\end{proof}
\else 
\begin{proof}
     The detailed proof is given in \cite{jiangneng2023lsfc}.
\end{proof}
\fi
\fi
\end{lemma}

%% file: sections/method4.tex
\section{Motivation and Method Overview}
\label{sec:method}

% piecewise-elementary SFC?
% SFCs with multiple BMPs

% m=3
% p=yxyyxx

% A bit merging pattern is a bitstring P of length 2m over the alphabet {x,y} such that it contains exactly m x's and m y's.

% Given any bitstring P = (p_1,p_2,..., p_2m), we define the SFC T_P((x1..xm),(y1..ym))=(z1,z2,...z_2m) as follows.
% Since P contains exactly m x's, let I={i_1,...,i_m} be the set of indices that such that p_{i_j}=x. Then we set z_{i_j}=x_j

% T_p(x1x2x3,y1y2y3)=(y1x1y2y3x2x3)

\subsection{Motivations and Challenges}
\label{sec:limitation}

% % \smallskip
\noindent\textbf{Motivation 1: Piecewise SFC Design.} 
%
%SFCs defined by BMPs in 
QUILTS and earlier SFCs based on BMPs only use
one BMP to compute SFC values for all data points,
% . 
% To the best of our knowledge, existing SFC methods include 
% Z-curve
% % , Hilbert curve, 
% and QUILTS follow the elementary SFC design (QUILTS selects one single bit merging pattern as the output curve).
% \change{, Hilbert curve also applies a global order generate rule to the whole space}
%
% The existing SFCs with per unique BMP 
which may not perform well for query processing.

%when data and queries are not uniformly distributed, which is often the case in a real-life situation. An %intuitive 
%example is 

\begin{figure}[h]
	% \vspace{-3mm}
	\centering
	\scriptsize
	\begin{tabular}{c c c}
        \hspace*{-3mm}
		\begin{minipage}{0.3\linewidth}
			\includegraphics[width=\linewidth]{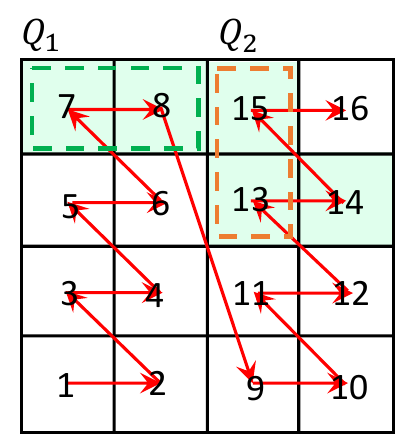}
		\end{minipage}
		&
  \hspace*{-3mm}
		\begin{minipage}{0.3\linewidth}
			\includegraphics[width=\linewidth]{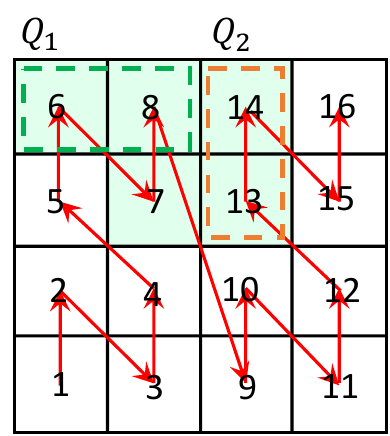}
		\end{minipage}
		&
  \hspace*{-3mm}
		\begin{minipage}{0.3\linewidth}
			\includegraphics[width=\linewidth]{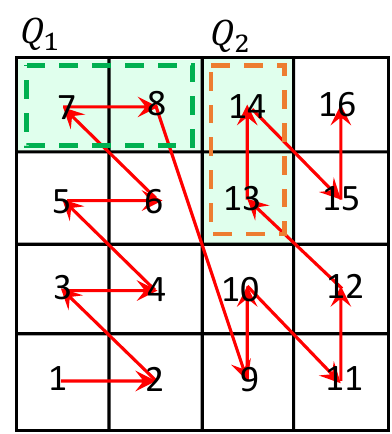}
		\end{minipage}
		\\
		(a) SFC-1 ({\tt XYYX})
		&
		(b) SFC-2 ({\tt XYXY})
		&
		(c) SFC-3 (piecewise)
		\\
		Grid scans: $2$ for $Q_1$, 
		&
		Grid scans: $3$ for $Q_1$, 
		&
		Grid scans: $2$ for $Q_1$, 
            \\
            $3$ for $Q_2$
            &
            $2$ for $Q_2$
            &
            $2$ for $Q_2$
	\end{tabular}
	% \vspace*{-3mm}
	\caption{Motivation for piecewise SFCs, 
    % grid scans number of the two queries are given in brackets. 
    SFC-1 is described by the BMP $\tt XYYX$
    % , 
    while SFC-2 
    % is described 
    by $\tt XYXY$. In contrast, SFC-3 (ours) is described by two BMPs: left by $\tt XYYX$ and right by $\tt XYYX$,
    % on the left, the SFC is described by $\tt XYYX$, and on the right, the SFC is described by $\tt XYXY$, 
    where the green shade highlights the 
    % grids that are scanned
    scanned grids.}\label{fig:motivation_non_universal}
	% \vspace*{-3mm}
\end{figure}

\begin{example}
 Figure~\ref{fig:motivation_non_universal} shows a $4\times4$ grid space, where the green and yellow dashed rectangles represent two window queries $Q_1$ (horizontal) and $Q_2$ (vertical), respectively. {\CHENG The red lines} represent the ordering of grid cells w.r.t. three SFCs.
For example, consider SFC-1, 
%Take SFC-1 for example, 
whose 
${\tt P}_1={\tt XYYX}$ and the computed value for input $\mathbf{x}=((x_1x_2)_2, (y_1y_2)_2)$ is $C_{{\tt P}_1}(\mathbf{x}) = (x_1y_1y_2x_2)_2$.
% (where $x_i$ (resp. $y_i$) represents the $i^{th}$ bit in the binary sequence of $x$ (resp. $y$)).
Notice that in SFC-1, $x_1$ is put as the first bit in the combined bit string, and thus any data point with $x_1 = 0$ (that resides in the left half of Figure~\ref{fig:motivation_non_universal} (a)) will have a smaller mapped value than any data point with $x_1 = 1$
{\CHENG (that resides in the right side of Figure~\ref{fig:motivation_non_universal}(a)).}
% (Figure~\ref{fig:motivation_non_universal}(b)). 
We label the grid ids based on the mapped values of grid cells computed by the SFC curves.
As discussed in Section~\ref{sec:prior}, a typical algorithm first locates the grid ids on the minimum (bottom-left corner) and the maximum (top-right corner) points of a query window.

Different SFCs will result in accesses of different grid cells for answering the two window queries $Q_1$ and $Q_2$. 
%
%
%
% organised by its curve in the figure. 
% Note that with SFC-1, the first bit 
%
For instance, with SFC-1,
$Q_1$ and $Q_2$ need 2 and 3 grid scans, respectively.
% 
% \walid{You may want to use the terminology by Jagadish et. al where they call them "runs". Each run is a consecutive sequence of points in the curve that is within the query window. The run ends when it exits the window and another run starts when the SFC reenters the window. The total number of runs some how reflects or approximates the number of I/Os made or contiguous memory accesses. They used the number of runs in the query window as a measure of goodness, and the target of course is to minimize the number of runs. }
\iffalse
% to answer  query $Q_1$,  
% we scan the range from the minimum point (grid 7) to the maximum point (grid 8), resulting in 2 grid scans. For $Q_2$, the grid ids for the minimum and maximum points are 13 and 15, respectively. Hence we need 3 grid scans ranging from  grid 13 to 15.
\fi
%
With SFC-2 (${\tt P}_2 = {\tt XYXY}$), 
$Q_1$ and $Q_2$ need 3 and 2 grid scans,
respectively. Detailed computation can be found in~\cite{li2023towards}.
Further, with SFC-1, cells in window of $Q_1$  are consecutive (cell $7$ and $8$ form a contiguous sequence, noted as 1 \emph{run} in \cite{jagadish1990linear}), while cell $13$ and $15$ in $Q_2$ are not consecutive (2 runs). Query with $1$ run means a contiguous memory access is available, which is preferred.
% we need  3 grid scans (from grid 6 to 8) for $Q_1$ and 2 grid scans (from grid 13 to 14) for $Q_2$.
% 
\end{example}

In the example, SFC-1 performs better for $Q_1$ while SFC-2 is better for $Q_2$.
%
% Therefore, 
A natural idea is whether we can 
combine the advantages from the two BMPs of SFC-1 and SFC-2, i.e., we use $\texttt{XYYX}$ to organize the data at the left hand side and $\texttt{XYXY}$ to organize the data at the right-hand side.
The design will result in  a \emph{piecewise} SFC, shown as SFC-3 in Figure~\ref{fig:motivation_non_universal}(c).
With SFC-3, we need 2 grid scans for both $Q_1$ and $Q_2$.
% , where the scanning ranges for $Q_1$ and $Q_2$ are from grid 7 to 8 (similar to SFC-1), and grid 13 to 14 (similar to SFC-2).
% , respectively.
%
This example motivates the need 
%of 
for
designing a piecewise SFC.

%other than an elementary one. 

% \begin{figure}
%     \centering
%     \scriptsize
%     \includegraphics[scale=0.57]{VLDB_Learned_SFC/figures/example_motivation.pdf}\\
%     \quad (a) SFC-1 ($x_1y_1y_2x_2$) \quad\quad  (b) SFC-2 ($x_1y_2x_2y_2$) \quad\quad\quad\; (c) SFC-3 (piecewise SFC) \\
%      \quad \;\;\; grid scans: $2$ for $Q_1$, $3$ for $Q_2$ \quad grid scans: $3$ for $Q_1$, $2$ for $Q_2$ \quad grid scans: $2$ for $Q_1$, $2$ for $Q_2$ \quad\quad
%     \caption{Motivation for piecewise SFC, 
%     % grid scans number of the two queries are given in brackets. 
%     SFC-1 is described by the BMP $\tt XYYX$, while SFC-2 is described by the BMP $\tt XYXY$. In contrast, SFC-3 (ours) is described by two BMPs: on the left, the SFC is described by $\tt XYYX$, and on the right, the SFC is described by $\tt XYXY$, where the green shade highlights the grids that are scanned.}
%     \label{fig:motivation_non_universal}
% \end{figure}

% % \smallskip
\noindent\textbf{Motivation 2: Learning-based Method for Piecewise SFC Construction.} 
Classic SFCs (Z-curve, Hilbert curve, etc.) are based on a single scheme and fail to utilize 
the underlying
database instance
% and workload 
to design the SFC.
In contrast, QUILTS proposes to utilize the given database and query workload to evaluate and select an SFC from an SFC set in which each SFC is described by a BMP.
However, QUILTS does not directly evaluate SFC w.r.t. query performance
%
% It designs 
but 
%design
uses
instead
heuristic rules to generate candidate SFCs. 
% clustering bits that would make grid cells intersect with a query rectangle together in the mapped space
The heuristic rules will select BMPs such that the resulting grid cells intersecting with a query would be continuous in the curve order, and hence results in fewer grid scans.
These heuristics only work for query workloads containing limited types of window queries (e.g., with 
% precisely 
the same aspect ratio), and are not effective under general situations (where more than one query type with different aspect ratios and region areas exist). Due to these limitations, it calls for more principled solutions to utilize database and query workloads for generating and selecting an SFC.
Learning-based methods would be promising for 
%the 
this
purpose.
%s.t. no need for expertise and experience in designing heuristics case by case.

%developing the SFC method without human-crafted heuristics will lead to a generic method that is effective in a realistic database.

\begin{figure}[ht]
    \begin{subfigure}[b]{.49\linewidth}
    \centering
    \hspace*{2mm}
    \includegraphics[width=0.965\linewidth]{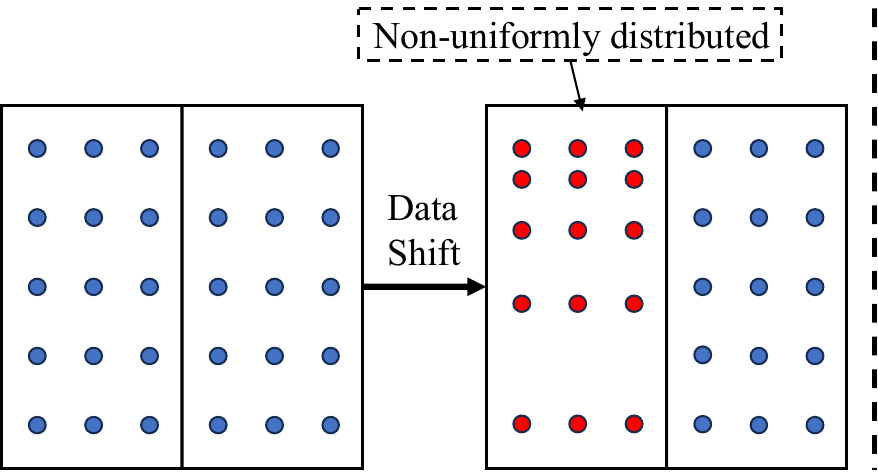}
    % \vspace*{-2.5mm}
    \caption{Illustration of a data shift.}
    \label{fig:data_shift}
    % \vspace*{-2mm}
    \end{subfigure} 
    % \hfill
    % 
    \begin{subfigure}[b]{.49\linewidth}
    \centering
    \includegraphics[width=0.94\linewidth]{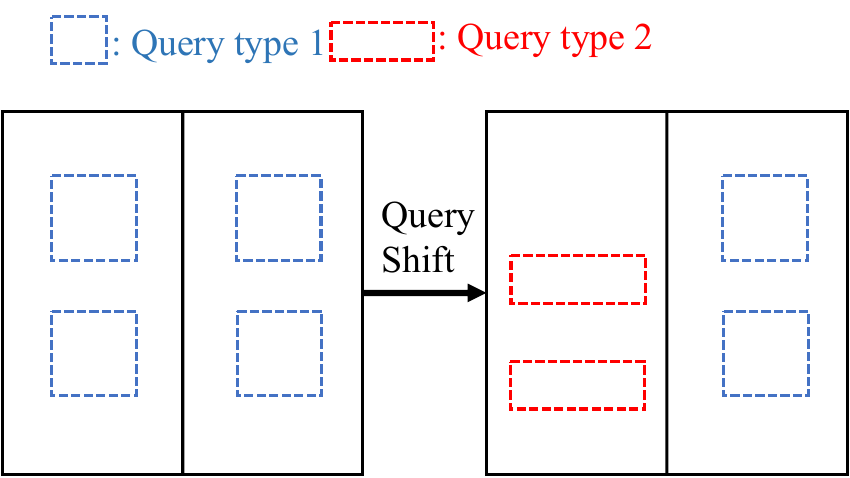}
    % \vspace*{-2.5mm}
    \caption{Illustration of a query  shift.}
    \label{fig:query_shift}
    % \vspace*{-2mm}
    \end{subfigure}
    % \vspace*{-2mm}
    \caption{\bluecolor{Examples of  distribution shifts of both data  and query workloads over different
    %among 
    subspaces.}}
    \label{fig:data_query_shift}
    % \vspace*{-5mm}
\end{figure}

\bluecolor{

\noindent \textbf{Motivation 3: Efficient Piecewise SFC Update.}
The distribution shift issue exists when maintaining data-driven learned indexes. 
When  distribution shift happens, 
the 
performance of 
the
learned modules can become suboptimal. A retraining procedure is preferred if the performance decreases to a certain degree. However, retraining the BMTree from scratch can be costly.  Moreover, in cases where distribution shifts occur unevenly across subspaces (e.g., some subspaces with significant distribution shifts while others with mild shifts), not all BMPs require redesign.
Two examples are given in Fig.~\ref{fig:data_query_shift} to illustrate the uneven shifts of data and query, respectively. 
% and Fig.~\ref{fig:query_shift} give examples of partial distribution shifts  of data and query, respectively.
In Fig.~\ref{fig:data_shift}, the data located in the left half of 
the 
space shifts from 
being
uniformly distributed to 
being
non-uniformly distributed. In Fig.~\ref{fig:query_shift}, the query located in the left half of the space has shifted not only the spatial distribution of the queries (considering the center point of each query rectangle) but also the categories of the queries (from Type 1 to Type 2 with different aspect ratios).
% 
% \walid{I do not recall your mentioning about different types of queries. I may have missed that though. Please recheck.}
%Such 
These
distribution shifts can lead 
%a performance decrease in the SFC learned for the historical database scenario.
to a decrease in performance of the SFC learned for the case of historical data and query distributions.
% \walid{What do you mean by "historical" databases?}
% Under this situation, 
The issue calls for a solution to provide an efficient way to update 
the
BMTree w.r.t.  distribution shifts. 
Additionally, it is preferable if the solution allows for the partial redesign of the BMTree while keeping a portion unchanged so 
that
only 
the 
data points located in the retrained subspaces need to update their SFC values to maintain 
the 
corresponding indexes.
% Further, it is preferred if the solution allows partially redesigning the BMTree and remains a certain portion unchanged, thus only a portion of the data points need to update their SFC values to maintain the corresponding indexes such as B+Tree.
% Further, the retraining procedure will be costly if the depth of BMTree becomes high. Instead, partial retraining of the piecewise SFC instead of fully retraining is an intuitive idea for resource saving. For instance, in Fig.~\ref{fig:data_shift} and \ref{fig:query_shift}, distribution shift only happens in the left half of the space, which means the SFC assigned to the right half side can remain unchanged.
% (the left side of Fig.~\ref{fig:data_shift} and \ref{fig:query_shift})

}

% \smallskip
\noindent\textbf{Challenges.} Piecewise SFC design 
%brings up 
gives rise to 
three challenges as discussed in 
the introduction section. 
(1)~How to partition the space and design an effective BMP for each subspace? The piecewise SFC design needs to consider both space partitioning and BMP generation. 
%It is critical in a piecewise SFC design to figure out whether partition the space and design distinct BMPs or design an identical BMP instead. It should also select an effective partition that primarily optimizes the performance.
(2)~How to design piecewise SFCs such that  
the
two desirable properties, 
namely,
monotonicity and injection, 
still 
hold? 
%
%Although an SFC described by a single BMP guarantees monotonicity and injection,  splicing multiple SFCs leads to property break risk. 
\iffalse
Combining different SFCs
for different subspaces to obtain a piecewise SFC for the whole space
may lead to the risk of breaking the properties. 
For instance, two data points with distinct BMPs in a piecewise SFC may end up with identical SFC values. 
\fi
%Thus, it requires us to design a method that collects BMPs can be spliced. 
(3)~How to design a data-driven approach to automatically build the BMTree, given a database and query workloads? 
(4)~How to identify the appropriate subspace for the partial retraining of the piecewise SFC without compromising its properties?
\subsection{Overview of the Proposed BMTree}

% % \smallskip

\noindent
\textbf{The Bit Merging Tree (BMTree) for Piecewise SFC Design.}
\revise{To address the first challenge, 
%to partition space and assign BMP, 
we propose a novel way of seamlessly integrating 
%the 
subspace partitioning and BMP generation by 
%building 
constructing
the BMTree; a binary tree
% structure
% , which is used 
% to 
that
models a piecewise SFC. 
%By building  a BMTree, we directly optimize the piecewise SFC w.r.t. the query performance.
Each node of the BMTree is filled with a bit from a dimension. The filled bit partitions the space into two subspaces corresponding to two child nodes. 
% The branching from a BMTree node to its children models a subspace partition. 
The left branch is the subspace where data points have a bit value of 0 and the right branch with 1.
% , respectively. 
%
The BMTtree partitions the whole data space into subspaces, each corresponding to a leaf node with its BMP being the concatenated bit sequence from the root to the leaf node. 
%path of BMTree corresponds to a BMP. 
% The concatenation of bits of each path from the root node to the leaf node represents a BMP. 
%To compute the SFC value, each data point can be mapped to one path of BMTree and thus get the related BMP. 
We present the BMTree structure in Section ~\ref{sec:why_bmtree}.
Furthermore, the BMTree mechanism guarantees that
%instead of directly designing 
the generated piecewise SFC satisfies the two properties that 
%addresses
address
the second challenge.
We prove 
that 
the piecewise SFC represented by a BMTree satisfies both monotonicity and injection
% properties 
in Section \ref{sec:theoretical}.}

% BMTree is a binary tree, where each node corresponds to a bit from a dimension, the left branch has a bit value of 0, and the right branch is 1.
% %
% Intuitively, the BMTree is used to design how different BMPs are combined. Each path of the BMTtree represents a BMP, which is the concatenation of the bit values of the path starting from the root node to the leaf node. Each data point can be mapped to a path of the BMTree, and the corresponding BMP is applied to the point. This is illustrated in Figure~\ref{fig:bmtree_example}. 
% We propose a learning based method to build the BMTree, i.e., determining the bit merging pattern represented by each path. 
% {\CHENG We will present the details of BMTree in Section~\ref{sec:why_bmtree} and theoretically prove that the BMTree satisfies both monotonicity and injection properties in Section~\ref{sec:theoretical}.}

% {\bf Gao the learning based part is not clear}

 %(corresponding to the Challenges).

% % \smallskip
{\CHENG
\noindent\textbf{RL-based Algorithm for Constructing a BMTree}.} \revise{To address the third challenge, we design a learning-based method that learns from 
data and query 
workloads
% a given dataset and query workload 
to build the BMTree. We model the building of the BMTree as a Markov Decision Process~\citep{puterman2014markov}. 
%Specifically, 
The process of building a BMTree comprises a sequence of actions to select bits for tree nodes by following 
a top-down order.
%To construct a BMTree, the learned policy selects bits to fill the tree nodes  following a top-down order.
%(a breadth-first filling from the root node to leaf nodes). 
%
To learn an effective policy for building the BMTree, we propose a new approach for integrating a greedy policy into the Monte Carlo Tree Search (MCTS) framework~\citep{browne2012survey}. Specifically, we develop a 
greedy
policy that selects
%greedy 
an action
%, which chooses actions 
to fill a bit 
%node by node
for each node during tree construction. For each node, the greedy policy 
%only 
chooses the bit that achieves the most significant reward 
%compared with other bits
among all the candidate bits.
% from different dimensions. 
%While deciding the current node's bit, other nodes remain unchanged. 
Afterwards, we apply the greedy policy as a guidance policy and use MCTS 
%based BMTree construction framework, which 
to optimize the BMTree with the objective of providing good query performance and avoiding local optima. Moreover, we introduce a fast computing metric, 
termed ScanRange, to speedup reward generation. We present the proposed solution in Section \ref{sec:mcts} and 
%the 
its
time complexity analysis in Section \ref{sec:theoretical}.}

\noindent
\bluecolor{\textbf{Partially Retraining a BMTree for Piecewise SFC Update.}
To address the last challenge, we develop a mechanism to efficiently update the BMTree. First, we propose a principal way that measures the shift of the subspace modeled by each BMTree node on query and data. A shift score measuring the distribution shift degree is introduced based on Jensen–Shannon (JS)
divergence~\cite{DBLP:conf/isit/FugledeT04} that is a widely used tool for measuring the similarity between distributions. Then, the mechanism detects the nodes with 
%large 
largest
performance optimization potentiality. 
%We then 
Then, we 
partially delete the nodes of the BMTree 
that need 
to be retrained, and apply an adapted RL framework to regenerate the deleted BMTree parts %w.r.t.
with respect to
the updated database scenario. The structure of the BMTree ensures that
the regenerated piecewise SFC 
%retains 
retrains
all desired properties, since the regenerated BMTree naturally models a piecewise SFC properly. We present the details of the retraining mechanism in Section~\ref{sec:retrain_bmtree}.
}

% To build the BMTree, we propose to model the process of building the tree, i.e., designing bit merging patterns as a decision making process. 
% %
% Specifically, to construct a BMTree, we make decisions on selecting bits to fill the tree nodes by following a top-down order (a breadth-first filling from the root node to leaf nodes).

% In this paper, we model this process as a branch Markov decision process~\cite{puterman2014markov, pliska1976optimization, hahn2021model}, and leverage the reinforcement learning technique to learn a query performance aware policy for selecting bits s.t. the generated BMTree optimizes the query performance.
% {\CHENG We will present the details of this learning based algorithm in Section~\ref{sec:mdp} and the time complexity analysis of the algorithm in Section~\ref{sec:theoretical}.}

% \subsection{BMTree: Bit Merging Tree}
\section{Designing Piecewise SFC: \\ Bit Merging Tree (BMTree) 
%for Maintaining a Piecewise SFC
}
\label{sec:why_bmtree}

We 
%proceed to 
present how to develop a piecewise SFC, modeled by the Bit Merging Tree (BMTree) that is a binary tree.

%We motivate our approach, discuss why should apply the deep learning method to construct the tree and the overall solution to learning a bit merging tree.

% there is at most one possible bit in dimension $i$ that can be filled to $b_t$ when the left $t-1$ bits have been decided (that is because when we choose bit in dimension $i$ for $b_t$ and $b^i_1$ to $b^i_{j-1}$ have been filled in the $t-1$ bits in $v_q$, the only choice to fill $b_t$ is $b^i_j$ to maintain the partial ordering. In other words, bits in one dimension should be selected one by one from left to right.). 
% Therefore, there are at most $n$ candidate bits that can be filled to $b_t$ unless some dimension's bits are exhaustively selected before $b_t$.

\noindent\textbf{Designing a BMP.} 
% First, we consider designing a bit merging pattern. Revisit definition \ref{def:bit_merging_pattern}, 
%
%QUILTS only considers a BMP as a basic unit and chooses one of them rather than designing a BMP character by character. 
%
To design a BMP ${\tt P}$, we need to decide which character ($\texttt{X}$ or $\texttt{Y}$ in the two-dimensional case) is filled in each position of ${\tt P}$. A left-to-right design procedure decides the filling characters in the order from $p_1$ to $p_{2m}$. 
%
% To satisfy monotonicity, when we fill two bits, $x_i$ and $x_j$, $i > j$,  from dimension 
%$i$ 
% $x$
%into $b_t$, 
% $b^i_{j^{\prime}}$ and $b^i_{j^{\prime\prime}}$ should be filled to $b_{t^{\prime}}$ and $b_{t^{\prime\prime}}$ in $v_q$ satisfying $t^{\prime} > t^{\prime\prime}$ iff $j^{\prime} > j^{\prime\prime}$, 
% into $b_s$ and $b_t$ in $v_q$, respectively, we must guarantee $s > t$.
%$x_i$ and $x_j$ in dimension $x$ should be filled to $b_s$ and $b_t$ in $v_q$ satisfying $s > t$ iff $i > j$.
%
%Under a left-to-right design, with n dimensions, bits in a dimension, say diemension $d_i$ should not be filled $v_q$ before all bits on the left side in $d_i$ are already been filled. Therefore, there is at most only one bit from a dimension eligible to be filled to $b_t$ and $n$ bits summing from all dimension. We call those bits the candidate bits for $b_t$. When designing RL algorithm next in this paper, we design action as choosing dimension instead of choosing bits since the above condition should be satisfied.
%
% Under a left-to-right design, a bit in a dimension, e.g., $x$ can be used to fill $v_q$ only if all bits before the bit on the left side in $x$ are already been filled. Therefore, 
The key to BMP design is to have a policy deciding which dimension ($\texttt{X}$ or $\texttt{Y}$) to fill into each position of ${\tt P}$.

% % \smallskip
\noindent\textbf{Designing piecewise SFC with multiple BMPs.} 
%Recall the bit merging pattern design in section \ref{sec:prior}. 
Next, we 
%We next 
discuss piecewise SFC design.
As discussed in 
% Introduction and 
Section~\ref{sec:limitation}, one challenge 
%of 
in
designing  a piecewise SFC is: How to handle two subtasks that are intertwined together, namely subspace partitioning and BMP design 
%for 
within
each subspace? 
%
% Moreover, i
It is also challenging to guarantee that 
the piecewise SFC comprising different 
BMPs for different subspaces 
%which are non-overlapping portions of $\mathcal{D}$'s domain, while the combined piecewise function 
still 
%works as an SFC that 
satisfies both 
the
injection and monotonicity properties. 
To address 
%the 
these
challenges, we 
%propose a novel 
introduce a new
%way designing BMPs for subspaces, in which we  
solution to simultaneously 
%generating 
generate
the subspaces and 
%designing
design
the 
BMPs for 
these 
subspaces. 

We follow a left-to-right BMP design approach, and start with an empty string $\tt P$. For example, if we fill $\texttt{X}$ in %to the empty 
the first position of $\tt P$, Bit $x_1$ will be filled to the $b_1$th position of $\tt P$; 
%when computing $T(\mathbf{x})$, 
Then, the whole data space 
is partitioned into two subspaces w.r.t. the value of Bit $x_1$, where one subspace corresponds to $x_1 = 0$ and the other corresponds to $x_1 = 1$. 
This partitioning enables us to separately design different BMPs for the two subspaces.
Notice that the BMPs for each subspace will share $\texttt{X}$ as the first character, but can have distinct filling choices for the next $2m - 1$ characters.
%The combined SFC is a piece-wise SFC with criteria guaranteed following this design procedure (see analysis in section \ref{sec:theoretical}). 
By recursively repeating this operation, we fill in the subsequent characters for each BMP for each subspace, thus generating multiple subspaces each with a different BMP.
% 
%An elegant perspective 
One advantage 
of 
%our idea 
this approach
is that 
%we integrate the 
it 
seamlessly
integrates 
subspace partitioning 
%and 
with BMP generation. %seamlessly. 

\begin{figure}
    \begin{subfigure}[b]{.49\linewidth}
    \centering
    \includegraphics[scale=0.3]{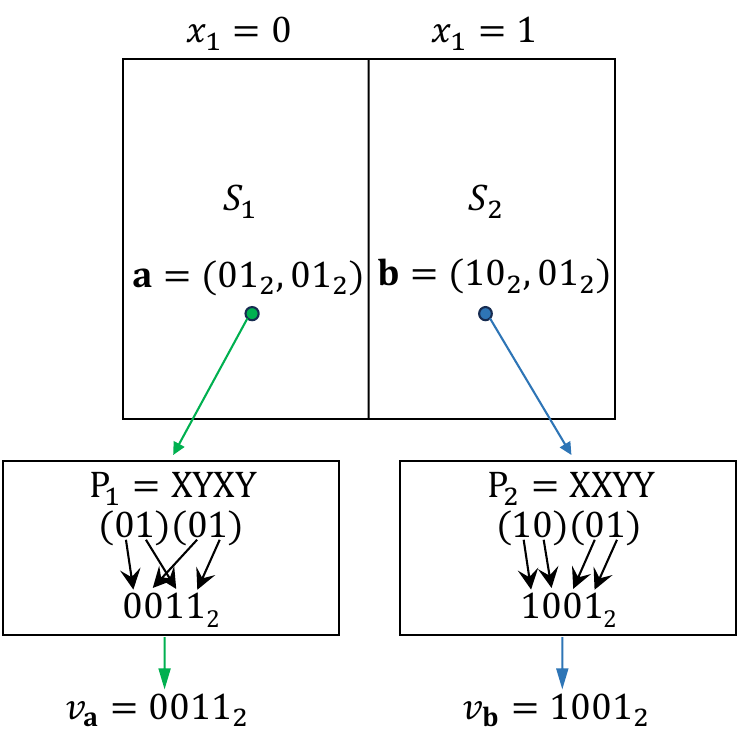}
    % \vspace*{-2.5mm}
    \caption{A piecewise SFC.}
    \label{fig:piecewise_sfc}
    % \vspace*{-2mm}
    \end{subfigure} \hfill
    \begin{subfigure}[b]{.49\linewidth}
    \centering
    \includegraphics[scale=0.35]{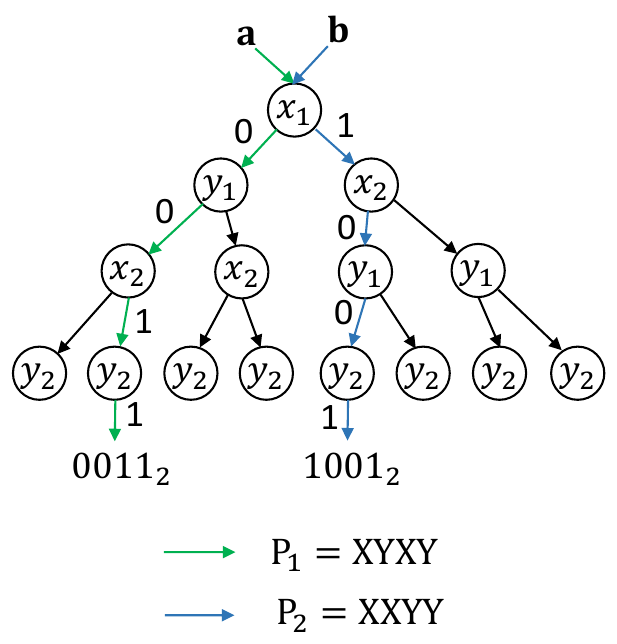}
    % \vspace*{-2.5mm}
    \caption{The BMTree.}
    \label{fig:bmtree_example}
    % % \vspace*{-2mm}
    \end{subfigure}
    % \vspace*{-2mm}
    \caption{(a) An example of a piecewise SFC that comprises two BMPs ${\tt P}_1$ and ${\tt P}_2$ for computing values of Data Points $\mathbf{a}$ and $\mathbf{b}$. (b) A BMTree that combines the two BMPs.}
    \vspace*{-2mm}
\end{figure}

% Our method applies the extended BitMergingFamily from Quilts but allows different subspaces have different bit merging functions while quilts only allows one unique bit merging function in the whole space. The main idea to design a non-universal SFC is to separately design bit merging functions by partitioning a dataset with respect to the value of a bit. 

% % \smallskip
\begin{example}
% To illustrate our idea, a
Figure \ref{fig:piecewise_sfc} gives 
an example of 
a 
piecewise SFC,
% In the example, d
where 
%dimension 
Dimensions 
$x$ and $y$ are bit strings of Length 2. First, $\texttt{X}$ is selected, and then the whole space is partitioned w.r.t. value of Bit $x_1$ into two subspaces where Subspace $S_1$ corresponds to $x_1 = 0$ and Subspace $S_2$ corresponds to $x_1 = 1$. 
%
%By filling $x_1$, 
Next, we separately design BMPs for $S_1$ and $S_2$, where all BMPs under $S_1$  share the first bit $x_1=0$ and BMPs under $S_2$ share the first bit $x_1=1$.
We  generate two example BMPs: ${\tt P}_1 = \texttt{XYXY}$ 
% is designed 
for $S_1$
% , which is the BMP of Z-curve 
and ${\tt P}_2 = \texttt{XXYY}$ 
% is 
for $S_2$.
% , which is the BMP of C-curve. 
Finally, we get a piecewise SFC that comprises  $C_{{\tt P}_1}$ for $S_1$ and $C_{{\tt P}_2}$ for $S_2$. This piecewise SFC represents the function:
    $
    C(\mathbf{x}) = \left\{ 
    \begin{array}{ c l }
(x_1y_1x_2y_2)_2 & \; \textrm{if } x_1 = 0 \\
    (x_1x_2y_1y_2)_2     & \; \textrm{if } x_1 = 1
    \end{array}
    \right.
    $.
    Therefore, if Data Point $\mathbf{a}$ is located in $S_1$, we  apply $C_{{\tt P}_1}$  to compute $\mathbf{a}$'s SFC value. 
    %otherwise, 
    Similarly, if Data Point
    $\mathbf{b}$ is in $S_2$, 
    we apply $C_{{\tt P}_2}$ to compute $\mathbf{b}$'s SFC value.
    %$T_{{\tt P}_2}$ is applied. 
\end{example}

%Considering a piece-wise SFC designed following our method, it is significant to efficiently store and maintain distinct bit merging patterns, which could ease tracking and applying corresponding patterns to compute SFC value. 

To facilitate the process of designing piecewise SFCs, we propose the Bit Merging Tree (BMTree) structure
%, which 
that
is used to simultaneously partition the space and 
to 
generate 
the 
BMPs. 
Figure \ref{fig:bmtree_example} gives the corresponding BMTree for the example piecewise SFC 
%in 
of
Figure \ref{fig:piecewise_sfc}. Since the example piecewise SFC is developed with only 2 BMPs, the left subtree of the root node shares ${\tt P}_1$ while the right subtree shares  ${\tt P}_2$. Next, we present the BMTree.
\noindent \textbf{The Bit Merging Tree (BMTree).}
    % \hanmao{A BMTree $T$ Is this $T$ the function? No right?} 
    The BMTree is a 
    %complete 
    binary tree 
    % for representing 
    %modeling
    that models
    a piecewise SFC $C_{\tt T}$,
    % mapping function , 
    and is denoted by $\tt T$. 
    %\walid{why denote the tree by the same name of the curve? You can refer to the BMTree corresponding to a sfc T by B(T) or something like that}
    The depth of a BMTree $\tt T$  equals the length of a BMP, and is denoted by $2m$ for the 2-dimensional space. Every node of $\tt T$ corresponds to a bit of $x_i$ or $y_i$, $1 \le i \le m$.
    The left (resp. right) child denotes the subspace with bit value 0 (resp. 1). Each path from the root node to a leaf node represents a BMP for the subspace of the leaf node, which is the concatenation of all the bits of the nodes in the path 
    from root to leaf. 
    % We represent the function defined by $\tt T$ as $T_{\tt T}$.
    %and combining all bits from the nodes. 
    % Therefore, the bit filled in one node should not be repeated to any of its ancestor nodes and should follow Quilts' pattern design rule. 
    The SFC value $C_{\tt T}(\mathbf{x})$ of a data point $\mathbf{x}$ is computed by traversing a path of $\tt T$ as follows. We start from the root node, and for each traversed node, denoted by $x_i$, if $x_i = 0$, we visit the left child node; otherwise 
    %go to 
    we visit
    the right child. 
    %node. 
    When we reach a leaf node, the corresponding BMP of the traversed path is used to compute 
    %for 
    $C_{\tt T}(\mathbf{x})$. 
    The green path in Figure \ref{fig:bmtree_example} is the path traversed for Point $\mathbf{a}$
    %, which 
    that
    represents BMP ${\tt P}_1$ while the blue path traversed for $\mathbf{b}$ represents ${\tt P}_2$. 

%  Every node of BMTree represents a subset of $\mathcal{D}$. Also each node stores two parameters $i$ $j$ which means bit $b^i_j$ is selected from dimension $i$ at $j^{th}$ bit. That the left(resp. right) child of a non-leaf node represent the $\mathcal{D}_l$(resp. $\mathcal{D}_u$) split according to $b^i_j$.

%Similar to Quilts' SFC set, BMTree also decides a set of piece-wise SFCs and the SFCs considered by Quilts is a subset of ours (when an identical bit merging pattern is assigned to the whole space). The size of Our considered SFC set grows way faster than the super-exponential increasing bit length of each dimension and is much larger than Quilts does.

% by solving a recursive function.
% : $F(m_1, \ldots, m_d) = \sum\limits_{1 \leq i \leq d} F^2(m_1, \ldots, m_i - 1, \ldots, m_d)$ where $m_i$ denotes for the bit number of dimension $i$ and the recursive stops at when no more than one input has non-zero value of function $F$: $F(0,\ldots n_i, \ldots, 0) = 1$. 
% The size grows much faster than a combinatorial number which is super-exponential.

To construct a BMTree, we develop a breadth-first construction algorithm to 
%fill 
assign
bits to 
the
BMTree's nodes. Details (pseudo-code and the corresponding illustration) can be found in \cite{li2023towards}.

\begin{figure*}[ht]
    \centering
    % \vspace*{-2mm}
    \includegraphics[width=\textwidth]{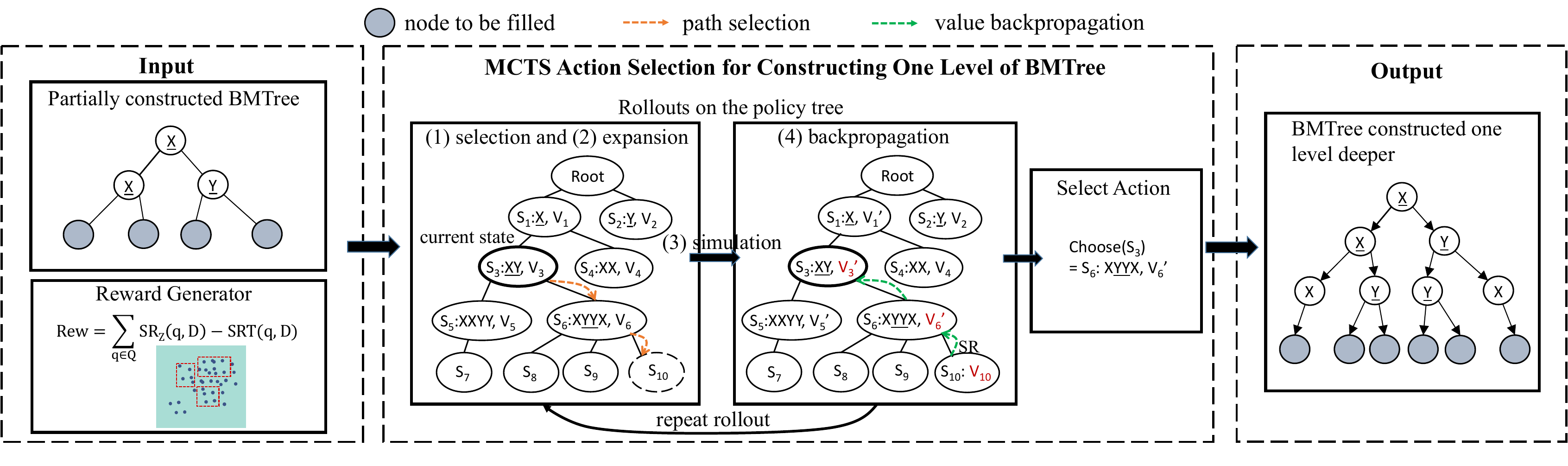}
    % \vspace{-2mm}
    
    \caption{Workflow of Monte Carlo Tree Search-Based BMTree construction. 
    % Values in red are updated values by the backpropagation step.
    }
    \label{fig:mcts_flow}
    \vspace{-2mm}
\end{figure*}

\section{Learning Piecewise SFC: \\ MCTS-based BMTree Construction} \label{sec:mcts}

% \revision{TODO: Motivate the MCTS method.}

% \revise{Argument why choose MCTS.}
% This subsection is to present our solution to Line~4 of Algorithm~\ref{alg:bfs_bmtree_construct}, i.e., learning a decision policy. 
%
It is difficult to design heuristic methods to construct the BMTree to optimize the querying performance for a workload on a database instance.
This could be observed from QUILTS that uses  
%the 
heuristic rules  for a workload containing specific types of window queries only, 
and fails to directly optimize query performance.
\revise{In contrast, we propose a reinforcement learning (RL) based method for learning a decision policy that builds the BMTree to optimize 
%the querying 
query
performance directly.
% To construct the BMTree to generate piecewise SFCs for  optimizing the querying performance, 
% we propose a reinforcement learning (RL) based method to learn the decision policy. 
%
%RL trains the policy to make decisions and interact with the environment based on the given reward signal.
}
% The goal of RL is to train an agent that provides an optimal policy and optimizes a reward. The agent takes the input state and learns to perform actions in the environment.
%
%
%
%

To allow an RL policy to construct the BMTree, we model the BMTree construction as a Markov decision process (MDP, for short). 
% Different from previous works~\citep{yang2020qd, liang2019neural}, which model the tree node as state input to build a tree structure with RL, we take the entire tree structure of BMTree as the state instead. This is because the BMTree is applied as the piecewise SFC value function and the querying performance is strongly bonded with the tree structure, which makes considering independent nodes as the state not reasonable. 
Then, we design a BMTree construction framework with a model-based RL method, termed Monte Carlo Tree Search (MCTS, for short). 
% MCTS is a light-weighted yet effective RL method that does not requires much parameter tuning compared with other deep RL methods. 
%
\revision{
% MCTS is a light-weighted RL algorithm which fits our problem. Compared with traditional algorithms, e.g., greedy or A*, MCTS has better ability of balancing the the exploration and exploitation, and thus, prevent from local optimality. On the other hand, compared with other RL algorithms, e.g., PPO, MCTS is a light weight algorithm which do not requires much parameter tuning and has a stable performance.
% MCTS algorithm is a lightweight reinforcement learning approach that is well-suited for our problem. Compared to traditional algorithms such as greedy or A*, MCTS demonstrates superior exploration-exploitation balance, which prevents from local optimal. In contrast to other reinforcement learning algorithms such as PPO~\cite{schulman2017proximal}, MCTS is a lightweight approach that boasts stable performance without necessitating extensive parameter tuning.
% \marginpar{\#1 R2W3D4}
Unlike traditional algorithms, 
e.g., 
%such as 
greedy or A*, MCTS is an RL approach that demonstrates superior exploration-exploitation balance, 
%preventing from 
{\chengr mitigating} the issue of local optimum.  MCTS is well-suited for 
%our problem, 
the problem at hand,
and offers stable performance without extensive parameter tuning, compared with other RL algorithms,
%such as 
e.g.,
PPO~\cite{schulman2017proximal}.
}

% Further, MCTS does not require representing the tree structure into a fixed-size embedding vector to feed as the input, which needs great effort and can be of high complexity~\citep{liang2019neural, yu2020reinforcement}. 

Figure~\ref{fig:mcts_flow} gives the workflow of the MCTS-based BMTree construction framework. We define one action that RL takes to be a series of bits that fill a level of nodes in the BMTree, and the nodes of the next level are then generated.
% , which need to be filled.
The action space size grows 
exponentially
with the 
%node number. 
number of nodes.
It becomes difficult for RL to learn a good policy with an 
%enormous action space size. 
enormously large action space.
To address this
issue, 
we design a greedy action selection algorithm
that
helps 
%to 
guide MCTS to search for good actions.
% when the action space is large. 
Moreover, we design a metric, 
termed
{\em ScanRange}
to speed up reward computing.

% First, we model the BMTree construction as a Markov decision process (MDP), and train the policy to make decisions and interact with the environment based on the given reward signal. 

% Recent work~\citep{yang2020qd, liang2019neural} models the tree structure construction as a branching MDP, in which each tree node is modeled as a state, ... . 
% They regard each tree node as a state, do not consider the relationship between nodes, and model the procedure as a multi-arm bandit problem instead of a long-term decision process. 

% \revise{In our problem, the environment is building the BMTree, and we aim to learn a policy to decide which dimension to choose for each BMTree node such that the query performance on the learned piecewise SFC is optimized. 
% %
% We tried multiple methods, both model-based methods (e.g., MCTS~\citep{browne2012survey}) and model-free methods (e.g., Q-learning~\citep{watkins1992q}, PPO~\citep{schulman2017proximal}). Empirical results show that the Monte Carlo Tree Search (MCTS) has provided stable effectiveness under different settings and requires fewer examples to gain good performance, thus can provide fast training compared with other methods.}

%

\noindent\textbf{BMTree Construction as a Decision Making Process.} We proceed to illustrate how we model 
%the 
BMTree construction as a MDP,
%in detail, 
including the design of the 
% model the BMTree construction as a Markov decision process to enable an RL-based solution, 
 \emph{states}, \emph{actions}, \emph{transitions} and \emph{reward}. 
 %design.

\noindent $\bullet$ \underline{\emph{States}}. 
Each partially constructed
BMTree structure $\tt T$ is represented by a state to map each tree with its corresponding query performance. 
The state of a BMTree 
%and be 
is
represented 
%with 
by
the bits filled to the BMTree's nodes. For example, in Figure \ref{fig:mcts_flow}, the current (partially constructed) BMTree's state is represented as ${\tt T} = \{(1:\underline{\tt X}), (2:\underline{\tt XY})\}$, where $\underline{\tt X}$ and $\underline{\tt XY}$ are bits filled to the nodes in Levels~1 and~2.
% , respectively.
%At the start of the BMTree construction, the beginning state of the decision sequence is an empty BMTree with only an unfilled root node. 
% Each BMTree represents a state. At the beginning of BMTree construction, the empty BMTree is the start state. When bits are filled to nodes of the current BMTree, the new BMTree represents a transited state.

% For example, in Figure \ref{fig:mcts_flow}, we have the current BMTree. We model the state of this BMTree as the filled nodes of different levels (start from 1), which we denote as ${\tt S} = \{(1:\underline{\tt X}), (2:\underline{\tt XY})\}$, where each tuple stores the bits filled to the corresponding BMTree level. The underlined bit means the node will split into two children nodes based on the filled bit. Otherwise, the node will not split.

\noindent $\bullet$ \underline{\emph{Actions}}. Consider a partially constructed BMTree $\tt T$ that currently has $N$ nodes to be filled. We define the actions as filling bits to these nodes. 
We aim to learn a policy 
%
% The policy 
that decides which bit to be filled for each node. Furthermore, the policy 
%also 
decides if the BMTree will split the subspace of one %specific 
tree node. If the policy decides to split, the tree node will generate two child nodes based on the filled bit $\tt b$, and the action is denoted 
%as 
by
$\tt \underline{b}$ (with an underline). Otherwise, 
%if the policy decides not to split, 
the tree node only generates one child node
%, which 
that
corresponds to the same subspace as its parent, the action is denoted 
%as 
by
$\tt b$. During the construction, the policy will assign bits to all $N$ nodes. The action is represented 
%as 
by
$A = \{a_1, \ldots, a_N\}, a_i = ({\tt b}_i, sp_i)$, where ${\tt b}_i$ denotes the bit for filling Node $n_i$, and $sp_i$ denotes whether to split the subspace. Given $\tt T$ with $N$ nodes to be filled, the action space size is $(2 n)^{N}$ where $n$ is the dimension number, and a factor of 2 comes from the decision of whether 
or not
to split the subspace.

% We aim to learn a policy denoted by $\pi$. During this process, we apply $\pi$ to make decisions of selecting a bit (or a character in the alphabet designing the BMP) for each BMtree node. 
% We enable the policy to decide whether to split the node or not. When a bit, $\tt X$ is selected for a node and the corresponding bit from the dimension is filled into the tree node, if the policy decide to split the node, the space at the node is split into two subspaces represented by two child nodes, as we explained earlier. We denote the splitting bit with a underline, $\tt \underline{X}$. 
% % Otherwise, we denote it without underlining: $\tt X$. 
% In our MCTS BMTree construction framework, we design the policy to fill BMTree's bits level by level to simplify the decision chain. Figure \ref{fig:mcts_flow} gives the current BMTree with the state ${\tt S}_3$. There are $4$ nodes to be filled at level $3$. The size of the action space is $(2 \times n)^{|Node|}$, where the $n$ denotes the dimension number (2 in this example), 2 means two choices on whether split the node and $|Node|$ is the node number (4 in this example). The algorithm selects $\tt X\underline{YY}X$ as the final action to construct the BMTree.

\noindent $\bullet$ \underline{\emph{Transition}}. With the selected action $A$ for 
the
unfilled nodes in $\tt T$, the framework will construct  $\tt T$ based on $A$. The transition is from the current partially constructed BMTree $\tt T$ to the newly constructed tree ${\tt T}'$, denoted 
%as 
by
${\tt T}' \leftarrow Transition({\tt T}, A)$. In our framework, 
%the state 
we start from an empty tree, and  construct the BMTree level 
%by level 
at a time
during the decision process. 
%Each time 
In each iteration,
the action generated by the policy will fill 
%precisely 
one level of BMTree nodes (starting from Level 1), and 
will 
generate nodes one level deeper.

\noindent $\bullet$ \underline{\emph{Rewards Design}}. After $\tt T$ 
%is transited 
evolves
into ${\tt T}'$, we design the reward that reflects the 
%querying 
expected query
performance of ${\tt T}'$ to evaluate the goodness of Action $A$. One might consider executing queries using the corresponding BMTree to see how well the SFC helps 
%to 
decrease 
%the 
I/O cost. However, 
%it 
this 
is time-consuming.
% to execute every query. 
%To this end, 
Thus,
we propose a metric,
%named 
that we term 
\emph{ScanRange (SR)}
%, which 
that
reflects the performance of executing a window query, and can be computed efficiently. We construct the reward based on 
%the $SR$ of 
${\tt T}'$'s SR. {We define the function $SR_{\tt T}(q,\mathcal{D})$ as taking a query $q$ and a dataset $\mathcal{D}$ as input, and outputs the ScanRange of $q$ over $\mathcal{D}$.}
%\walid{Define the signature of the SR function as you use if below: $SR_T(x,y)$ and explain each parameter x,y.}
% We then construct the reward based on the different of ScanRange computed with $\tt T$ and ${\tt T}'$. 

\noindent
%\textbf{Computing Rewards Efficiently.} 
\textbf{Efficient Reward Computing.}
% When applying the RL method, one of the bottlenecks is computing reward, which is costly. 
% Instead of directly executing the query, we develop a metric named \emph{ScanRange} (SR) to reflect how much data will be scanned. It is cheap to compute SR, and it is calculated as follows.
%
$SR$ is 
%calculated as follow. 
computed as follows.
Given a BMTree $\tt T$, we randomly sample data points 
% $\mathbf{x}$
from $\mathcal{D}$ with a sampling rate $r_s$. Then, the sampled data points are sorted according to their SFC values.
% , calculated by $v = T_{\tt T}(\mathbf{x})$.
\revise{To compute 
the 
SFC values on 
% half-constructed
a partially constructed BMTree, we apply a policy extended from the Z-curve to the unfilled portions of the BMP in each subspace.} Sorted data points are then evenly partitioned into $\frac{r_s |\mathcal{D}|}{|B|}$ blocks, where $|B|$ denotes 
%\# 
the number 
of points per block.
For a given Window Query $q$ represented by its minimum point $\mathbf{q}_{min}$ and maximum point $\mathbf{q}_{max}$, we 
%first 
calculate the SFC value of the minimum (resp. maximum) point as $v_{min} = C_{\tt T}(\mathbf{q}_{min})$ (resp. $v_{max} = C_{\tt T}(\mathbf{q}_{max})$). 
%Then, (this is not a step). It is just notation.
We denote the blocks that $v_{min}$ and $v_{max}$ fall into 
%are denoted as 
by
$ID_{min}$ and $ID_{max}$, 
respectively. 
We calculate 
% the $SR$ of $q$ 
 $q$'s $SR$
 given $\tt T$ and $\mathcal{D}$ as $SR_{\tt T}(q,\mathcal{D})= ID_{max} - ID_{min}$. 
The calculation of SR is 
%much 
way
cheaper than 
%executing queries.
actually evaluating Query $q$.
%(because the former does not involve executing actual queries).

We develop a reward generator based on the 
%defined 
evaluated
$SR$. We take the performance of the Z-curve as a baseline. Given the dataset $\mathcal{D}$ and a query workload $\mathcal{Q}$. The generator sorts the data points based on their SFC values, and 
%compute 
computes
the reward as:

\begin{equation}\label{eq:reward_compute}
    Rew = \sum\limits_{q \in \mathcal{Q}} \left(SR_{\texttt{Z}}(q, \mathcal{D}) - SR_{\texttt{T}}(q, \mathcal{D}) \right)
\end{equation}

Intuitively, the reward is positive if the BMTree constructed by the policy achieves a lower SR than the Z-curve.
{This design allows the agent to assess the actual performance of the constructed BMTree compared to Z-curve. Empirical studies show that this choice helps the agent efficiently identify good actions.}
% This aligns with our objective to minimize 
%the 
% SR.
% We apply the SR of Z curve $SR_{\texttt{Z}}(q, \mathcal{D})$ so that the reward reflects how better 
%\walid{You need to add better  intuition as to why you compare against the z-curve. For example, Why not have the reward is computed based on T' vs T in contrast to z-curve vs T?}
% and thus the I/O cost. 
We normalize the reward by dividing the reward 
% with the SR 
of the Z-curve.

% to construct a BMTree:

%
% and (4) \underline{\emph{Reward}}: The transition is from the old BMTree to the new one after filled with bits, and new child nodes are generated correspondingly. We want to compare the performance difference between BMTrees. To enable half-constructed BMTree to compute SFC value to test the performance, we apply a policy extended from Z-curve to guide the none-designed portion of BMP in each subspace. We propose a method of evaluating the reward based on a sampled dataset to optimize its policy to maximize the \emph{reward}. We  then introduce the MCTS algorithm part and the reward design part in detail.
%
\begin{example}
%We give an example of how the decision process works in 
Refer to 
Figure \ref{fig:mcts_flow}. The partially constructed BMTree is represented 
%with 
by
the bits filled to different levels, denoted by ${\tt T} = \{(1:\underline{\tt X}), (2:\underline{\tt XY})\}$,
where each tuple is the bits filled 
%to 
in
the corresponding BMTree level. The learned policy selects the action $A = {\tt X\underline{YY}X}$. 
%The framework then builds the
% new BMTree 
The next level of the BMTree is constructed based on $A$. The reward signal is computed based on the performance of the 
BMTree's newly added level.
%one level deeper constructed BMTree. 
% Next, 
The BMtree will continue to be 
provided as 
input for constructing 
its 
next level.

\end{example}
\revise{We proceed to present the proposed MCTS framework, including a BMTree $\tt T$ under construction, {a policy tree that helps to decide the action and
gets updated gradually,}
and a reward generator that generates the reward based on $\tt T$.}

\noindent\textbf{Policy Tree.}
MCTS~\citep{browne2012survey, zhou2021learned} is a model-based RL method. 
The high-level idea of MCTS is to search in a tree structure, where each node of the tree structure denotes a state. Given the current state, the objective of MCTS is to find the optimal child node (i.e., the next state) that potentially achieves an optimal reward. {The  structure that records and updates the historical states and their associated rewards is named \emph{policy tree}~\citep{zhou2021learned}, where each child node denotes a next possible action (the design choice for the next level of the BMTree),} and we 
%formally 
define it as follows:
% MCTS learns to model the environment by building a tree structure named \emph{policy tree}. 
% By building and updating the policy tree, MCTS learns to model the environment and provide actions that tend to provide optimal performance. We define the policy tree as follows.

\revise{
\begin{definition}[Policy Tree]
    The policy tree is 
    a tree structure that
    %that helps to model 
    models
    the environment. Each node of the policy tree corresponds to a state, {representing a current partially constructed BMTree.}
    % (or a partially constructed BMTree).
    Moreover, every node stores: (1)~Action $A$  
    %transits 
    % maps
    % the parent BMTree to itself 
    {that constructs one more level of the BMTree,}
    %\walid{The above is unclear. Please rephrase. Also, all of a sudden you switched from T and T' gradual construction one level at a time to a "policy tree". The connection is unclear. State why you need such a new term (Policy Tree). This is unclear. Is it a tree of evolving BMTrees that helps with the search process? Please state clearly.}
    and (2)~A reward value that reflects the goodness for choosing a node. The root node of the policy tree corresponds to an empty BMTree, and each path of the policy tree from the root node to the leaf node corresponds to a decision procedure of constructing a BMTree. {The middle section of Fig.~\ref{fig:mcts_flow} illustrates an example of policy tree.}
    % The child nodes of the policy tree stores the BMTree constructed based on the stored 
\end{definition}
}

% MCTS is a policy-optimization algorithm for a markov decision process. MCTS optimizes the policy by building and updating a policy tree~\citep{browne2012survey} with four phases: selection, expansion, simulation, and backpropagation. We introduce how to use MCTS to construct BMTree.

\noindent\textbf{Rollouts.} \label{sec:rollouts} To choose an action,
% for a given state, 
MCTS 
%first 
checks the reward that different action choices can achieve. To achieve this, MCTS will make several attempts in which it simulates several paths in the policy tree, and then checks if the attempted path results in 
%a good 
better
performance. 
%MCTS then 
Then, MCTS
updates the policy tree based on the simulations, {referred to as \emph{rollout} in~\cite{browne2012survey}, indicating the operation that involves repeatedly selecting different actions and ultimately choosing the optimal one.} 
%\walid{What process is named a rollout? This is unclear from the above. Please make it clearer.}
A rollout consists of four phases: (1)~{\em Selection} that 
%, which 
selects the attempted path corresponding to a BMTree construction procedure, (2)~{\em Expansion}
%, which 
that 
adds the unobserved state node to the policy tree, (3)~{\em Simulation}
%, which 
that 
tests the selection's 
%the 
query 
performance,
% of the selection, 
and (4)~{\em Backpropagation}
%, which 
that 
updates the 
% respective 
%
reward value. We proceed to present 
%our 
the 
design of 
each of these 
four steps.
% We introduce how MCTS builds and updates the policy tree with four phases, selection, expansion, simulation and backpropagation.

% \underline{\emph{policy tree}} and \underline{\emph{Rollouts}}: A policy tree consists of state nodes and directed edges. Each path of a policy tree represented a Markov process example. 
% % MCTS will balance the exploitation as well as exploration during the tree search procedure. 
% In our framework, the policy tree starts with a root node representing an empty BMTree. Given the current state node in a policy tree, the action space of the node includes all bits filling combinations to the current deepest BMTree nodes. The transited state represents a new BMTree with one level deeper filled bits (as shown in Figure \ref{fig:mcts_flow}, from the current BMTree to the new BMTree). When applying MCTS to construct BMTree, an important step is to do rollouts to predict the rewards that different actions can provide before making a final choice for a state node. The rollout step tries several decision paths and simulates them to test this path's reward. The algorithm updates the corresponding node's value based on the simulation.

\noindent (1) \underline{\emph{Selection}}. The selection step aims to select a path in the policy tree that potentially achieves good performance. Starting from the current state ${\tt S}_t$ with the initialized path: ${\tt Path} = \{{\tt S}_t\}$, 
%the framework will 
we first check if all child nodes have been observed in the %historical 
previous rollouts. If there are unobserved nodes, 
%the framework will 
we choose one of them
% the unobserved children 
and add it to the path. 
Otherwise,
% If all child nodes have been observed, 
%the framework applies 
we apply the Upper Confidence bounds applied to Trees (UCT) action selection algorithm~\citep{kocsis2006bandit} to select a child node that balances 
%the 
exploration 
%and 
vs.
exploitation. Specifically, UCT selects the child node with the maximum 
%recomputed 
 value %computed as: % follows:
% \begin{equation} \label{eq:uct}
    $v_{uct} = \frac{{\tt V}_{t + 1}}{num({\tt S}_{t + 1})} + c \cdot \sqrt{\frac{\ln(num({\tt S}_t))}{num({\tt S}_{t + 1})}}$. ${\tt S}_{t + 1} = Transition({\tt S}_t, A)$,
% \end{equation}
where ${\tt V}_{t+1}$ is the reward value of the child node ${\tt S}_{t+1}$ 
%transited 
evolving
from ${\tt S}_t$ by Action $A$; $num({\tt S}_{t+1})$ and $num({\tt S}_t)$ denote the times of observing Nodes ${\tt S}_{t+1}$ and ${\tt S}_t$, respectively, during rollouts;
% , respectively;
$c$ is a factor defaulted by $1$. %
Then, the selected node ${\tt S}_{t + 1}$ is added to the path: ${\tt Path} =\{{\tt S}_t \rightarrow {\tt S}_{t + 1}\}$. The selection step continues until the last node of $\tt Path$ 
%
%if 
is
an unobserved node {(the policy does not know the expected value of this node).} 
%\walid{unobserved or observed? a bit confusing.}
%It then 
Then, it
returns the $\tt Path$ for the next step.

\noindent (2) \underline{\emph{Expansion}}. 
% The selected path $\tt Path$ from the selection step is then added to the tree nodes. 
In the expansion step, the unobserved nodes in $\tt Path$ are added to the policy tree. 
%The observing time 
{The number of observations of each node 
% The times of observing each node 
${\tt S}_t$ 
in $\tt Path$, denoted as $num({\tt S}_t)$, is incremented by $1$. Then, this value is 
%then 
used to compute the average reward.}
%\walid{What do you mean by "recorded for the algorithm"? Please rephrase to make it clear what do you mean.}

\noindent (3) \underline{\emph{Simulation}}. 
%The framework 
We simulate the performance of the selected $\tt Path$ by  constructing the BMTree based on the actions stored in the nodes of the path. Then, 
the
constructed BMTree is 
%then 
input to the reward generator to compute the 
tree's 
%ScanRange 
SR
metric.

\noindent (4) \underline{\emph{Backpropagation}}. In this step, 
%the MCTS framework 
we update the value of each node in $\tt Path$. 
% To enable MCTS judiciously selects the best action, w
We apply the maximum value update rule that updates the value of a state ${\tt S}_t$ with the maximum reward it gains from simulation, computed by $\texttt{V}'_t = \max(\texttt{V}_t, Rew)$, where $\texttt{V}_t$ is  
%former 
State ${\tt S}_t$'s old value, $Rew$ is the reward gained during the simulation, and $\texttt{V}'_t$ is the updated value.

% \underline{\emph{Simulation}} and \underline{\emph{Backpropagation}}: MCTS then simulates constructing BMTree based on the selected path during the \emph{simulation} step. After that, the algorithm inputs the result BMTree into the reward generator to compute the reward. In the \emph{backpropagation} step, MCTS updates the values based on the rollout. We involve the max rule, which updates the value of a state $\texttt{S}$ with the maximum reward it could gain from children, which is computed as:
% \begin{equation}
%     \texttt{V}'_t = \max(\texttt{V}_t, Rew)
% \end{equation}
% where $\texttt{V}_t$ is the former value of state node $t$, $Rew_{ro}$ is the reward gained during the rollout and $\texttt{V}'_t$ is the updated value.

%

\begin{example}
Refer to Figure \ref{fig:mcts_flow}. State $\tt \texttt{S}_3$ corresponds to the input partially constructed BMTree. During the rollouts, we select Path $\tt  \{\texttt{S}_3 \rightarrow \texttt{S}_6 \rightarrow \texttt{S}_{10} \}$ in the selection step. 
%It then 
Then, it
expands the new observed State $\tt \texttt{S}_8$ to the policy tree in the expansion step. 
%The framework 
We construct the BMTree based on the selected path and 
%testify the 
compute SR. In the backpropagation step, the values of $\tt \texttt{S}_3$, $\tt \texttt{S}_6$ and $\tt \texttt{S}_{10}$ are  updated whose values are listed in red, based on the  computed SR in $\tt \texttt{S}_{10}$. 

% The algorithm then select several paths to do the rollout, such as $\{\texttt{S}_3 \rightarrow \texttt{S}_5 \rightarrow \texttt{S}_7 \}$ and $\{\texttt{S}_3 \rightarrow \texttt{S}_6 \rightarrow \texttt{S}_8 \}$. These sampled paths will then be expanded to the policy tree during the \emph{expansion} phase. The algorithm will then construct the BMTree based on the path and verify the performance of the BMTree. 
\end{example}

% \underline{\emph{Selection}}: 

After the rollouts procedure, the algorithm selects the action with the highest reward, and then BMTree $\tt T$ is 
%then 
constructed %correspondingly. 
accordingly.
In the example, $\tt {\tt S}_6$ is selected with the largest value $\tt {\tt V}_6'$ compared 
%with 
to the 
other child nodes. 
%It then 
Then, it
returns 
%the action 
Action
$\texttt{X\underline{YY}X}$ 
%as the action 
to build the BMTree one level deeper.
% The algorithm then selects the action of ${\tt S}_6$: $\texttt{X\underline{YY}X}$ since it gains the highest value ${\tt V}_6'$ compared with other child nodes of ${\tt S}_3$.

\noindent\textbf{Greedy Action Selection.} 
We design the greedy action selection algorithm (GAS, for short) for the selection step in  rollouts for MCTS to find a good action for a partially constructed BMTree. Given $\tt T$ with $N$ nodes to be filled, GAS generates an action $A_g$ by greedily assigning a bit to each BMTree node that achieves the minimum SR compared to other bits when $\tt T$ is filled with that bit. 
To
summarize, the MCTS-based BMTree construction has two steps: BMTree initialization and MCTS-based BMTree learning. Its detailed pseudo-code can be found in \cite{li2023towards}.

%% file: sections/method_update.tex
\bluecolor{

\section{Updating Piecewise SFC} \label{sec:retrain_bmtree}

% Here add the method of SFC fast update.

In this section, we proceed to handle BMTree maintenance. As described in Section~\ref{sec:limitation}, with the change in distribution of the data and/or query workload, 
%shifted, 
the performance of the BMTree is no longer optimal. Retraining the whole BMTree from scratch 
%can bring issues on efficiency and high resource consumption. 
would be inefficient and would consume resources.
% To prevent from retraining piecewise SFC from scratch, w
To achieve efficient piecewise SFC update, we design a mechanism that partially retrains a BMTree based on the pre-trained BMTree instead of fully retraining a BMTree from scratch, while notably improving 
%the 
query performance under the 
%updated 
new
data and query 
%distribution. 
distributions.
We proceed to introduce: %(1)~Distribution shift degree measurement
(1)~Measurement of the degree of distribution shift 
%, which 
that 
determines whether the BMTree nodes should be retrained; (2)~Detection of 
which 
BMTree nodes to be retrained
%, which 
that 
identifies the nodes to be retrained for an optimal effectiveness-efficiency trade-off; and (3)~Partial BMTree retraining
%, which 
that 
enables 
%the 
partial retraining of the selected BMTree nodes while maintaining the rest of the BMTree unchanged. 
% The partial retraining workflow is a 2-step procedure, including \emph{retraining domain detection} and \emph{BMTree retraining procedure}. 
% Partial retraining of BMTree brings a challenge on how 
% It first detects suitable sub-spaces that can potentially bring notable performance improvement while keeping the retraining workload acceptable. We then design a method that partially retrains the subtree of the BMTree w.r.t. the detection results with an adapted RL training environment.
% how to retrain the BMTree partially with the updated scenario based on a RL environment. 

% We follow the structure of BMTree to conduct the piecewise SFC update.

\subsection{Assessment of the  Distribution Shift} 
%\subsection{Distribution Shift Measurement} \label{sec:retrain_detect}

Partial retraining of a piecewise SFC requires retraining a portion of the piecewise SFC while maintaining the overall structure still a piecewise SFC. Furthermore, detecting the subspace that could improve 
%the 
query performance after being trained is non-trivial.
To address these 
%problems, We 
issues, we
follow the pre-designed BMTree structure modeling, where we split the domain w.r.t. the structure of 
the 
BMTree. As 
%described 
in Section~\ref{sec:why_bmtree}, different nodes of BMTree represent different subspaces. To achieve effective and efficient 
%retraining subspace detection, 
detection of the 
retraining subspace,
we 
%introduce the data and query shift modeling 
model the degree by which data and query distributions drift
%with 
within
the BMTree structure 
%utilized 
as follows.

%Suppose the node $\tt N$ denotes 
Let $\tt N$ be 
a BMTree node
%, note that $\tt N$ can represent 
that represents 
a subspace of the whole data space domain. Suppose an action is 
%acted 
applied
to $\tt N$, and $\tt N$ is split into the two child nodes 
%noted as 
$\tt N_1$ and $\tt N_2$, 
where
each child node denotes half of the original subspace. With actions further assigned to $\tt N_1$ and $\tt N_2$, the grandchild nodes of $\tt N$ split from $\tt N_1$ and $\tt N_2$ (4 grandchild nodes if $\tt N_1$ and $\tt N_2$ are all split) denote more fine-grained subspaces, respectively. We measure the distribution difference of the subspace denoted by $\tt N$ before and after the data or query is updated. We 
% then proceed to introduce how to 
model the data and query distribution shifts of Node $\tt N$ as follows:

\begin{figure}%[ht]
    \centering
    \begin{subfigure}[b]{.49\linewidth}
    \centering
        \includegraphics[width=1\linewidth]{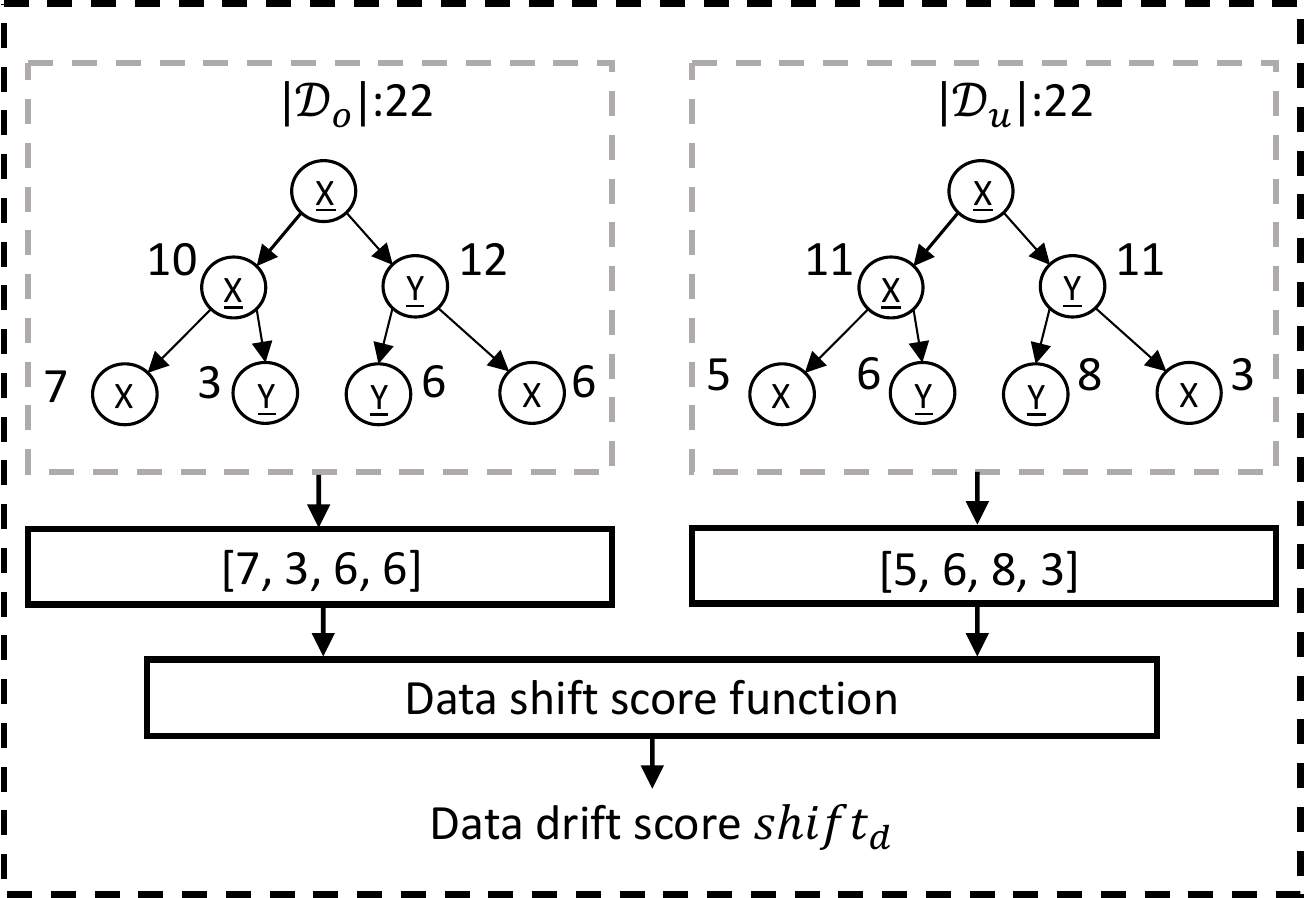}
        \caption{Measuring data shift}
        %data shift measuring.}
    \label{fig:retrain_detect_data}
    \end{subfigure}
    \begin{subfigure}[b]{.49\linewidth}
    \centering
        \includegraphics[width=1\linewidth]{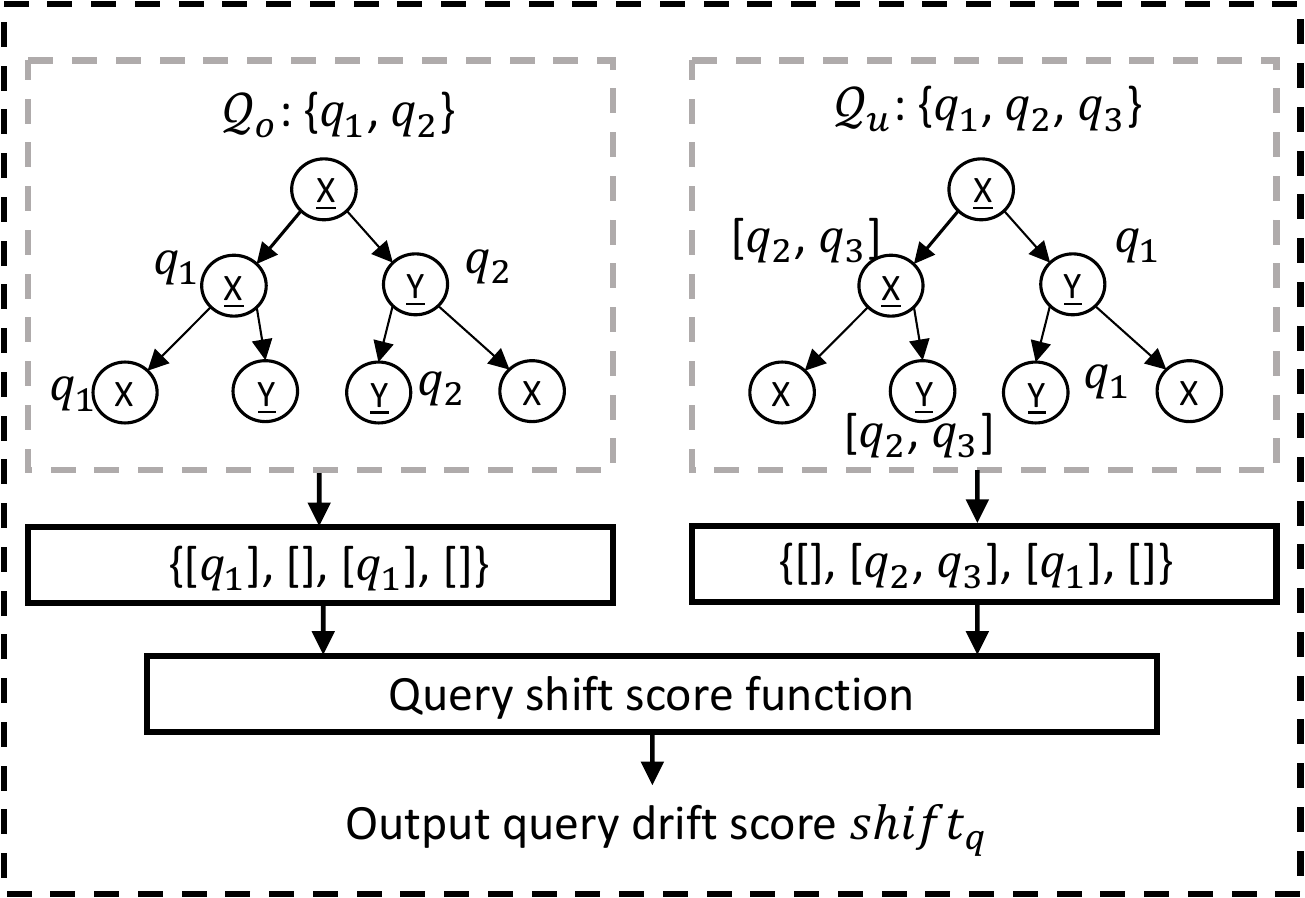}
        \caption{
        %Query shift measuring.}
        Measuring query shift.}
    \label{fig:retrain_detect_query}
    \end{subfigure}
    % \vspace*{-2mm}
     \caption{\bluecolor{
     %Distribution shift measuring w.r.t. BMTree structure.}}
     Measuring distribution shifts within the BMTree.}}
     
    \label{fig:retrain_detect}
    \vspace*{-2mm}
\end{figure}

\smallskip
\noindent
%\textbf{Data shift modeling.} 
\textbf{Modeling Data Shift.} 
Suppose the split level is set to $2$ that denotes that the data shift of a BMTree node $\tt N$ is computed w.r.t. the granchild nodes $2$ levels deeper than $\tt N$ (4 grandchild nodes by default).
% is split into 4 nodes, and t
The historical and updated datasets are denoted as $\mathcal{D}_o$ and $\mathcal{D}_u$, respectively. 
%An example is given in the 
Refer to 
Fig.~\ref{fig:retrain_detect_data}. 
% In Fig.~\ref{fig:retrain_detect_data}, 
The data points are split w.r.t. the nodes. On the left side (resp. right side) of the figure, the $\mathcal{D}_o$ (resp. $\mathcal{D}_u$) is split into four parts at Level $2$ in the subtree, and we represent the distribution as $ l^d_o = [\frac{7}{22},\frac{3}{22},\frac{6}{22},\frac{6}{22}]$ (resp. $ l^d_u = [\frac{5}{22},\frac{6}{22},\frac{8}{22},\frac{3}{22}]$), where the distribution is normalized by the cardinality of the dataset. Note that the representation of the data will change if the action assigned to the sub-tree is different. 
Then, we apply the Jensen–Shannon (JS) divergence~\cite{DBLP:conf/isit/FugledeT04} to measure the data shift, defined as follows:
% 
% Then, we apply the KL divergence to measure the data shift: 
\begin{equation}
    shift_d \triangleq D_{\tt JS}\left(l^d_o \bigl|\bigr| l^d_u\right) = 
    \frac{1}{2} \Big( D_{\tt KL}\left(l^d_o \bigl|\bigr| l^d_{mix}\right) + 
     D_{\tt KL}\left(l^d_u \bigl|\bigr| l^d_{mix}\right)\Big),
    % \sum_{i} \frac{l^d_o[i]}{|\mathcal{D}_o|} \cdot \log \left(\frac{l^d_o[i]}{|\mathcal{D}_o|} \cdot  \frac{|\mathcal{D}_u|}{ l^d_u[i]}\right)
\end{equation}
where $D_{\tt KL}$ denotes the 
% based on the
Kullback–Leibler (KL) divergence function,
% which is 
defined as $D_{\tt KL}(l^d_o \bigl|\bigr| l^d_u) = \sum_{i} 
l^d_o[i]
% \frac{l^d_o[i]}{|\mathcal{D}_o|}
\cdot \log  \left(
l^d_o[i]
% \frac{l^d_o[i]}{|\mathcal{D}_o|}
/  l^d_u[i]
% \frac{|\mathcal{D}_u|}{ l^d_u[i]}
\right)$. 
$l^d_{mix}$ denotes the mixed distribution of $l^d_o$ and $l^d_u$: $l^d_{mix} = \frac{1}{2}(l^d_o + l^d_u)$.  The larger the KL divergence $D_{\tt JS}(l^d_o \bigl|\bigr| l^d_u)$ is, the greater the difference between $\mathcal{D}_o$ and $\mathcal{D}_u$.}

% where $l^d_o[i]$ (resp. $l^d_u[i]$) denotes the $i$-th element of $l^d_o$ (resp. $l^d_u$). The larger the KL diversity $D_{\tt KL}(\mathcal{D}_o || \mathcal{D}_u)$ is, the more diverse $\mathcal{D}_o$ and $\mathcal{D}_u$ are.}

% We apply KL divergence to modeling the data distribution change.

\bluecolor{

\smallskip
\noindent
%\textbf{Query shift modeling.} 
\textbf{Modeling Query Shift.} 
For the query shift, we 
%also 
split the query set w.r.t. the BMTree nodes. Specifically, 
%the algorithm computes  
we compute
the center point of each query, and divide the query set according to the center point, e.g., if the query is denoted by: $(x_{min}, y_{min}, x_{max}, y_{max})$, the center point is computed as: $(\frac{x_{min} + x_{max}}{2}, \frac{y_{min} + y_{max}}{2})$. Then, according to the grandchild nodes of $\tt N$, the queries are split into 4 parts. 
%An example is given in 
Refer to Fig.~\ref{fig:retrain_detect_query}. On the left side of the figure, the old queryset $\mathcal{Q}_o = \{q_1, q_2\}$ is split into a list of four subsets $l^q_o = \{\{q_1\}, \{\}, \{q_2\}, \{\} \}$, while on right side the updated queryset $\mathcal{Q}_u = \{q_1, q_2, q_3\}$ is split as $l^q_u = \{\{\},\{q_2,q_3\}, \{q_1\}, \{\}\}$. Unlike the 
%shift measure of data which 
measure of data shift that
directly compares the 
%data point number 
number of data points
of each list element, 
%we first cluster the queries in each subset of the in the list 
%\walid{above sentence is incomplete. Do you mean: cluster the queries in each of the nodes? please verify if my fix below is correct.}
we cluster the queries in each node 
into different clusters w.r.t. the area and the aspect ratio, and thus compute the JS divergence of each list element. Then, the JS divergence is 
%then 
averaged across the list elements:
\begin{equation}
    shift_q  \triangleq \frac{1}{|l^q_o|} \sum_i { D_{\tt JS}}\left(l^q_o[i] \; \bigl|\bigr| \; l^q_u[i]\right) 
\end{equation}

% will modeling the query distribution by first assigning the queries into different sub-spaces.

After modeling the distribution shifts of 
both
data and 
%query, we then 
queries, next we 
introduce how to decide the subspaces to be retrained. When the retraining procedure begins, the method will recursively compute the data and query shifts of nodes in the BMTree $\tt T$ under a Breadth-First Search (BFS) order. We restrict the distribution shift computation 
%in 
to 
a limited depth of nodes in $\tt T$, since the nodes with 
%large 
larger
depth represent small data subspaces, and will contribute limited 
%performance improvement. 
improvement in performance.
%Under the situation when 
When
both data and query are shifted, the shift scores of data and query are weight summed 
%up 
as the final shift score: $shift_m = \alpha \cdot shift_d + (1 - \alpha) \cdot shift_q$, 
% \purplecolor
{where $\alpha$ is the weight parameter and is set by default to $0.5$.} 
%\walid{State here what alpha is and how its value is determined.}
Then, the nodes to be retrained are 
%then 
filtered based on 
%the 
this
score. 
% \redcolor{TODO, describe the actual weighted sum up mechanism.}
% 
% 
% We sort the BMTree nodes w.r.t. the weighted shift score.
% The nodes are then selected w.r.t. the shift score. 
% To be specific, 
A shift threshold $\theta_s$ is set to filter the BMTree nodes. Nodes with shift score lower than $\theta_s$ are not 
%considered to be 
retrained. 

%\subsection{Partial Retraining BMTree Nodes Detection}
\subsection{Deciding Which BMTree Nodes to Partially Retrain}

% the are then selected if the score is larger than the threshold.
 %We observe 
 Observe
 that the performance optimization potential of each node does not 
 %necessarily pure dependent 
 solely depend 
 on the distribution shift degree. Instead, we  propose to introduce a score based on  the change 
 %of 
 in
 average ScanRange  ($SR_{\tt T}$) before and after the data and/or queries change,  to 
 %fast measure 
 measure fast
 the possible optimization potentials 
 %to retrain the node. 
 when retraining a node.
 The optimization potentials score $\tt OP$  on Node $\tt N$ is computed 
 %w.r.t. the $SR$ 
 based on $SR$
 as follows:
\begin{equation} \label{eq:os_computation}
    \begin{aligned}
    {\tt OP}({\tt N}, \mathcal{D}_o, \mathcal{D}_u, \mathcal{Q}^\prime_o, \mathcal{Q}^\prime_u, {\tt T}) = 
    & \underset{q_u \in \mathcal{Q}^\prime_u}{\operatorname{avg}} SR_{\tt T}(q_u, \mathcal{D}_u)  \\
    & -
    \underset{q_o \in \mathcal{Q}^\prime_o}{\operatorname{avg}} SR_{\tt T}(q_o, \mathcal{D}_o),
    \end{aligned}
\end{equation}
where $\mathcal{Q}^\prime_o$ (resp. $\mathcal{Q}^\prime_u$) denotes the subset of historical (resp. updated) query workloads that the BMTree node $\tt N$ contains.  $\underset{q_u \in \mathcal{Q}^\prime_u}{\operatorname{avg}} SR_{\tt T}(q_u)$ and $\underset{q_o \in \mathcal{Q}^\prime_o}{\operatorname{avg}} SR_{\tt T}(q_o)$ denote the average $SR$ of $\mathcal{Q}^\prime_o$ and $\mathcal{Q}^\prime_u$ on BMTree $\tt T$, respectively. 
%We then 
Then, we 
compare the filtered BMTree nodes 
%level by level 
level at a time,
and select the nodes with 
%maximized 
maximum
$\tt OP$ score as the node to be retrained.

During 
%the 
retraining,
%procedure, 
to ensure a certain degree of efficiency improvement, a \emph{retraining constraint ratio} $R_{rc}$ is set to limit the retraining area of 
the 
retrained subspaces  compared to a 
%fully retraining 
full retrain
(e.g., if $r = 0.5$, the accumulated  area of the retrained subspaces should not reach half the whole space). 

% \redcolor{TODO}

\begin{algorithm}[ht]
    % \caption{Learned Space-Filling Curve Algorithm}
    \caption{
    %Retraining BMTree nodes detection.}
    Deciding on Which BMTree Nodes to Retrain.}
	\label{alg:bmtree_retrain_detect}
    \SetKwInOut{Input}{input}\SetKwInOut{Output}{output}
	\small
		\Input{$n$-dimensional old and updated dataset $\mathcal{D}_{o}$, $\mathcal{D}_{u}$,  old and updated training workload $\mathcal{Q}_{o}$, $\mathcal{Q}_{u}$, BMTree $\tt T$;}
		\Output{Retrain nodes of $\tt T$;}

        Initialize $\mathcal{L}, \mathcal{R} \leftarrow \emptyset, \emptyset$ \;
        $\mathcal{N} \leftarrow \{{\tt T}.root\}$\;
        \While{$(\mathcal{N} \not= \emptyset)$ $\wedge$ $(\mathcal{N}{\tt [0]}.depth < d_m$)}{
            ${\tt N} \leftarrow \mathcal{N}.pop(0)$\;
            $s \leftarrow {\tt ShiftScore}({\tt N}, \mathcal{D}_{o}, \mathcal{D}_{u}, \mathcal{Q}_{o}, \mathcal{Q}_{u}, {\tt T})$\;
            \If{$(s \geq \theta)$}{
            $\mathcal{L}.\operatorname{append}\bigl(( {\tt OP_N}, {\tt N})\bigr)$\;
            }

            $\mathcal{N}.\operatorname{append}(N.child\_nodes)$\;

            \If{$\mathcal{N}{\tt [0]}.depth > {\tt N}.depth$}{
                $\mathcal{L}.sort({\tt OP_N})$\;
                \For{$({\tt OP_N}, {\tt N}) \in \mathcal{L}$}{
                    \If{ $({\tt SpaceRatio}(\mathcal{R} + {\tt N}) < R_{rc})$}{
                    $\mathcal{R}.\operatorname{append}({\tt N})$\;
                
            }
            $\mathcal{L} \leftarrow \emptyset$
        }
            }
        }
        % \For{${\tt N} \in {\tt T}$, ${\tt N}.depth < d_{m}$}{
        %     $s \leftarrow {\tt shiftScore}({\tt N}, \mathcal{D}_{o}, \mathcal{D}_{u}, \mathcal{Q}_{o}, \mathcal{Q}_{u}, {\tt T})$\;
        %     $\mathcal{L}$.append($({\tt N}, s)$)\;
        % }
        % $\mathcal{L}$.sort()\;
        % $\mathcal{R} \leftarrow \emptyset$\;
        % \For{$(s, {\tt N}) \in \mathcal{L}$}{
        %     \If{$(s > \theta)$ $\wedge$ $({\tt SpaceRatio}(\mathcal{R} + {\tt N}) < r)$}{
        %         $\mathcal{R}.\operatorname{append}({\tt N})$\;
                
        %     }
            
        % }

        \Return $\mathcal{R}$
\end{algorithm}

The algorithm for 
%retraining BMTree nodes detection 
detecting the BMTree nodes that need retraining 
is Listed in Algorithm~\ref{alg:bmtree_retrain_detect}. 
% 
%It first 
First, it
initializes a queue $\mathcal{N}$ with the root node of $\tt T$ (Line 1). 
Then, the algorithm 
%is processed 
processes 
level 
%by level 
at a time
of the BMTree (Lines~3 --~14). The leftmost node  of $\mathcal{N}$ is popped (Line~4), then the shift score $s$ is computed by the $\tt ShiftScore$ function (with 
%the dataset update information and BMTree 
information about the dataset update and the BMTree
given, as described before) (Line~5). Then, the nodes satisfying the threshold $\theta$ 
%is then 
are
added to $\mathcal{L}$, and the $\tt OP$ of $\tt N$ is computed w.r.t. Eq.~\ref{eq:os_computation} (Lines~6 --~7). If a level of BMTree is evaluated (Line~9), the algorithm sorts $\mathcal{L}$ w.r.t. $\tt OP_N$ (Line~10). Then, if the nodes with greater $\tt OS$ score satisfy the retraining constraint ratio $R_{rc}$ (Line~12), 
%it is 
they are 
added to the retraining nodes list (Line~13).
 
% The shift score of each node is then computed by the $\tt ShiftScore$ function (with the dataset update information and BMTree given, as described before), under a BFS order with a constraint of maximum depth $d_m$ (lines 3--6). The nodes are then selected by detecting whether the score is larger than threshold $\theta$ and added to the retrain node list $\mathcal{R}$ if it does not break the space ratio constraint (lines 7--9).
% The nodes are then sorted w.r.t. $s$ (line 7). 

% In the experimental study, three settings are evaluated: (1) data shifts with query fixed, (2) query shifts with data fixed, and (3) both data \& query shift.
}

\bluecolor{

\subsection{BMTree Reconstruction and Retraining} \label{sec:bmtree_reconstruct_retrain}

We proceed to introduce the BMTree reconstruction and 
%the following 
retraining
procedures. 
First, we initialize the BMTree w.r.t. the 
%pre-retrained 
pre-trained
BMTree and the BMTree nodes needing retraining that have resulted from the retraining detection procedure. 
%to be retrained detection results.

% 
When the retrain domain (i.e., the to-be-retrained BMTree nodes) is decided, how to retrain the piecewise SFC while maintaining the rest of the designed piecewise SFC portion unchanged is non-trivial.
% , hereby facilitating the cost of performance maintainence.
Revisiting the design of seamless partitioning and BMP generation introduced in Section~\ref{sec:why_bmtree}, we propose to partially maintain the BMTree structure, and conduct the retraining procedure.

\begin{figure}[ht]
    \centering
    \includegraphics[width=1\linewidth]{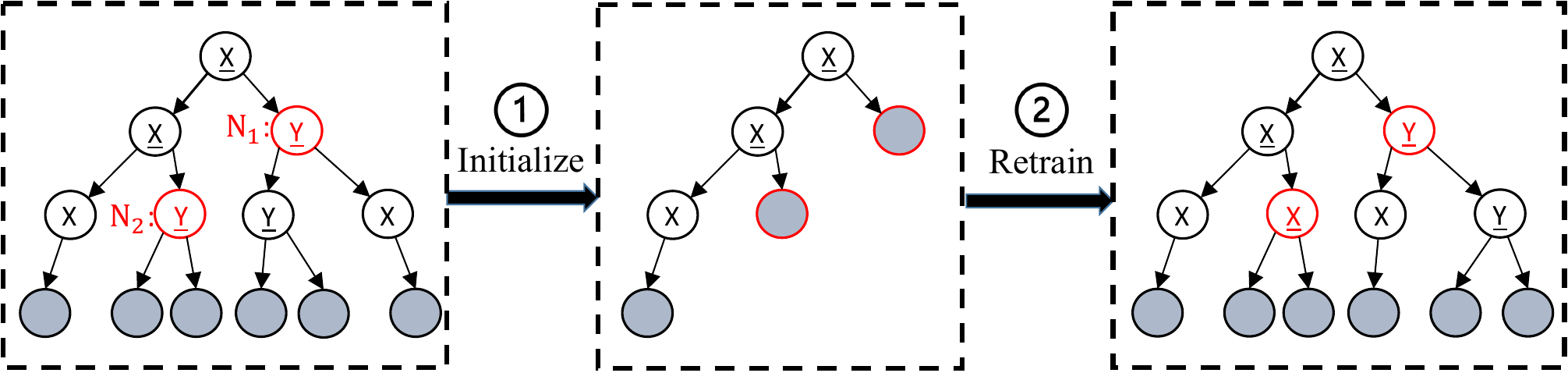}
    \caption{Partial BMTree retraining procedure.}
    \label{fig:bmtree_retrain_procedure}
\end{figure}

% The subspace that Section~\ref{sec:retrain_detect} decided to retrain can be modeled by nodes in the BMTree $\tt T$.
To conduct the retraining procedure, 
%we  first 
first, we 
manipulate the BMTree so that the partial retraining procedure could be finished by regenerating the BMTree structure. As 
%described 
in Fig.~\ref{fig:bmtree_retrain_procedure}, suppose $\tt N_1$ and $\tt N_2$ are two nodes that represent the subspaces to be retrained. In the initialization of the retraining  procedure, we delete the child nodes of $\tt N_1$ and $\tt N_2$ as well as the actions assigned to $\tt N_1$ and $\tt N_2$ while 
%remaining 
the other nodes remain unchanged. The 
%partially remained 
BMTree's
unchanged portion is 
%shown 
in the middle 
%side 
of Fig.~\ref{fig:bmtree_retrain_procedure}.

}

\bluecolor{

% \subsection{Retraining procedure}

%We then 
Then, we 
input the partially deleted BMTree $\tt T$ into an RL environment for retraining. We apply the MCTS method for retraining. Different from the environment introduced for training the  BMTree from scratch (Section~\ref{sec:mcts}), here the environment 
%here 
is 
%redesigned 
% \purplecolor
{developed to support partially training the BMTree, and is designed as follows:}
% designed
% according to the partial BMTree structure regenerate purpose as follows: 
%\walid{The previous sentence is unclear. Please rephrase. What do you mean by "partial BMTree structure regenerate purpose"?}

(1) \underline{\emph{State}}. We design the state of the retraining RL environment as the nodes to be retrained. In Fig.~\ref{fig:bmtree_retrain_procedure}, $\tt N_1$ and $\tt N_2$ are initialized in the state: $\tt S = \{ (N_1: None ), (N_2: None ) \}$, where $\tt None$ denotes that no action 
%is taken 
has been taken 
yet. (2) \underline{\emph{Action}}. Then, we design the action space as the actions assigned to $\tt N_1$ and $\tt N_2$. 
% \purplecolor
{After the action is decided, the child nodes of $\tt N_1$ and $\tt N_2$ are generated,}
% acted to $\tt S$, 
%\walid{What do you mean by: acted to S? Also, what is S? you need to have clearer decription.}
and the child nodes of $\tt N_1$ and $\tt N_2$ represent the transited state. 
%Here, different from the former RL environment, t
The nodes of $\tt T$ in one state do not require to have identical depth. (3) \underline{\emph{Reward Design}}. 
%As for the reward generation, w
We apply the updated dataset and query workloads to generate the reward 
%which is computed following the reward generation introduced 
as in Section~\ref{sec:mcts}. The RL policy training environment will 
%then 
produce a regenerated BMTree noted as $\tt T^{\prime}$ w.r.t. the redesigned state, action, and reward. Further, to improve 
%the 
efficiency, during the partial retraining procedure, we only generate reward w.r.t. the queries in $\mathcal{Q}_u$ that fall in the retrained nodes.
}

\bluecolor{
%\subsection{Partial Retraining Workflow}
\subsection{Workflow for Partial Retraining}

We summarize the partial retraining procedure 
as follows: 
(1)~If there exists BMTree node satisfying the shift score filter, the node to retrain detection procedure is conducted (Alg.~\ref{alg:bmtree_retrain_detect}). Then, the retraining RL environment is initialized and the BMTree is regenerated based on the updated data and query.
%as introduced above. 
Particularly, if the retraining result does not meet 
%the 
expectation (e.g., with optimization ratio less than $1\%$), the procedure will select and retrain more untrained nodes 
%following 
as in
Section~\ref{sec:bmtree_reconstruct_retrain}.

\begin{algorithm}[ht]
    % \caption{Learned Space-Filling Curve Algorithm}
    \caption{BMTree Structure Partial Retraining.}
	\label{alg:bmtree_retrain}
    \SetKwInOut{Input}{input}\SetKwInOut{Output}{output}
	\small
		\Input{$n$-dimensional old and updated datasets $\mathcal{D}_{o}$, $\mathcal{D}_{u}$,  old and updated training workloads $\mathcal{Q}_{o}$, $\mathcal{Q}_{u}$, BMTree $\tt T$;}
		\Output{Retrained BMTree $\tt T^{\prime}$;}

        \If{$\exists {\tt N} \in {\tt T}$ with $shift_m({\tt N}) \geq \theta_s$}{
            $\mathcal{R} \leftarrow {\tt Retrain\_Detector}({\tt T},\mathcal{D}_{o},\mathcal{D}_{u},\mathcal{Q}_{o},\mathcal{Q}_{u})$\;
        $S, {\tt T_p} = {\tt Initial}(\mathcal{R},{\tt T})$\;
        $\mathcal{Q}^\prime_u \leftarrow \{ q| q \in \mathcal{Q}_u \wedge \exists {\tt N} \; \text{s.t. }  {\tt N} \text{ contains } q\}$\;
        ${\tt T^{\prime}} \leftarrow {\tt MCTS}({\tt T_p}, \mathcal{D}_{u}, \mathcal{Q}^\prime_{u})$\;
        } 
        if limited optimization then Retrain $\tt T^\prime$\;
		    
        \Return $\tt T^{\prime}$
\end{algorithm}

The pseudo-code of the retraining procedure is 
%also 
given in Alg.~\ref{alg:bmtree_retrain}. If there exists a BMTree node that satisfy the shift score requirement (Line~1), it first detects the BMTree nodes to be retrained (Line~2). Then,  the RL retrain environment, the partially deleted BMTree $\tt T_p$, and the initial state for RL training are initialized (Line~3), and the queries contained by the 
%retrained 
to-be-retrained
BMTree nodes are selected for retraining (Line~4). The MCTS algorithm with the environment redesigned as above is applied to regenerate the BMTree w.r.t. the updated database and the query workload (Line~5). If $\tt T^\prime$ has limited 
%optimization, 
performance
enhancement, 
%where we set 
when $\tt T^\prime$ 
optimizes 
%the 
ScanRange 
by
less than $1\%$ 
improvement
compared with 
the original 
$\tt T$ (Line~6).

% \smallskip
% \noindent
% \textbf{Discussion.} 
After retraining the BMTree nodes is complete, data is needed to update the SFC values. With BMTree 
being 
partially retrained, only the data located in the retrained subspaces should be updated. The retraining procedure also reduces the cost of the following index update procedure.

}

%% file: sections/analysis.tex
\section{Analysis and Discussion}
\label{sec:theoretical}

\bluecolor{
\noindent \textbf{Injection and Monotonicity.} We prove that piecewise SFCs modeled by the BMTree satisfy both injection and monotonicity properties. The proof is detailed in \cite{li2023towards}.
}

\noindent\textbf{Time Complexity Analysis.} 
We provide time complexities for SFC value computation and MCTS-based BMTree construction. The time complexity for computing the SFC value of $\mathbf{x}$ using the constructed BMTree is $O(M)$, where $M$ is the length of $C_{\tt T}(\mathbf{x})$. This complexity is comparable to other SFCs described by BMPs.
% We present time complexities for SFC value computing and MCTS based BMTree construction time, respectively. 
%  Once we construct the BMTree, we use it for computing the SFC value of $\mathbf{x}$ with the time complexity $O(M)$, where 
%  $M$ denotes the length of $T_{\tt T}(\mathbf{x})$. We note that the complexity is in line with other SFCs described by BMPs for query processing. 
% 
 %Each computation costs less than 0.01ms under our experiment setting.
% 
For BMTree construction, the complexity of 
% Algorithm \ref{alg:mcts}
the MCTS BMTree construction
is $O\left( M \cdot \left(N + |\mathcal{D}_s| \left(M +\log|\mathcal{D}_s|\right) + |\mathcal{Q}|\right) \right)$, 
% where $M$ is the bit length of the 
% %merged 
% SFC value, 
where $N$ is the child node size of the policy tree,
%each time takes action, 
$|\mathcal{D}_s|$ and $|\mathcal{Q}|$ correspond to the size of the sampled data and query workloads. 
It takes at most $M$ actions to construct the BMTree. In each step of choosing an action, the selection step is bounded by the child node size $O(N)$; the simulation time corresponds to the computation of ScanRange
%, which 
that
takes $O(M \cdot |\mathcal{D}_s|)$ for SFC value computing, $O(|\mathcal{D}_s| \cdot \log(|\mathcal{D}_s|)$ to sort data, and $O(|\mathcal{Q}|)$ to compute ScanRange for each query.
% 
% it costs $|\mathcal{D}_s| \log(|\mathcal{D}_s|)$ time to sort $\mathcal{D}_s$ based on the computed SFC value (costs $O(M \cdot |\mathcal{D}_s|)$ where $O(M)$ is SFC computation complexity.
% as analyzed below). %The reward is computed based on $\mathcal{Q}$.
% 
% the complexity of training a BMTree costs $O\left(N \left(K (|D| \log{|D|} + |Q||D|) + 2^K \right)\right)$, where $|D|$ and $|Q|$ denote the cardinality of database and query workload, respectively, $N$ denotes the number of training episodes, and $K$ denotes the max depth that RL policy takes actions. The analysis is given below. For each episode, the time cost of training consists of (1) that of computing reward (i.e., ScanRange) at each depth, where the algorithm first sorts the data by the SFC values with the complexity of $O(|D| \log{|D|})$. Then $|Q|$ queries are applied to compute the ScanRanges on the sorted data, with the complexity of $O(|D||Q|)$. The sorting and reward computing happens in each depth with $K$ times in total; (2) that of constructing states and actions at each node, which is bounded by the total number of nodes in the first $K$ levels, with the complexity of $O(2^K)$. We note that the training procedure only executes once, and the best BMTree during the training is collected.}
% 
\bluecolor{For BMTree update, suppose the BMTree construction complexity is noted as $T({\tt BMT\_Train})$. The BMTree retraining time is bounded by $R_{rc} \cdot T({\tt BMT\_Train})$ with $R_{rc}$ as the retraining constraint ratio, since the ratio will limit the retrain nodes and the depth of the BMTree that needs to be generated by the MCTS algorithm.
}

%% file: sections/evaluation2.tex
\section{Evaluation}
\label{sec:evaluation}

\bluecolor{Experiments aim to evaluate the following:
% \marginpar{\#2 D2}
(1)~The effectiveness of BMTree's design, including (i)~Evaluating the proposed piecewise SFC method vs. existing SFCs when applied 
%for 
to the 
SFC-based indexes vs. 
the 
other indexes,
% (1) evaluate if the proposed piecewise SFC method outperforms the existing SFCs for query processing, and how the SFC based indexes perform over other indexes,
% 
(ii)~The BMTree under different settings (e.g., scalability, dimensionality, and aspect ratio), and
% (2) evaluate how BMTree is affected under different settings (varying data and query size, dimensionality, distribution shift, etc.),
%necessity of 
(iii)~Components of the BMTree by evaluating different BMTree variants.
(2)~The effectiveness of partial retraining, including (i)~Evaluating the performance of partial retraining 
while 
varying distribution shift settings, and (ii)~{Evaluating the choice of parameters, 
%by varying parameters such as 
e.g., the 
retraining constraint and the shift score threshold study during partial retraining.}
%\walid{Unclear. What does this say? Please rephrase to make clearer.}
}
% (3) evaluate different BMTree variants to identify the necessity of designed components,
\iffalse
and (5) suitability of ScanRange ($SR$) as an I/O replacement.
\fi

\subsection{Experimental Setup}
\label{sec:setup}

\begin{table}%[ht]
	\centering
	% \vspace*{-2.5mm}
	\caption{Experiment Parameters.}
	\label{tbl:experiment_parameters}
	% \vspace*{-3.5mm}
	\begin{tabular}{c|c}
	\hline
	Parameter  & Value \\
	\hline \hline
	Data & GAU \textbf{UNI} OSM-US TIGER \\
	Query & \texttt{GAU} \textbf{\texttt{SKE}} \texttt{UNI} \\
	Sampling rate & 0.01 0.025 \textbf{0.05} 0.075 0.1 \\
	\change{$\#$ Training $Q$} & \change{100 500 \textbf{1000} 1500 2000} \\
	Max depth & 1 5 \textbf{10} 15 20 \\
% 	Rollouts & 1 5 \textbf{10} 15 20 \\
% 	Training episodes & 100 200 \textbf{300} 400 500 \\
 	\hline
	\end{tabular}	
    \vspace*{-2mm}
\end{table}

% \smallskip
\noindent\textbf{Dataset.} We conduct experiments on both synthetic and 
%two 
real datasets. For synthetic datasets, we generate data points in the two-dimensional data space with a granularity size of $2^{20} \times 2^{20}$
%, which 
that 
follow either uniform  (denoted as UNI) or Gaussian distributions (denoted as GAU) with $\mu_d$ as the center point of the space domain.
%
%Real world datasets are obtained from OpenStreetMap API~\citep{osm} (denoted as OSM-US) and the US Census Bureau TIGER project~\citep{tiger} (denoted as TIGER). 
Real data OSM-US contains about 100 
%millions of 
Million
spatial objects in the U.S. extracted from OpenStreetMap API~\citep{osm}, and TIGER~\citep{tiger} contains 2.3 
%millions of 
Million
water areas in North America cleaned by SpatialHadoop~\citep{eldawy2015spatialhadoop}.

% \smallskip
\noindent\textbf{Query Workload.} 
We follow~\cite{nishimura2017quilts,qi2020effectively} to generate query workloads. 
%Specifically, 
We generate 
%different 
various
types of window queries, 
%and each type of queries 
where each query type
has a fixed area selected 
%out of 
from 
$\{2^{30}, 2^{32}, 2^{34}\}$ and a fixed aspect ratio selected 
%out of 
from
$\{4, 1, 1/4\}$; Each workload comprises multiple 
%types of queries, which have
query types with
different combinations of  areas and ratios.
% Since our method requires the query workload during the training, we consider the following different query workloads, and use them to guide the training. By following~\cite{nishimura2017quilts}, we consider window queries with different areas ($\{2^{30}, 2^{32}, 2^{34}\}$) and height-width ratios ($\{4, 1, 1/4\}$), and generate different combinations of the areas and height-width ratios for querying. 
%In addition, w
We generate 
%query 
queries
with Uniform (\texttt{UNI}) and Gausian (\texttt{GAU}) distributions (same as  
%by following work like 
in~\cite{ding2020tsunami,qi2020effectively}),
%Specifically, the different types of queries in one workload follow the same distribution, either 
%including the uniform distribution (denoted as ) and the Gaussian distribution (denoted as \texttt{GAU}). 
%\walid{Omitted the above text as it is repeated from the begin of this section under Datasets subtitle.}
We also generate a skewed workload (denoted by \texttt{SKE}), in which queries follow Gaussian distributions with different $\mu$ values.
%\walid{Why didn't you mention that at the begin of the experimental setup section? Please move it there to be with UNI and GAU.}

% In addition, we explore different distributions for the query workload, including uniform distribution (denoted as \texttt{UNI}), Gaussian distribution (denoted as \texttt{GAU}), and skew distribution (denoted as \texttt{SKE}).
% % and designed distribution (denoted as DES).
% For \texttt{SKE}, we design three query types with (height, width) at: $\{(2^{14},2^{16}),(2^{16},2^{16}), (2^{18},2^{16})\}$, and we locate different query types with certain degree of query skew following Tsunami~\cite{ding2020tsunami}.

% For DES, we generate the query workload following the heuristic restriction of Quilts~\cite{nishimura2017quilts}, where the restriction means that each dimension in a small query class (e.g., $(2^{14},2^{16})$) should be dominated by each corresponding dimension in a large query class (e.g., $(2^{18},2^{16})$). Note that the three query classes we designed can be dominated one by one and satisfied the pruning algorithms in Quilts. In addition, we also test the performance of $k$NN query in the experiments.

% \smallskip
\noindent\textbf{Index Structures.} To evaluate the performance of the proposed piecewise SFC compared with the existing SFCs, we integrate the proposed piecewise SFC and the baseline SFCs into both traditional indexes and learned index structures.
%so that they can be compared in the same framework. 
%to support query processing. For traditional index, 
First, we integrate the piecewise SFC (and baseline SFCs) into the PostgreSQL database system and a built-in B$^+$-Tree variant in PostgreSQL is employed with SFC values as key values. Second, we use a learned spatial index, RSMI~\cite{rsmi2020code}, to compare the performance of the piecewise SFC 
%and 
against the 
baseline SFCs 
%when they are used in 
within
RSMI. 
%Here, t
The B$^+$-Tree of \revision{PostgreSQL} is a disk-based index 
%and 
while
the released implementation of RSMI~\cite{rsmi2020code} is memory based. We choose them to evaluate the performance of the piecewise SFC under 
%different 
various
scenarios. 
%
\iffalse
We 
%also 
combine the BMTree into ZM~\citep{wang2019learned}, another SFC-based learned index, to further demonstrate BMTree’s applicability.
\fi

%\revision{We also evaluate how the SFC based indexes (RSMI, ZM~\citep{wang2019learned}), {\jn with and without BMTree}, perform compared with traditional indexes, including {\jn STR}~\cite{DBLP:conf/icde/LeuteneggerEL97} {\jn (a bulk-loaded R Tree)}, R* Tree~\cite{beckmann1990r}, Grid-File~\cite{nievergelt1984grid}, and Quad-tree~\cite{finkel1974quad}.}

%We use two different types of indexes to evaluate the generalization of the learned SFC.  

% \smallskip
\noindent\textbf{SFC Baselines.} 
We choose the following SFC methods as our baselines:
(1)~Z-curve~\citep{mokbel2003analysis, skopal2006new};
%is a popular SFC, whose mapping function is given in Equation~\ref{eq:z_value}. 
(2)~Hilbert Curve~\citep{moon2001analysis}; 
%is a hierarchically constructed curve, which recursively splits spaces and generates the curve order. 
% Note that the Hilbert curve is not monotonic, and thus the query results based on the Hilbert curve are approximate.\\
(3)~QUILTS~\citep{nishimura2017quilts}.
%(Section 2.2).
%extends the bit interleaving technique of Z-curve and takes more curves into consideration. QUILTS considers SFCs with different BMP and selects the best one validated on some pre-defined queries for evaluation. 
% \\
% $\bullet$ R* Tree~\citep{beckmann1990r}. We compare the well known spatial index R* Tree with SFCs based on RSMI.
% $\bullet$ BMTree-G is a variation of the proposed BMTree, where the BMTree-G only gets a \underline{G}lobal reward when a complete tree is built. Note that in order not to exceed memory, we bound the tree depth up to 20, and then the heuristic rule as Z-curve is adopted to build the whole tree.

% \smallskip
\noindent\textbf{Evaluation Metrics.}
For experiments conducted with PostgreSQL, we use the I/O cost (I/O) recorded by PostgreSQL 
%system 
and 
the 
Query Latency (QL). For experiments under RSMI, we report the
%node access number 
number of node accesses
of its tree structure and QL for a fair comparison by following~\cite{rsmi2020code, gu2021rlr}.
% by following~\cite{rsmi2020code}, where we perform window query and $k$NN query with the two index structures (i.e., B+Tree and RSMI), and BA and RT are evaluated accordingly. 

% \smallskip
\noindent\textbf{Parameter Settings.} Table~\ref{tbl:experiment_parameters} lists the parameters used in our experiments, and the default settings are in bold. %\revise{
{We set 
% the rollout number 
the number of rollouts (as described in Section~\ref{sec:rollouts})}
%\walid{what is the rollout number? explain. Do you mean: The number of rollouts? What is a rollout? explain.}
in MCTS at 10 by default. The max depth is the depth of 
the
BMTree built via the RL model; the sampling rate (0.05 by default) is the rate of sampling training data for computing the ScanRange.
% for reward,

%}

% We generate 300 episodes for policy learning, and train the BMTree on given databases with  uniform distribution as the default, where the max depth denotes the maximum depth of BMTree built via the RL model; sampling rate denotes the sampling rate of the training data for computing the ScanRange for reward,  with 0.05 as the default; 
% 1000 training window queries are used. For the actor network, we use a feed forward neural network with 2 hidden layers followed by one output layer, and each hidden layer has 256 neurons with the Tanh activation layer. The output layer has $n$ neurons corresponding to the action space, and the Softmax layer is adopted normalizing probability. For the critic network, it follows the identical model structure as the actor network except that the output layer has only one neuron to provide an expected cumulative reward value. Two networks do not share parameters. We adopt Adam stochastic gradient descent for training, and the learning rates for the actor and critic networks are set to 0.001 and 0.05, respectively. The reward discount factor is set to 0.99 and  $\epsilon$ for PPO algorithm is set to 0.2. 

% \input{VLDB Learned SFC/sections/experiment_query_execute}

% \smallskip
\noindent\textbf{Evaluation Platform.} We train the BMTree with PyTorch 1.9, Python 3.8. 
% All SFCs and indexes are coded in C++ for fair evaluation.
The experiments are conducted on an 
%80-cores 
80-core
server with an Intel(R) Xeon(R) Gold 6248 CPU@2.50GHz 64.0GB RAM, no GPU resource is leveraged.
% to train the model. 
% The datasets and code are available via the link\footnote{\url{https://github.com/gravesprite/Learned-BMTree}} to reproduce our work.
% \footnote{\url{https://www.dropbox.com/scl/fo/9q1qqopggkbz17h1g8it2/h?dl=0&rlkey=vw0e7hjyke5etlnkg9unks3fr}} to reproduce our work.

%% file: sections/result2.tex
% \subsection{Experimental Results}

%\noindent\textbf{Effectiveness results.
%varying data and query workload.
%}
%Figures~\ref{fig:bptree_window_query} and \ref{fig:rsmi_effect} report the result of window query evaluation. 
\subsection{Evaluation of the BMTree}
%design} 
\label{sec:results}
%\noindent 

\subsubsection{Effectiveness}
%study}
%{\bf Effectiveness study}. 
This experiment is to compare the effectiveness of the learned piecewise SFC in query processing 
%with 
against the
other SFCs 
under both the PostgreSQL and the RSMI environments. 
%To evaluate window queries, 
We also compare 
{an SFC-based index combined with the BMTree}
%\walid{What is "optimized SFC-based index"? Please state explicitly. Do you mean the "learned piecewise SFC"? if so, you need to stick to one name.}
%with 
against the 
other indexes. 
% including R-tree variants, Grid-File, and Quad-tree.
For each experiment, we use 1000 windows queries, {\chengr which} are randomly generated by following respective distributions for training, and another 2000 different window queries following the same distribution for 
%performance 
evaluation.
% , {\chengr which} are generated  {\chengr by following the same distribution}. 

\input{sections/experiment_query_execute}

\noindent 
\textbf{Results on PostgreSQL.} Figures~\ref{fig:bptree_block_access} and~\ref{fig:bptree_response_time} show the I/O and QL on window queries. 
\change{ 
To ensure PostgreSQL conducts \texttt{indexscan} during querying,
both the \texttt{bitmapscan} and \texttt{seqscan} in PostgreSQL are disabled.}
%
%Difficult to provide accurate query with breaking of monotonicity,
% where Hilbert breaks monotonicity, and thus it cannot provide the exact query results. W
% Since the algorithm for processing windows query cannot return accurate results on the Hilbert curve without additional structure, 
% which violates the monotonicity property, 
We do not include the Hilbert curve for this experiment 
%since
as
the Hilbert curve requires additional structure and dedicated algorithm for returning accurate results for window queries, and PostgreSQL does not support them for the Hilbert curve. 
%querying with Hilbert curve with only SFC values provided. 
% Note that in the previous work~\cite{nishimura2017quilts} QUILTS is not compared with Hilbert curve too.
%However,  we include the Hilbert curve for the experiments on RSMI and index comparison study.
% , which returns approximate results.  
% Aslo, it is hard to compare PostgreSQL's build-in B+ tree and R-tree fairly. We omit the R-tree here and leave it to the RSMI experiment.

%
Observe that
 the proposed BMTree
 % that the proposed piecewise SFC modeled by BMTree
 consistently outperforms the baselines in all the combinations of data and query distributions
in terms of both I/O and QL. 
Between the two baselines, QUILTS performs worse for 
the 
\texttt{SKE} workload, and performs similarly as the Z-curve for \texttt{UNI} and \texttt{GAU} workloads.
\revision{
%This is because 
The reason is that our query workload contains queries with 
%very 
different aspect ratios (e.g., $4$ and $1/4$), 
%by following~\cite{qi2020effectively}, 
rather than queries with similar aspect ratios as it is used in QUILTS~\cite{nishimura2017quilts}. 
%which does not meet the requirement of QUILTS' algorithm. 
QUILTS can only choose queries with a particular aspect ratio to optimize,  and 
%thus
this
results in poor performance for queries with different aspect ratios.}
% results in bad performance among other queries.
% QUILTS can only optimize limited queries with particular aspect ratio under our experiment setting where distinguish queries with different aspect ratios exists and breaking the requirement of QUILTS' requirement. 
% QUILTS will perform better than Z-curve under workload with similar ratio based on our experimental results.
% We also conducted experiments where the queries are with similar ratio, and QUILTS performs better than Z-curve. 
% }
% This might be because
% (1)~the heuristic rules used in QUILTS for selecting a BMP may not work for many cases, and (2)~QUILTS easily overfits with the training queries and is not working on the evaluation queries. 
%
%\change{
The BMTree outperforms the Z-curve by 5.2\%--39.1\% (resp. 7.7\%--59.8\%, 6.3\%--29.8\% and 25.1\%--77.8\%) in terms of I/O 
%and similar of query time 
on UNI (resp. GAU, OSM-US ,and TIGER) datasets across 
%different types of 
the various workloads. The results in terms of QL are consistent with those of I/O. 
%
%In particular, BMTree outperforms Z-curve under the \texttt{SKE} workload by 39.1\%, 59.8\%, 20.5\%, 29.51\% in UNI, GAU, OSM-US, and TIGER datasets.
%
BMTree's superior performance is because 
%it utilizes both the data property and the workload property to learn BMPs to generate more effective SFCs. %
(1)~The BMTree 
%is able to generate 
generates
piecewise SFCs
% , i.e., generating different BMPs 
to handle distinct query distributions,
% in different subspaces
and (2)~The BMTree is equipped with effective learning 
%technique 
to generate BMPs and subspaces. 
%
%We n
Notice that under the \texttt{UNI} workload, the BMTree outperforms the Z-curve by  25.1\%  on TIGER while it only outperforms Z-curve slightly on the other three datasets. This is 
%as 
expected: Under 
%a uniform 
the \texttt{UNI}
query workload, the BMTree can only make use of the data 
%distribution, 
but not the query distributions to optimize  performance; TIGER is very skewed and the BMTree can capture TIGER's skewed data 
%feature of . 
nature.

\noindent
\textbf{Results on RSMI.} 
%RSMI, as a learned index, returns approximate  results for window queries.
 %
The original RSMI~\cite{qi2020effectively} uses the Hilbert curve, and we include it as a baseline for this experiment as RSMI returns approximate results for all curves.  All the curves achieve comparable recall (99.5\% or above) using RSMI's algorithms for window queries.
Figures~\ref{fig:rsmi_window_access} and~\ref{fig:rsmi_window_time} 
show the 
%node access number 
number of node accesses
and QL for all 
%the 
curves 
when
using RSMI. 
%We o
Observe that 
the
BMTree consistently outperforms all 
%the 
baselines. 
%For example,  
The 
BMTree outperforms the Z-curve by 18.2\%--29.0\% (resp. 13.7\%--28.4\%, 13.5\%--26.5\%, and 2.8\%--25.3\%) in terms  of 
%node access number 
the number of node accesses 
on the UNI (resp. GAU, US-OSM, and TIGER) datasets.
%Still, BMTree is likely to have good optimize performance when query and/or data are under certain degree of skewness.
%We also o
Also, observe that the Hilbert curve achieves similar performance 
%with 
to that of the 
BMTree on 
the
GAU dataset
%, which 
that
could be %contributed
attributed
to its good 
%toleration 
tolerance 
to 
%the skewness
data skew~\citep{xu2014optimality}.

%in some cases (with \texttt{SKE} workload, GAU or OSM-US dataset). This is because of Hilbert curve's good toleration of the skewness~\citep{xu2014optimality 

\noindent  {\bf 
%Comparing 
Comparison 
with Other Indexes.}
% \marginpar{\#1 R1W2D7}
% \marginpar{\#2 R2O3}
% \marginpar{\#3 E2}
% We compare different indexes under the same settings (in memory, same node size). The result is reported in Figure ...
{We compare 
against
the performance of two SFC-based indexes, RSMI and ZM combining our BMTree,}  with baseline indexes including (1) two R-tree variants: {\jn STR}~\cite{DBLP:conf/icde/LeuteneggerEL97} and R* Tree~\cite{beckmann1990r};
and (2) two partition-based methods: Grid-File and Quad-Tree. 
%\walid{Where is the BMTree here?}
\bluecolor{The 
%experimental 
results are given in \cite{li2023towards}
%, which 
that
reveal the generality of 
the 
BMTree on enhancing query performance 
combining different indexes.
%\walid{What do you mean by "Combining different indexes? Where is that combination? Please highlight and explain in the  text above.}

}

\input{VLDB_Learned_SFC/figures/knn_query_performance.tex}

\noindent 
\textbf{Effect on $k$NN Queries.}
The piecewise SFC is learned to optimize window queries. 
% and we would like t
%To see 
Here, we investigate 
%whether it has negative
its
influence on the performance of 
%the 
$k$NN queries.
We generate 1,000 $k$NN query points 
% randomly by 
following the  data distribution, and we apply the $k$NN algorithm ~\citep{qi2020effectively} in PostgreSQL with $k$ set 
%at 
to
25.
%We make the following observations. 
%
 % show the I/O cost and query latency for the $k$NN query. 
% 
We report the I/O and QL ratios in Figure~\ref{fig:bptree_knn_block_access} and \ref{fig:bptree_knn_response_time}
%, which 
that
are the ratio of results of 
the
different curves divided by the result of the Z-curve.
% 
%We observe that the BMTree is comparable with the baselines: 
The BMTree performs slightly better than the baselines on GAU and OSM-US while the Z-curve is slightly better on UNI and TIGER. Thus, 
%although 
while
the piecewise SFC 
% (or BMTree)
is optimized for window queries, 
%the performance of the 
its $k$NN query performance
is not compromised.

% \noindent
\revision{
% TODO The result of jointly optimizing window query and $k$NN query is reported in Figure \ref{fig:exp_optimize_knn}. 
%
\noindent {\bf Optimizing Window 
and $k$NN queries.}
% \marginpar{\#2 R1O1}
% \marginpar{\#3 W2}
We evaluate the performance when window queries and $k$NN queries are optimized together. To optimize 
%our BMTree method 
the BMTree
for $k$NN queries, we convert $k$NN queries into window queries by following \cite{qi2020effectively} and 
% thus generate the $k$NN training workload. 
{\chengr include them in the training workload.}
%We then 
Then, we 
vary the weight of the {\chengr objective based on $k$NN queries} relative to {\chengr 
% that based on the 
window queries} from $0\%$ to $100\%$ during training. 
% The results are reported in 
Figures~\ref{fig:knn_optimize_window} and~\ref{fig:knn_optimize_knn} 
%reports 
give the 
%reports the 
window 
and $k$NN query I/Os. 
%We o
Observe that as the weight 
% of the $k$NN objective 
{\chengr increases}, the window query I/O {\chengr tends} to increase while the $k$NN query I/O {\chengr tends} to decrease.
%We a
Also, observe that when the weight is between $25\%$ and $75\%$, the performance of the window query only mildly degrades, while the {\chengr performance of $k$NN query is better than that based on the Z-curve.}
% better I/O cost than the Z-curve.
The results show {\chengr the potential of
%our method
the BMTree} to 
%optimzie 
optimize
the two query types 
%of queries 
together. 

%achieve good trade off {\chengr between performances of handling} different {\chengr types of} queries.

% We evaluate the performance when window query and $k$NN query are jointly optimized. To enable our BMTree method to optimize $k$NN, we convert $k$NN queries into window queries following \cite{qi2020effectively} and thus generate the $k$NN training workload. 
% We then varied the weight of the $k$NN query objective relative to the window query objective from $0\%$ to $100\%$ during training. 
% % The results are reported in 
% Figure \ref{fig:knn_optimize_window} reports the window query I/O cost and Figure \ref{fig:knn_optimize_knn} reports the $k$NN query I/O cost. We observe that as the weight of the $k$NN objective increased, the window query I/O cost tended to increase while the $k$NN query I/O cost tended to decrease. We also observe that when the $k$NN objective weight is between $25\%$ and $75\%$, the performance of the window query only mildly degraded, while the $k$NN query results in a better I/O cost than the Z-curve.
% The result shows a potential to achieve good trade off when jointly optimizing different queries.
}

\subsubsection{Effect of Varying the Settings}
%\noindent{\bf  Varying settings.}
\revision{ 
We evaluate the performance of the BMTree under various settings: dataset/query size,
\iffalse
distribution shifting, \fi
dimensionality, and window aspect ratio. More settings can be found in \cite{li2023towards}. }

\begin{figure}[ht]
% \vspace*{-3mm}
\hspace*{-4mm}
\begin{subfigure}{.49\linewidth}
    \centering
    \hspace*{1mm}
    \begin{tikzpicture}[scale=0.55]

    \begin{axis}[
    width=8.5cm,
    height=4cm,
        xlabel= Dataset Size (million),
        ylabel= I/O Cost, %ymode=log,
        every axis y label/.style={
        at={(-0.05, 1.3)},
        anchor=north west,
        },
        xmin=0, xmax=150,
         ymax=110000,
        xtick={0.1,  5, 10, 20, 50, 100, 150},
        ytick={10000, 50000, 100000},
        % y tick label style={/pgf/number format/sci, scaled ticks=base 10:0},
        legend style={at={(0.0,1.0)},,anchor=north west}
        ]
    
    \addplot[smooth,mark=*,red] plot coordinates {
        (0.1, 48.073888888888895)
        % (1, 453.9511111111111)
        (5, 2249.9694444444444)
        (10, 4498.448333333334)
        % (15, 6748.995000000001)
        (20, 8995.433333333332)
        (50, 22485.096666666668)
        (100, 44964.79277777777)
        (150, 67441.42333333332)
    };
    \addlegendentry{Z-curve}

    \addplot[smooth,color=blue,mark=star]
        plot coordinates {
        (0.1,  65.83722222222222)
        % (1, 633.151111111111)
        (5, 3142.0922222222225)
        (10, 6281.713333333333)
        % (15, 9423.906111111111)
        (20, 12560.812777777777)
        (50, 31398.509444444444)
        (100, 62786.31055555555)
        (150, 94181.79)
        };
    \addlegendentry{QUILTS}

    \addplot[smooth,color=black!30!green,mark=o]
        plot coordinates {
        (0.1,  27.734444444444442)
        % (1, 251.47333333333333)
        (5, 1238.7427777777777)
        (10, 2475.269444444444)
        % (15, 3711.4900000000002)
        (20, 4945.955)
        (50, 12361.997222222222)
        (100, 24718.880555555555)
        (150, 37074.515)
        };
    \addlegendentry{BMTree}
    
    \end{axis}

\end{tikzpicture}
% \vspace*{-6mm}
    \caption{I/O Cost}
    \label{fig:scalability_blk}
    % \vspace*{-4mm}
    \end{subfigure}
    \begin{subfigure}{.49\linewidth}
    \centering
    \hspace*{1mm}
    \begin{tikzpicture}[scale=0.55]

    \begin{axis}[
    width=8.5cm,
    height=4cm,
        xlabel= Dataset Size (million),
        ylabel= Query Latency (s), %ymode=log,
        every axis y label/.style={
        at={(-0.05, 1.3)},
        anchor=north west,
        },
        xmin=0, xmax=150,
         ymax=1600,
        xtick={0.1,  5, 10, 20, 50, 100, 150},
        % ytick={10000, 50000, 100000},
        % y tick label style={/pgf/number format/sci, scaled ticks=base 10:-2},
        % y tick label style={scaled ticks=base 10:-2},
        legend style={at={(0.0,1.0)},,anchor=north west}
        ]
    
    \addplot[smooth,mark=*,red] plot coordinates {
        (0.1, 4.072898152139452)
        (5, 40.728981521394516)
        (10, 81.45796304278903)
        (20, 162.91592608557806)
        (50, 407.2898152139452)
        (100, 781.8818909592098)
        (150, 1148.1491759088303)
    };
    \addlegendentry{Z-curve}

    \addplot[smooth,color=blue,mark=star]
        plot coordinates {
        (0.1,  5.239982081784141)
        (5, 52.399820817841416)
        (10, 104.79964163568283)
        (20, 209.59928327136566)
        (50, 523.9982081784142)
        (100, 1000.2653127776252)
        (150, 1495.9954761134254)
        };
    \addlegendentry{QUILTS}

    \addplot[smooth,color=black!30!green,mark=o]
        plot coordinates {
        (0.1,  2.73999804523256)
        (5, 27.39998045232561)
        (10, 54.79996090465122)
        (20, 109.59992180930244)
        (50, 273.9998045232561)
        (100, 512.4268595377604)
        (150,  753.2030438052284)
        };
    \addlegendentry{BMTree}
    
    \end{axis}

\end{tikzpicture}
% \vspace*{-6mm}
    \caption{Query Latency}
    \label{fig:scalability_time}
    % \vspace*{-4mm}
    \end{subfigure}
\caption{Performance vs dataset size.}
\label{fig:scalability}
% \vspace*{-4mm}

\end{figure}

% \smallskip
\noindent
\textbf{Scalability of the Learned SFCs.} To evaluate the scalability of the BMTree,
% the learned SFCs,
we evaluate the 
% querying 
performance of the SFCs 
% from BMTree and baselines 
by varying data size from 0.1 to 150 Million. {We construct the BMTree 
using $10^5$ sampled data points as input for RL training}
% based on the 1 Million data, 
%\walid{What do you mean by: construct the BMTree based on the 1 Million data? Please clarify.}
and the others follow the default settings. The 
%result is shown 
results are 
in Figure~\ref{fig:scalability}. 
%We o
Observe that the BMTree displays a linear trend for 
%the 
both
I/O and QL when data size increases. We observe similar trends for baselines.

\noindent
\textbf{Effect of Higher Dimensionality.} To evaluate the effect of dimensionality on the effectiveness of the learned SFC, we vary the dimensionality from 2 to 6  on the datasets for both the uniform and normal distributions. 
\ifnum\extend=1
{\jn We report the IO in Figure~\ref{fig:high_dim}.}
\else \fi
The BMTree consistently outperforms the baselines, and saves up to 54\% of the I/O cost compared to the best Z-curve baseline. This demonstrates that 
%our method 
the BMTree
generalizes well on data with more than two dimensions. 
\ifnum\extend=1
\else
Due to  space limitation, the detailed results can be found in~\cite{jiangneng2023lsfc}.
\fi

%To evaluate the effect of dimensionality on the effectiveness of the learned SFC, we vary the dimensionality from 2 to 6  on the dataset following both uniform and normal distributions. The size and granularity (number of grid cells in the data space) of data remain the same, while the workload remains the same selectivity. Figure \ref{fig:high_dim} shows the I/O cost result of BMTree and baselines.  BMTree consistently outperforms the baselines and saves up to 54\% of I/O cost compared with the best baseline Z-curve. This demonstrates that our method generalizes well on data with more than 2 dimensions. We also observe that the I/O cost of all methods increases with the increase of dimensionality.

\ifnum\extend=1

{
\begin{figure}[ht]
% \vspace*{-3mm}
\begin{subfigure}{.485\linewidth}
    \centering
    \begin{tikzpicture}[scale=0.55]

    \begin{axis} [ybar, %ymode=log,
bar width = 5pt,
width=8.5cm,
height=4cm,
ymax=250,
%symbolic x coords={UNI,SKE,GAU,DES},
xtick={2,3,4, 5,6},
xticklabels = {2, 3, 4, 5, 6},
ylabel = I/O Cost,
xlabel = Dimensionality,
every axis y label/.style={
    at={(-0.05, 1.2)},
    anchor=north west,
},
% extra x ticks={2, 5, 8, 11},
% extra x tick labels={UNI, GAU, OSM-US, TIGER},
extra x tick style={tick label style={yshift={ifthenelse(\ticknum==0, "-3mm","-3mm")}}},
legend style={
                    at={(0.0,1.0)},
                    anchor=north west,
                    legend columns=3,
                    /tikz/every even column/.append style={column sep=0.0cm}
                        },
]

\addplot[draw = black, 
] coordinates {
    (2, 100) 
    (3, 100) 
    (4, 100)
    (5, 100)
    (6, 100) 
};
\addplot[draw = black, 
postaction={
        pattern=north east lines
    }
] coordinates {
    (2, 153.5420968)
    (3, 140.183252)
    (4, 183.0529209)
    (5, 197.9445944)
    (6, 199.1899003)
};
\addplot[draw = black, 
postaction={
        pattern= crosshatch
    } 
% fill = black!30!green
] coordinates {
    (2, 54.72657661)
    (3, 68.89763439)
    (4, 52.46085058)
    (5, 50.67551236)
    (6, 52.08463777)
};

\legend {Z-curve, QUILTS, BMTree};

\end{axis}

\end{tikzpicture}
% \vspace*{-2mm}
    \caption{Uniform Data}
    \label{fig:dimension_uniform}
    % \vspace*{-4mm}
    \end{subfigure}
    \begin{subfigure}{.485\linewidth}
    \centering
    \begin{tikzpicture}[scale=0.55]

    \begin{axis} [ybar, %ymode=log,
bar width = 5pt,
width=8.5cm,
height=4cm,
ymax=250,
%symbolic x coords={UNI,SKE,GAU,DES},
xtick={2,3,4, 5,6},
xticklabels = {2, 3, 4, 5, 6},
ylabel = I/O Cost,
xlabel = Dimensionality,
every axis y label/.style={
    at={(-0.05, 1.2)},
    anchor=north west,
},
% extra x ticks={2, 5, 8, 11},
% extra x tick labels={UNI, GAU, OSM-US, TIGER},
extra x tick style={tick label style={yshift={ifthenelse(\ticknum==0, "-3mm","-3mm")}}},
legend style={
                    at={(0.0,1.0)},
                    anchor=north west,
                    legend columns=3,
                    /tikz/every even column/.append style={column sep=0.0cm}
                        },
]

\addplot[draw = black, 
] coordinates {
    (2, 100) 
    (3, 100) 
    (4, 100)
    (5, 100)
    (6, 100) 
};
\addplot[draw = black, 
postaction={
        pattern=north east lines
    }
] coordinates {
    (2, 153.4881576)
    (3, 132.3239528)
    (4, 181.6658188)
    (5, 197.588653)
    (6, 199.0613734)
};
\addplot[draw = black, 
% fill = black!30!green
postaction={
        pattern= crosshatch
    } 
] coordinates {
    (2, 54.64881597)
    (3, 47.01478877)
    (4, 46.43604932)
    (5, 52.48392641)
    (6, 48.02065562)
};

\legend {Z-curve, QUILTS, BMTree};

\end{axis}

\end{tikzpicture}
% \vspace*{-2mm}
    \caption{Normal Data}
    \label{fig:dimension_normal}
    % \vspace*{-4mm}
    \end{subfigure}
\caption{I/O cost vs dimensionality.}
\label{fig:high_dim}
\vspace*{-1mm}
\end{figure}
}
\else \fi

\revision{
\noindent {\bf Effect of Varying Query Aspect Ratio and Selectivity.} 
% \marginpar{\#1 R3D6}
\ifnum\extend=1
(1)~We evaluate the BMTree performance   by varying query aspect ratios from \{$4$ , $\frac{1}{4}$\} to \{$128$ , $\frac{1}{128}$\},
and the results are reported in Figure~\ref{fig:varying_ratio}. {\chengr Observe} that the BMTree {\chengr performs consistently better than the other SFCs across different aspect ratios including very wide ones.}
\else
(1) We evaluate the performance of BMTree by varying query aspect ratios from \{$4$ , $\frac{1}{4}$\} to \{$128$ , $\frac{1}{128}$\}. 
% Due to the space limit, the detailed results are in \cite{jiangneng2023lsfc}.
\fi
\ifnum\extend=1
(2) We %evaluate the performance of BMTree by
vary query selectivity from $0.0001\%$ to $1\%$, and report the results in Figure~\ref{fig:varying_selectivity}. 
Observe that the %optimization 
improvement
of the BMTree is subtle under very small query range. 
{\chengr This is due to the fact that for small query ranges, the points that are within a range are few,
% and tend to have similar values based on a SFC curve 
and the index tends to perform similarly 
for different SFCs.}
\else
(2) We %evaluate the performance of BMTree by
vary query selectivity from $0.0001\%$ to $1\%$. Observe that the %optimization 
improvement
in the BMTree is subtle under very small query ranges (See  \cite{jiangneng2023lsfc} for details).
% Due to the space limit, the detailed results are given in \cite{jiangneng2023lsfc}.
\fi
% The reason could be because BMTree fed with queries of very small range overfits easily, and requires more queries to be effective.

% We evaluate how our MCTS based BMTree method perform varying different query aspect ratio. The generated query have different query ratio from \{$4$ \& $\frac{1}{4}$\} to \{$128$ \& $\frac{1}{128}$\}. The result is reported in Figure \ref{fig:exp_wide_ratio}. We observe that BMTree still perform well under very wide aspect ratio (e.g., \{$128$ \& $\frac{1}{128}$\}). QUILTS tends to have better performance when aspect ratio increases. This could because that very wide ratio queries could be largely optimized with instance based optimization. Even though QUILTS can only optimize certain types of query ($128$ or $\frac{1}{128}$), the performance increasing is already obvious.}
}

\iffalse
\noindent
\textbf{Effect of max depth.}
We evaluate the effect of the max depth parameter which is the depth of BMTree that are built using RL. We vary the max depth from 1 to 20.
%, and report the average $SR$ for testing and the corresponding BMTree construct time.
\ifnum\extend=1
{\jn The results are reported in Table.~\ref{table:parameter_depth}.}
\else \fi
The result shows that as the max depth increases, $SR$ drops and then tends to be stable after 10. 
\ifnum\extend=1
\else
Due to  space limitation, the detailed results are in~\cite{jiangneng2023lsfc}. 
\fi

\ifnum\extend=1
\begin{table}[ht]
% \setlength{\tabcolsep}{9pt}
% \vspace*{-3mm}
\caption{The effect of the max depths.}
	\label{table:parameter_depth}
	% \vspace*{-3mm}
\centering
	\begin{tabular}{ |l|m{0.66cm}|
	m{0.66cm}|
	m{0.66cm}|m{0.66cm}|m{0.66cm}|}
	\hline
	Max depth & 1 &5 & 10 & 15 & 20 \\ \hline \hline
	ScanRange & 967.6 & 671.1 & 532.9 & 527.8 & 527.6  \\ \hline
    \revision{Training Time (min)} & 2.0 & 8.8 & 100.1 & 254.8 & 409.5 \\ \hline
	\end{tabular}	
	% \vspace*{-3mm}
\end{table}
\else \fi

\fi

\ifnum\extend=1
\begin{figure}[ht]
    \centering
    \begin{subfigure}{.485\linewidth}
    \centering
    % \hspace*{-3mm}
        \includegraphics[scale=0.55]{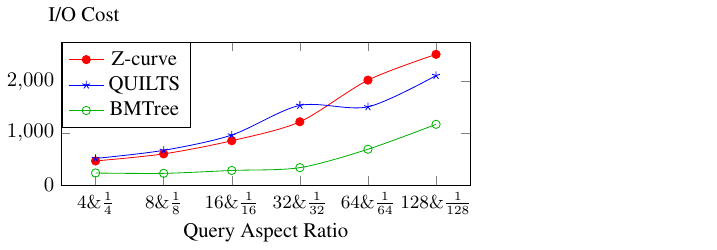}
        % \vspace*{-6mm}
        \caption{Varying aspect ratio}
        \label{fig:varying_ratio}
    \end{subfigure}
    \begin{subfigure}{.485\linewidth}
    \centering
    % \hspace*{-12mm}
        \includegraphics[scale=0.55]{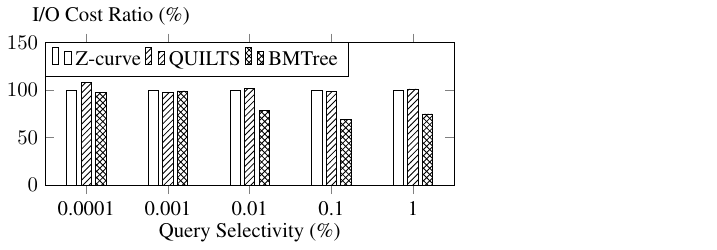}
        % \vspace*{-6mm}
        \caption{Varying selectivity}
        \label{fig:varying_selectivity}
    \end{subfigure}
    % \vspace*{-4mm}
    \caption{\revision{Varying query aspect ratio and selectivity.}}
    \vspace*{-1mm}
\end{figure}
\else \fi

\revision{
\subsubsection{Evaluating BMTree Variants}
We study 4 BMTree variants: 
% \marginpar{\#2 R4O2 R3O5}
BMTree-\underline{D}ata \underline{D}riven (with dataset only), BMTree-noGAS (with no GAS algorithm), BMTree-greedy (pure greedy), and BMTree-LMT (with limited BMPs). Results are in Figure~\ref{fig:bmtree_more_exp}.
% We designed four BMTree variants to evaluate the effectiveness of different components of MCTS-BMTree method architecture, including (1) BMTree-\underline{D}ata \underline{D}riven (BMTree-DD) which only dataset is given, (2) BMTree-noGAS which doesn't use GAS, (3) BMTree-greedy which is a pure greedy algorithm without MCTS, and (4) BMTree-lmt with limited BMPs applied. The result is reported in Figure~\ref{fig:bmtree_more_exp}.
\input{VLDB_Learned_SFC/figures/figure_bmtree_variants.tex}
\noindent {\bf (1) BMTree-DD.} We evaluate the BMTree performance when the query workload is not available. We generate training queries for the BMTree by following the dataset's distribution. From Figure~\ref{fig:bmtree_more_exp},  observe that BMTree-DD performs comparable to the BMTree on the \texttt{UNI} workloads for all datasets. However, on the \texttt{SKE} workload, the BMTree performs generally much better.
{\bf (2) BMTree-noGAS.} We evaluate the effectiveness of the GAS algorithm. Observe the performance drop compared 
%with 
to
the MCTS using GAS. This shows the effect of GAS.
{\bf (3) BMTree-greedy.} 
We apply GAS for all action selections, and build a {\chengr purely} greedy based %built 
BMTree.
% , denoted by BMTree-greedy.
%   Results are reported in Figure~\ref{fig:bmtree_more_exp}.
Observe that MCTS with GAS outperforms both BMTree-noGAS and BMTree-greedy. This indicates a synergistic improvement {\chengr of} MCTS {\chengr over} GAS.
% We apply GAS for all action selections and build a pure greedy based built BMTree, denoted by BMTree-greedy. 
% We observe that MCTS with GAS outperforms both the BMTree-noGAS and BMTree-greedy. This indicates a synergistic improvement between MCTS and GAS.
{\bf (4) BMTree-LMT.} We consider all BMPs, and design a baseline BMTree where only the Z- and C-curves are allowed to be assigned to the subspaces.
% ., denoted as BMTree-LMT. 
% In Figure~\ref{fig:bmtree_more_exp}. 
% we observe that BMTree-LMT only achieves very limited optimization compared with other BMTree variants, which demonstrates the necessity of arbitrary BMP design.
We {\chengr observe a significant improvement in the BMTree} over using the Z- and C-curves alone. This demonstrates the necessity of considering all BMPs.

}

\begin{figure*}[ht]
    \centering
    \begin{subfigure}{0.32\linewidth}
    \centering
        \input{VLDB_Learned_SFC/figures/exp_data_shift_io}
        % \vspace*{-3mm}
        \caption{Performance of I/O Cost.}
    \label{fig:exp_data_shift_io}
    \end{subfigure}
    \begin{subfigure}{0.32\linewidth}
    \centering
        \input{VLDB_Learned_SFC/figures/exp_data_shift_time}
        % \vspace*{-3mm}
        \caption{Performance of Query Latency.}
        \label{fig:exp_data_shift_time}
    \end{subfigure}
    \begin{subfigure}{0.32\linewidth}
    \centering
        \input{VLDB_Learned_SFC/figures/exp_data_shift_train}
        % \vspace*{-3mm}
        \caption{Performance of Training Latency.}
        \label{fig:exp_data_shift_train}
    \end{subfigure}
    \vspace*{-1mm}
    \caption{\bluecolor{Evaluation of data shift 
    %with query fixed. 
    while fixing the queries.
    The data is shifted from a {\tt GAU} to {\tt UNI} with varying {\tt UNI} percentage from $10\%$ to $90\%$.}}
    \label{fig:exp_data_shift}
    \vspace*{-3mm}
\end{figure*}
\begin{figure*}
    \centering
    \begin{subfigure}{0.32\linewidth}
    \centering
        \input{VLDB_Learned_SFC/figures/exp_query_shift_io}
        \caption{Performance of I/O Cost.}
    \label{fig:exp_query_shift_io}
    \end{subfigure}
    \begin{subfigure}{0.32\linewidth}
    \centering
        \input{VLDB_Learned_SFC/figures/exp_query_shift_time}
        \caption{Performance of Query Latency.}
        \label{fig:exp_query_shift_time}
    \end{subfigure}
    \begin{subfigure}{0.32\linewidth}
    \centering
        \input{VLDB_Learned_SFC/figures/exp_query_shift_train}
        \caption{Performance of Training Latency.}
        \label{fig:exp_query_shift_train}
    \end{subfigure}
    \vspace*{-1mm}
    \caption{\bluecolor{Query shift 
    %with data fixed. 
    while fixing the data.
    Query workload is shifted from $10\%$ to $90\%$ of the new query workload.
    % Query workload is shifted while data is unchanged.
    }}
    \vspace*{-3mm}
    \label{fig:exp_query_shift}
\end{figure*}

\bluecolor{
\smallskip
\subsection{Effectiveness of Partial Retraining of the BMTree} 

\subsubsection{Varying Distribution Shift Settings} We evaluate the effectiveness of our proposed partial retraining mechanism. Specifically, we evaluate three situations: data shift while the queries are fixed, query shift while the data is fixed, and the composed scenario. We compare three methods: (1)~Keeping the original BMTree unchanged (noted as {\tt {BMT}-O}), (2)~Fully retrained BMTree (noted as {\tt {BMT}-{FR}}), and (3)~Partially retrained BMTree (noted as {\tt {BMT}-{PR}}). We restrict the partial retrain constraint ratio to $0.5$, where at most half of the space area can be retrained. Also, %We also 
we
evaluate the performance 
%with varying retrain constraints.
while varying the retrain constraints.

% \noindent \textbf{Effectiveness varying different settings.}

\smallskip
\noindent \textbf{Evaluation of Data Shift.} We evaluate 3 different metrics: the I/O Cost and Query Latency of the constructed BMTree, and the training time 
%which is consumed 
needed 
to retrain the BMTree. The data 
%is shifted 
shifts 
from the $\tt GAU$ to the $\tt UNI$ distribution. The results are 
%shown 
in Fig.~\ref{fig:exp_data_shift}. 
In Fig.~\ref{fig:exp_data_shift_io},
the partially retrained BMTree {\tt {BMT}-{PR}} achieves a performance increase on I/O cost compared 
%with 
to
the original BMTree {\tt {BMT}-{O}}, 
%with 
while
optimizing 
the 
percentage from $9.2\%$ to $12.1\%$ %with 
and
varying the shift percentage. Compared 
%with 
to
the fully retrained BMTree {\tt {BMT}-{FR}}, {\tt {BMT}-{PR}} achieves an average of $90.6\%$ 
%of 
performance improvement achieved by {\tt {BMT}-{FR}}. Under 
%the situation of 
a $90\%$ data shift, {\tt {BMT}-{PR}} outperforms {\tt {BMT}-{FR}} and achieves over $2.3\times$ 
%the I/O Cost reduction 
reduction in I/O cost
compared with {\tt {BMT}-{FR}}.  %This is because 
The reason is 
that {\tt {BMT}-{PR}} allows the agent to focus on optimizing the subspace with vital distribution changes, which 
%gives the potential 
allows
for the partially retrained BMTree to achieve a better performance on this focused subspace.
% \redcolor{TODO}
% 
The results of the query latency are generally consistent 
%to 
with
the results of I/O Cost, in which  {\tt {BMT}-{PR}} achieves  an average of $91.7\%$ 
%of 
performance improvement achieved by {\tt {BMT}-{TR}},
% {\tt {BMT}-{PR}} performs similarly results on query latency compared with other methods,
as  in Fig.~\ref{fig:exp_data_shift_time}.

As for the training time (Fig.~\ref{fig:exp_data_shift_train}), {\tt {BMT}-{FR}} costs from $7857 s$ to $8314 s$ ($8125.5s$ on average) to retrain the BMTree from scratch, while {\tt {BMT}-{PR}} costs from $559.2s$ to $3778.8s$ ($1833.3s$ on average).  {\tt {BMT}-{PR}} achieves approximately $4.4\times$ reduction of training time compared with {\tt {BMT}-{FR}}, that is aligned with the time complexity estimation, where with a retraining constraint ratio of $R_{rc}$, the training time of  {\tt {BMT}-{PR}} is upper bounded by $R_{rc} \cdot {\tt time}(\text{{\tt {BMT}-{PR}}})$. 
% Under a data shift of $10\%$, the partial retraining algorithm returns that no BMTree node should to be retrained, thus costs no extra expenses on retraining.
}

\smallskip
\bluecolor{
\noindent \textbf{Evaluation of Query Shift.} We proceed to evaluate 
%the shift of query.
the effect of query workload shift.
Under the $\tt GAU$ data, we shift the distribution of the query workload. Specifically, we vary the $\mu$ values of the Gaussian distributions of queries, and generate 2 skew query workloads, namely $\tt SKE_1$ and $\tt SKE_2$, respectively. The query workload is shifted from  $\tt SKE_1$ to $\tt SKE_2$. We estimate the retrain performance by varying the shift percentage.
% Similar to data shift evaluation, we evaluate I/O cost, query latency, and training time.
The results are  in Fig.~\ref{fig:exp_query_shift}. 
%In Fig.~\ref{fig:exp_query_shift_io}, b
Both the fully and partially retrained BMTrees {\tt {BMT}-{ FR}} and  {\tt {BMT}-{PR}} have limited optimization compared to the original BMTree (less than $1\%$ on I/O Cost) before the shift reaches $50\%$ percentage. However, when the shift percentage reaches $70\%$, the optimizing potential becomes vital.  {\tt {BMT}-{PR}} 
%achieves from reduce of 
reduces
I/O Cost from $7.3\%$ to $16.7\%$ compared with the {\tt {BMT}-{O}}. We observe that {\tt {BMT}-{PR}} achieves better performance on I/O Cost compared with {\tt {BMT}-{FR}} from shift percentage $70\%$ to $90\%$ (over $1.8\times$ at $90\%$). This reveals that retraining a subspace with a significant change in query workload  may 
%achieve a significant performance enhancing 
significantly enhance performance
compared with training a BMTree for the whole data space. The query latency results are similar 
%to 
in
I/O Cost (as  in Fig.~\ref{fig:exp_query_shift_time}) 
%which in 
that is 
consistent 
%in data shift situation. 
with the data shift situation.

The training time (Fig.~\ref{fig:exp_query_shift_train}) also aligns with the time complexity evaluation. {\tt {BMT}-{FR}} spends from $7295.5s$ to $ 8485.8s$ (on average, $7716.9s$) to retrain the BMTree, while {\tt {BMT}-{PR}} spends from $255.9s$ to $1571.1s$ (on average, $1237.1s$), achieves approximately over $6.2\times$ 
%training time reduction.
reduction in training time.
}

\smallskip
\bluecolor{
\noindent \textbf{Evaluation of Mixed Shift of Data and Query.}  We evaluate the scenario when both data and query shift, and the shift settings of data and query 
%follows 
follow
the former experiments. We select  shift percentages from: $\{25\%, 50\%, 75\%\}$ for both data and query.
% , and shift settings of data and query follow the former experiments.
The results are  in Fig.~\ref{fig:exp-mixed-shift}.
\begin{figure}[ht]
    \centering
    \includegraphics[width=1\linewidth]{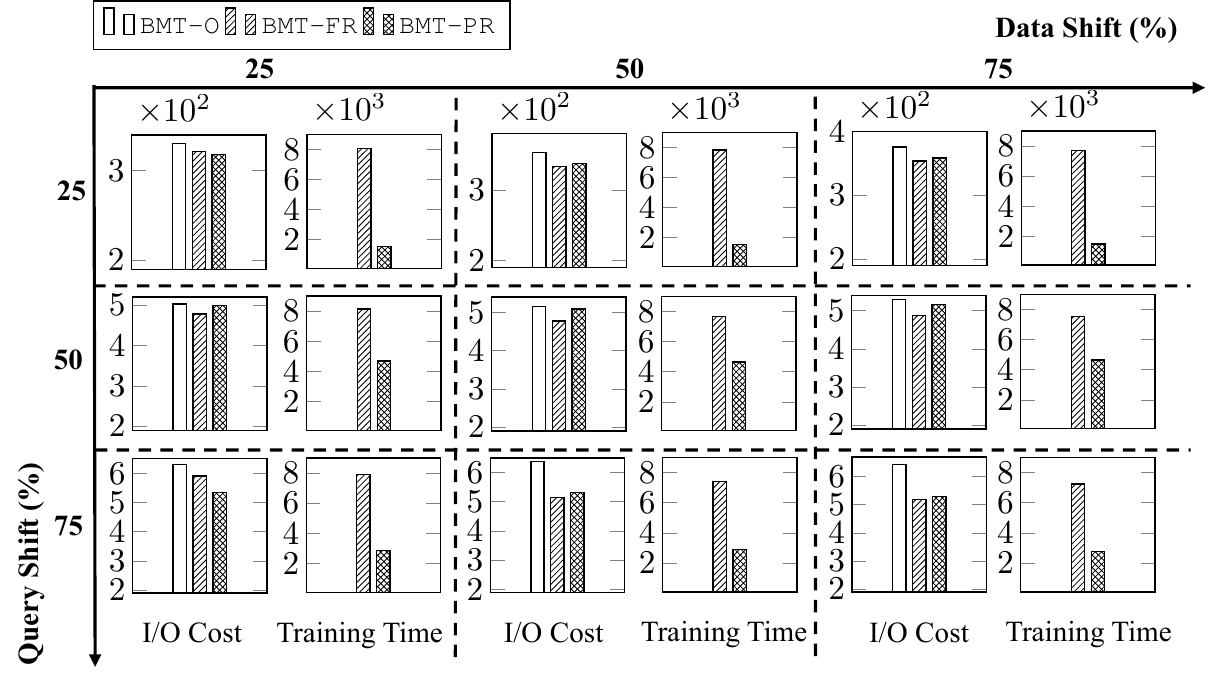}
    \vspace*{-2mm}
    \caption{\bluecolor{Evaluation of mixed shift of data and query.}}
    \label{fig:exp-mixed-shift}
\end{figure}
%In Fig.~\ref{fig:exp-mixed-shift},
The partially retrained BMTree {\tt {BMT}-{PR}} achieves different levels of optimizations 
when 
varying data and query shift percentages. {\tt {BMT}-{PR}} achieves remarkable I/O Cost reduction when query or data shift reaches $75\%$ (the 3rd row and the 3rd column in Fig.~\ref{fig:exp-mixed-shift} denote when data and query shift reaches $75\%$, respectively), which achieves on average $8.3\%$ (resp. $16.5\%$) reduction 
%of 
in
I/O cost under $75\%$ data shift (resp. query shift) compared with {\tt {BMT}-{O}}. {\tt {BMT}-{PR}} outperforms {\tt {BMT}-{FR}} in certain cases (e.g., $25\% \times 25\%$ data-query shift), and achieves competitive performance in the majority of situations.  {\tt {BMT}-{PR}} remains efficient compared with {\tt {BMT}-{FR}} on training time. Particularly, under $50\%$ query shift,
%situation, 
the training time reduction of  {\tt {BMT}-{PR}} is less than $2\times$ compared with {\tt {BMT}-{FR}}. 
%We o
Observe that 
the
second partial retraining is triggered since the first partial retraining does not achieve notable optimization. Compared with {\tt {BMT}-{FR}}, a full retraining is preferred 
%in this situation 
as the action in the root node of the BMTree needs to be modified.}

% which achieves over $20\%$ percentage of I/O Cost reduction when query shift percentage reaches $75\%$ compared with {\tt {BMT}-{O}}. {\tt {BMT}-{PR}} maintains its effectiveness under the situation where the shift is mild and outperforms {\tt {BMT}-{FR}} (in $20\%\times20\%$ situation), while the training time optimization ratio ($2\times$) remains.}

\bluecolor{
\subsubsection{Varying the Retraining Hyperparameters} 
We 
%proceed to 
evaluate how the retraining hyperparameters affect 
%the performance, specifically the 
I/O Cost. We consider 
%two parameters: 
the retraining constraint ratio and the shift score threshold. We conduct the hyperparameter estimations under 
%the setting with 
$75\%  \times 75\%$ data and query shifts. The results are 
%shown 
in Fig.~\ref{fig:retrain_vary_parameter}. }

\begin{figure}[ht]
    \centering
    
    \begin{subfigure}{.49\linewidth}
	\centering
 \hspace*{-2mm}
\resizebox{1.05\columnwidth}{!}{
 \input{VLDB_Learned_SFC/figures/retrain_vary_constraint}
      }  
	\caption{\bluecolor{Retrain constraint ratio.}}
\label{fig:retrain_vary_constriant}
	% \vspace*{-3mm}
	\end{subfigure}
	\begin{subfigure}{.49\linewidth}
	\centering
\resizebox{1.05\columnwidth}{!}{
    \input{VLDB_Learned_SFC/figures/retrain_vary_threshold}
    }
	\caption{\bluecolor{Shift score threshold.}}
	\label{fig:retrain_vary_threshold}
	% \vspace*{-3mm}
	\end{subfigure}
    \vspace*{-1mm}
    \caption{\bluecolor{Varying  retrain  constraint ratio \& shift score threshold.}}
    \label{fig:retrain_vary_parameter}
    \vspace*{-1mm}
\end{figure}

\bluecolor{\noindent
\textbf{(1) {The Retraining Constraint Ratio $R_{rc}$}.}
We vary $R_{rc}$ from $0.1$ to $1$ to study how $R_{rc}$ affects the retraining performance. As in Fig.~\ref{fig:retrain_vary_constriant}, {\tt {BMT}-{PR}} achieves almost no performance enhancement compared to {\tt {BMT}-{O}}, as the very small retraining constraint ratios ($0.1$ and $0.2$) limit the ability of the retrain method to retrain the important nodes that do not satisfy the constraint. {\tt {BMT}-{PR}} achieves optimal performance with a constraint ratio of $0.5$ that allows retraining nodes that mostly affect  query performance.

\noindent
\textbf{(2) The Shift Score Threshold.} We vary the shift score threshold from $0.1$ to $0.5$. As in Fig.~\ref{fig:retrain_vary_threshold},  observe that the threshold from $0.1$ to $0.35$, {\tt {BMT}-{PR}} hassimilar performance compared to a full retrain. When the threshold is $0.4$ and above, it filters nodes that  achieve notable performance improvement (with threshold lower than $0.4$) and become ineffective. This identifies the best choices for the shift score threshold.
}

%% file: sections/experiment_query_execute.tex
\begin{figure*}
	\begin{subfigure}{.485\linewidth}
	\centering
        \hspace*{-20mm}
        \includegraphics[]{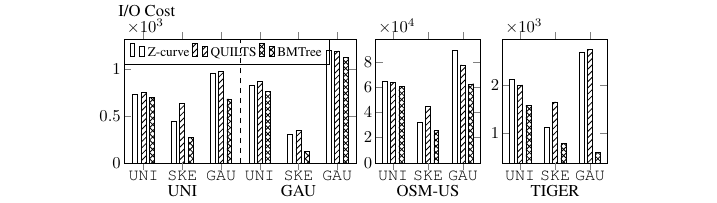}
        \vspace*{-6mm}
	\caption{Performance of I/O Cost.}
\label{fig:bptree_block_access}
	% \vspace*{-3mm}
	\end{subfigure}
	\begin{subfigure}{.485\linewidth}
	\centering
 \hspace*{-15mm}
	\includegraphics[]{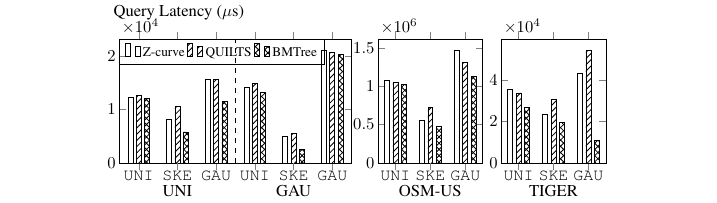}
    \vspace*{-6mm}
	\caption{\revision{Performance of Query Latency.}}
	\label{fig:bptree_response_time}
	% \vspace*{-3mm}
	\end{subfigure}

\vspace*{-1mm}
\caption{Results
%evaluation on I/O cost and query time, 
under PostgreSQL, where the first and the second lines under the x bar denote the query and data distribution.}
\label{fig:bptree_window_query}
\vspace*{-3mm}

\end{figure*}

\begin{figure*}
\begin{subfigure}{.485\linewidth}
\centering
\hspace*{-20mm}
    \includegraphics[]{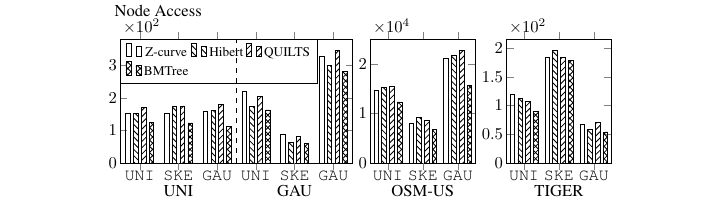}
    \vspace*{-6mm}
\caption{Performance of Node Access.}
\label{fig:rsmi_window_access}
% \vspace*{-3mm}
    \end{subfigure} 
    \begin{subfigure}{.485\linewidth}
    \centering
     \hspace*{-15mm}
    \includegraphics[]{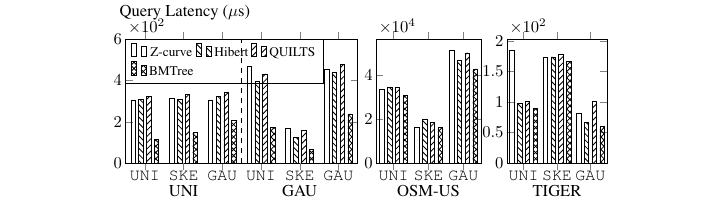}
    \vspace*{-6mm}
\caption{\revision{Performance of Query Latency.}}
\label{fig:rsmi_window_time}
% \vspace*{-3mm}
    \end{subfigure}
    
\vspace*{-1mm}
\caption{Results using RSMI learned index structure. 
}
\label{fig:rsmi_effect}
\vspace*{-3mm}

\end{figure*}

%% file: VLDB_Learned_SFC/figures/knn_query_performance.tex
\begin{figure}[ht]

\ifnum\extend=1
\else 
% \vspace*{-3mm}
\fi

\begin{subfigure}{.485\linewidth}
    \centering
    \begin{tikzpicture}[scale=0.55]

\begin{axis} [ybar, %ymode=log,
bar width = 5pt,
width=8.5cm,
height=4cm,
ymin=0,
ymax=170,
xmin=0, xmax=5,
%symbolic x coords={UNI,SKE,GAU,DES},
xtick={1,2,3,4},
xticklabels = {UNI, GAU, OSM-US, TIGER},
ylabel = I/O Cost Ratio (\%),
every axis y label/.style={
    at={(-0.05, 1.2)},
    anchor=north west,
},
% extra x ticks={2, 5, 8, 11},
% extra x tick labels={UNI, GAU, OSM-US, TIGER},
% extra x tick style={tick label style={yshift={ifthenelse(\ticknum==0, "-4mm","-4mm")}}},
legend style={
                    at={(0.0,1.0)},
                    anchor=north west,
                    legend columns=-1,
                    /tikz/every even column/.append style={column sep=0cm}
                        },
]

% \addplot[draw = black, fill = purple] coordinates {
%     (1, 79.71433333333334) 
%     (2, 77.57766666666667) 
%     (3, 63.165)
%     (4, 28.846999999999998)
% };
\addplot[draw = black, 
] coordinates {
    (1, 100) 
    (2, 100) 
    (3, 100)
    (4, 100)
};

\addplot+[
        draw = black, 
        fill=none,
        postaction={
        pattern=north east lines
    },
            error bars/.cd,
                y dir=both,
                % (changed from `y explicit` so the error bars are (clearly) visible
                y explicit relative,
        ] coordinates {
            (1, 110.2255116) +- (0, 0.08349409651292639)
            (2, 94.453832) +- (0, 0.03495365413958053)
            (3, 98.64823228) +- (0, 0.0562881007553679)
            (4, 134.5105605)+- (0,0.1617719378687047)
        };

\addplot+[
        draw = black, 
        fill=none,
        % fill = black!30!green,
        postaction={
        pattern= crosshatch
    } ,
            error bars/.cd,
                y dir=both,
                % (changed from `y explicit` so the error bars are (clearly) visible
                y explicit relative,
        ] coordinates {
            (1, 115.6092187) +- (0, 0.12604780402646776)
            (2, 89.82509106) +- (0, 0.0649825401552657)
            (3, 88.72401736) +- (0, 0.11732778659859121)
            (4, 101.813318)+- (0,0.19773324083849386)
        };

% \draw [dashed] (25,1) -- (25,16);

% \draw [dashed] (55,1) -- (55,16);

% \draw [dashed] (85,1) -- (85,16);

\legend {Z-curve,  QUILTS, BMTree};

\end{axis}

\end{tikzpicture}
% \vspace*{-6mm}
    \caption{I/O Cost}
    \label{fig:bptree_knn_block_access}
    % \vspace*{-4mm}
    \end{subfigure}
\begin{subfigure}{.485\linewidth}
    \centering
    \begin{tikzpicture}[scale=0.55]

\begin{axis} [ybar, %ymode=log,
bar width = 5pt,
width=8.5cm,
height=4cm,
ymin=0,
ymax=170,
xmin=0, xmax=5,
%symbolic x coords={UNI,SKE,GAU,DES},
xtick={1,2,3,4},
xticklabels = {UNI, GAU, OSM-US, TIGER},
ylabel = Query Latency Ratio (\%), %Nanoseconds($\times 10^9$) Secs,
every axis y label/.style={
    at={(-0.05, 1.2)},
    anchor=north west,
},
% extra x ticks={2, 5, 8, 11},
% extra x tick labels={UNI, GAU, OSM-US, TIGER},
% extra x tick style={tick label style={yshift={ifthenelse(\ticknum==0, "-4mm","-4mm")}}},
legend style={
                    at={(0.0,1.0)},
                    anchor=north west,
                    legend columns=-1,
                    /tikz/every even column/.append style={column sep=0cm}
                        },
]

% \addplot[draw = black, fill = purple] coordinates {
%     (1, 0.8959558716666667) 
%     (2, 0.8274081016666667) 
%     (3, 0.818397248)
%     (4, 0.34921156000000003)
% };
\addplot[draw = black, 
] coordinates {
    (1, 100) 
    (2, 100) 
    (3, 100)
    (4, 100)
};
% \addplot[draw = black, fill = blue] coordinates {
%     (1, 2.064626932) 
%     (2, 5.606841882333334) 
%     (3, 10.286432190333333) 
%     (4, 27.865600266666664)
% };
\addplot+[
        draw = black, 
        fill=none,
        postaction={
        pattern=north east lines
    },
            error bars/.cd,
                y dir=both,
                % (changed from `y explicit` so the error bars are (clearly) visible
                y explicit relative,
        ] coordinates {
            (1, 101.61325) +- (0, 0.10613681909798893)
            (2, 96.86837634) +- (0, 0.05077610941812236)
            (3, 97.25396582) +- (0, 0.03402057525147451)
            (4, 125.1391275)+- (0,0.12471287745925895)
        };

\addplot+[
        draw = black, 
        % fill = black!30!green,
        fill=none,
        postaction={
        pattern= crosshatch
    } ,
            error bars/.cd,
                y dir=both,
                % (changed from `y explicit` so the error bars are (clearly) visible
                y explicit relative,
        ] coordinates {
            (1, 111.0511071) +- (0, 0.09172261202210705)
            (2, 92.32384166) +- (0, 0.04104757637173383)
            (3, 93.68812319) +- (0, 0.06942971524065437)
            (4, 105.6567972)+- (0,0.1394033584828265)
        };

% \draw [dashed] (25,1) -- (25,16);

% \draw [dashed] (55,1) -- (55,16);

% \draw [dashed] (85,1) -- (85,16);

\legend {Z-curve,  QUILTS, BMTree};

\end{axis}

\end{tikzpicture}
% \vspace*{-6mm}
    \caption{Query Latency}
    \label{fig:bptree_knn_response_time}
    % \vspace*{-4mm}
    \end{subfigure}

\caption{Performance of $k$NN queries.}
\label{fig:bptree_knn_query}
\vspace*{-3mm}

\end{figure}

\begin{figure}[ht]
    \centering
    \begin{subfigure}{.485\linewidth}
    \centering
    \hspace*{-12mm}
        \includegraphics[scale=0.55]{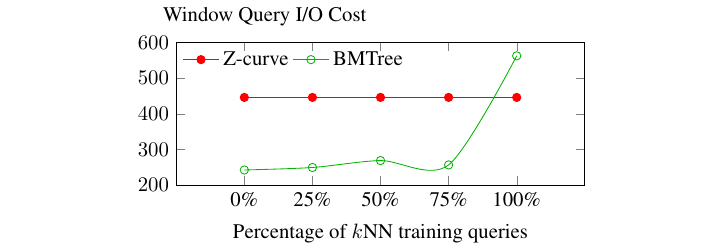}
        % \vspace*{-5mm}
        \caption{\revision{Window Query I/O}}
        \label{fig:knn_optimize_window}
    \end{subfigure}
    \begin{subfigure}{.485\linewidth}
    \centering
    \hspace*{-12mm}
        \includegraphics[scale=0.55]{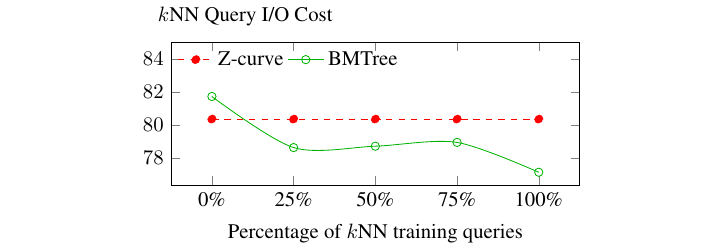}
        % \vspace*{-5mm}
        \caption{\revision{$k$NN Query I/O}}
        \label{fig:knn_optimize_knn}
    \end{subfigure}
    % \vspace*{-4mm}
    \caption{\revision{Optimization of window query \& $k$NN query.}}
    \label{fig:exp_optimize_knn}
    \vspace*{-2mm}
\end{figure}

%% file: VLDB_Learned_SFC/figures/figure_bmtree_variants.tex
\begin{figure}[ht]
    \centering
    % \vspace*{-3mm}
    \begin{tikzpicture}[scale=0.8]

\begin{axis} [ybar, %ymode=log,
bar width = 2pt,
width=11.5cm,
height=4.2cm,
ymax=140,
%symbolic x coords={UNI,SKE,GAU,DES},
xtick={1,2,3,4, 5,6,7,8, 9, 10,11,12},
xticklabels = {\texttt{UNI}, \texttt{SKE}, \texttt{GAU}, \texttt{UNI}, \texttt{SKE}, \texttt{GAU}, \texttt{UNI}, \texttt{SKE}, \texttt{GAU},  \texttt{UNI}, \texttt{SKE}, \texttt{GAU}},
x tick label style={yshift=1mm},
ylabel = I/O Cost Ratio (\%),
every axis y label/.style={
    at={(-0.05, 1.2)},
    anchor=north west,
},
extra x ticks={2, 5, 8, 11},
extra x tick labels={UNI, GAU, OSM-US, TIGER},
extra x tick style={tick label style={yshift={ifthenelse(\ticknum==0, "-3mm","-3mm")}}},
legend style={ scale=0.7,
font=\small,
fill=none,
draw=none,
                    at={(0.0,1.03)},
                    anchor=north west,
                    legend columns=4,
                    /tikz/every even column/.append style={column sep=0.0cm}
                        },
]

\addplot[draw = black, 
bar shift=-6 pt,
% fill = black!30!green
postaction={
        pattern=crosshatch,
    }
] coordinates {
    (1, 94.81324599)
    (2, 60.87783526)
    (3, 71.55158419)
    (4, 92.25153703)
    (5, 40.15413867)
    (6, 94.00390189)
    (7, 93.72209692)
    (8, 79.51019503)
    (9, 70.24136722)
    (10, 74.93481441)
    (11, 70.49891519)
    (12, 22.21077034)
};
\addplot[draw = black, 
% fill = black!30!green, 
fill = white!90!gray,
bar shift=-3 pt,
% postaction={
%         pattern=crosshatch
%     }
    ] coordinates {
    (1, 94.81324599)
    (2, 70.64934547)
    (3, 96.69018419)
    (4, 92.17450837)
    (5, 48.9417042)
    (6, 94.00390189)
    (7, 84.45020805)
    (8, 97.77273275)
    (9, 73.83358553)
    (10, 82.41640392)
    (11, 103.1853447)
    (12, 80.82123507)
};

\addplot[draw = black, 
% fill = black!30!green, 
bar shift=0 pt,
fill = white!60!gray,
% postaction={
%         pattern=crosshatch
%     } 
    ] coordinates {
    % (1, 94.81324599)
    % (2, 63.23423553)
    % (3, 78.88135467)
    % (4, 92.65212655)
    % (5, 42.43465259)
    % (6, 98.09675027)
    % (7, 93.72209692)
    % (8, 85.31657996)
    % (9, 74.62369193)
    % (10, 82.36311159)
    % (11, 70.49891519)
    % (12, 32.5629812)
    
    (1, 96.37421902)
    (2, 63.34541275)
    (3, 74.58565988)
    (4, 91.22122807)
    (5, 51.83434246)
    (6, 92.89092131)
    (7, 95.38885478)
    (8, 94.13772864)
    (9, 74.46143447)
    (10, 74.15722608)
    (11, 71.3405404)
    (12, 23.86505817)
};

\addplot[draw = black, 
% fill = black!30!green, 
fill = white!30!gray,
bar shift=3 pt,
% postaction={
%         pattern=crosshatch
%     } 
    ] coordinates {
    (1, 94.81324599)
    (2, 63.23423553)
    (3, 78.88135467)
    (4, 92.65212655)
    (5, 42.43465259)
    (6, 98.09675027)
    (7, 93.72209692)
    (8, 85.31657996)
    (9, 74.62369193)
    (10, 82.36311159)
    (11, 73.49891519)
    (12, 32.5629812)
    
    % (1, 96.37421902)
    % (2, 63.34541275)
    % (3, 74.58565988)
    % (4, 91.22122807)
    % (5, 51.83434246)
    % (6, 92.89092131)
    % (7, 95.38885478)
    % (8, 94.13772864)
    % (9, 74.46143447)
    % (10, 74.15722608)
    % (11, 71.3405404)
    % (12, 23.86505817)
};

\addplot[draw = black, 
% fill = black!30!green, 
fill = white!0!gray,
bar shift=6 pt,
% postaction={
%         pattern=crosshatch
%     } 
    ] coordinates {
    (1, 99.7991014)
     (2, 99.385894)
     (3, 99.9820728)
     (4, 100.1608698)
     (5, 99.3251126)
     (6, 99.93999954)
     (7, 100.2912579)
     (8, 102.3012125)
     (9, 100.0088157)
     (10, 97.1647268)
     (11, 98.88772717)
     (12, 100.3413326)
    
};

\draw [dashed] (25,0) -- (25,200);

\draw [dashed] (55,0) -- (55,200);

\draw [dashed] (85,0) -- (85,200);

\legend {BMTree, BMTree-DD, BMTree-noGAS, \revision{BMTree-greedy}, \revision{BMTree-LMT}};

\end{axis}

\end{tikzpicture}
    \vspace*{-1mm}
    \caption{\revision{I/O Cost on BMTree Variants.}}
    \vspace*{-2mm}
    \label{fig:bmtree_more_exp}
\end{figure}

%% file: VLDB_Learned_SFC/figures/exp_data_shift_io.tex
    \begin{tikzpicture}[scale=0.7]

    \begin{axis}[
    width=8.5cm,
    height=4cm,
    xlabel= (\%),
    ylabel= I/O Cost, %ymode=log,
    every axis y label/.style={
    at={(-0.05, 1.34)},
    anchor=north west,
    },
    every axis x label/.style={
        at={(0.94, 0.01)},
        anchor=north west,
        },
    xmin=0, xmax=10,
    ymax=270,
    xtick={1, 2, 3, 4, 5, 6, 7, 8, 9 },
    xticklabels = {10, 20, 30, 40, 50, 60, 70, 80, 90},
    scale ticks above exponent=1,
    % y tick label style={/pgf/number format/sci, scaled ticks=base 10:0},
    legend style={
    fill=none,
    draw=none,
    at={(0.0,1.0)},anchor=north west},
    legend columns=2
    ]
    
    \addplot[smooth,mark=*,red] plot coordinates {
        (1, 135.63613613613612)
        (2, 148.69619619619618)
        (3, 162.1951951951952)
        (4, 175.52552552552552)
        (5, 188.81231231231232)
        (6, 202.15115115115114)
        (7, 215.3878878878879)
        (8, 228.69819819819818)
        (9, 242.27777777777777)
        
    };
    \addlegendentry{{\tt {BMT}-{O}}}

    \addplot[smooth,color=blue,mark=star]
        plot coordinates {
        (1, 116.4474474)
        (2, 127.8408408)
        (3, 149.7942943)
        (4, 149.0695696)
        (5, 160.1706707)
        (6, 170.4814815)
        (7, 181.2967968)
        (8, 193.0685686)
        (9, 232.6116116)
        };
    \addlegendentry{{\tt {BMT}-{FR}}}

    \addplot[smooth,color=black!30!green,mark=o]
        plot coordinates {
        (1, 131.6583584)
        (2, 131.7247247)
        (3, 142.47998)
        (4, 159.0595596)
        (5, 169.7322322)
        (6, 181.6816817)
        (7, 194.4714715)
        (8, 207.458959)
        (9, 219.0990991)

        % (1, 135.4614615)
        % (2, 132.4944945)
        % (3, 142.4384384)
        % (4, 157.2342342)
        % (5, 170.0625626)
        % (6, 182.3183183)
        % (7, 195.6281281)
        % (8, 207.505005)
        % (9, 219.6111111)
        };
    \addlegendentry{{\tt {BMT}-{PR}}}
    
    \end{axis}

\end{tikzpicture}

%% file: VLDB_Learned_SFC/figures/exp_data_shift_time.tex
    \begin{tikzpicture}[scale=0.7]

    \begin{axis}[
    width=8.5cm,
    height=4cm,
    xlabel= (\%),
    ylabel= Query Latency ($\mu$s), %ymode=log,
    every axis y label/.style={
    at={(-0.05, 1.34)},
    anchor=north west,
    },
    every axis x label/.style={
        at={(0.94, 0.01)},
        anchor=north west,
        },
    xmin=0, xmax=10,
    ymax=6000,
    xtick={1, 2, 3, 4, 5, 6, 7, 8, 9 },
    xticklabels = {10, 20, 30, 40, 50, 60, 70, 80, 90},
    scale ticks above exponent=1,
    % y tick label style={/pgf/number format/sci, scaled ticks=base 10:0},
    legend style={at={(1.0,1.0)},anchor=north east}
    ]
    
    \addplot[smooth,mark=*,red] plot coordinates {
        (9, 5.515141052765412 * 1000)
        (8, 5.143136949510546 * 1000)
        (7, 4.818827540308863 * 1000)
        (6, 4.608406796231045 * 1000)
        (5, 4.148721217631817 * 1000)
        (4, 3.7375025085739426 * 1000)
        (3, 3.573405134069311 * 1000)
        (2, 3.241041401127079 * 1000)
        (1, 2.864772015744382 * 1000)
    };
    % \addlegendentry{Original}

    \addplot[color=blue,mark=star]
        plot coordinates {
        (1, 2.617058453 * 1000)
        (2, 2.925477467 * 1000)
        (3, 3.184785356 * 1000)
        (4, 3.321106847 * 1000)
        (5, 3.624685056 * 1000)
        (6, 4.213558542 * 1000)
        (7, 4.205407442 * 1000)
        (8, 4.67022439 * 1000)
        (9, 4.9399216117 * 1000)
        };
    % \addlegendentry{Total Retrain}

    \addplot[smooth,color=black!30!green,mark=o]
        plot coordinates {
        (1, 2.686809491 * 1000)
        (2, 2.752346081 * 1000)
        (3, 3.024243855 * 1000)
        (4, 3.539896703 * 1000)
        (5, 3.813683688 * 1000)
        (6, 3.99375821 * 1000)
        (7, 4.308387443 * 1000)
        (8, 4.600079807 * 1000)
        (9, 4.891281371 * 1000)        

        % (1, 2.890781478 * 1000)
        % (2, 2.770791779 * 1000)
        % (3, 3.250857015 * 1000)
        % (4, 3.54255797 * 1000)
        % (5, 3.843682068 * 1000)
        % (6, 4.186963415 * 1000)
        % (7, 4.616869343 * 1000)
        % (8, 4.807855393 * 1000)
        % (9, 5.084441469 * 1000)
        };
    % \addlegendentry{Partial Retrain}
    
    \end{axis}

\end{tikzpicture}

%% file: VLDB_Learned_SFC/figures/exp_data_shift_train.tex
    \begin{tikzpicture}[scale=0.7]

    \begin{axis}[
    width=8.5cm,
    height=4cm,
    xlabel= (\%),
    ylabel= Training Time (s), %ymode=log,
    every axis y label/.style={
    at={(-0.05, 1.34)},
    anchor=north west,
    },
    every axis x label/.style={
        at={(0.94, 0.01)},
        anchor=north west,
        },
    xmin=0, xmax=10,
    ymax=9000,
    xtick={1, 2, 3, 4, 5, 6, 7, 8, 9 },
    xticklabels = {10, 20, 30, 40, 50, 60, 70, 80, 90},
    scale ticks above exponent=1,
    % y tick label style={/pgf/number format/sci, scaled ticks=base 10:0},
    legend style={at={(0.0,0.0)},anchor=south west}
    ]
    
    % \addplot[smooth,mark=*,red] plot coordinates {
    %     (1, 8041.246569871902)
    %     (2, 8041.246569871902)
    %     (3, 8041.246569871902)
    %     (4, 8041.246569871902)
    %     (5, 8041.246569871902)
    %     (6, 8041.246569871902)
    %     (7, 8041.246569871902)
    %     (8, 8041.246569871902)
    %     (9, 8041.246569871902)
    % };
    % \addlegendentry{Original}

    \addplot[smooth,color=blue,mark=star]
        plot coordinates {
        (9, 7857.85293889045)
        (8, 7953.655436038971)
        (7, 7976.9796278476715)
        (6, 8279.312395572662)
        (5, 8314.634133815765)
        (4, 8184.730911254883)
        (3, 8052.908196687698)
        (2, 8201.454149723053)
        (1, 8308.12857913971)
        };
    % \addlegendentry{Total Retrain}

    \addplot[smooth,color=black!30!green,mark=o]
        plot coordinates {
        (1,559.2266428)
        (2,3778.763054)
        (3, 3764.748218)
        (4, 1419.16016)
        (5, 1401.870786)
        (6, 1422.856793)
        (7, 1396.37336)
        (8, 1377.609946)
        (9, 1378.674264)

        % (1, 0)
        % (2, 4154.608017)
        % (3, 4141.300584)
        % (4, 4372.017914)
        % (5, 4228.063336)
        % (6, 4241.944699)
        % (7, 4254.403685)
        % (8, 4198.14254)
        % (9, 4160.371317)
        };
    % \addlegendentry{Partial Retrain}
    
    \end{axis}

\end{tikzpicture}

%% file: VLDB_Learned_SFC/figures/exp_query_shift_io.tex
    \begin{tikzpicture}[scale=0.7]

    \begin{axis}[
    width=8.5cm,
    height=4cm,
        xlabel= (\%),
        ylabel= I/O Cost, %ymode=log,
        every axis y label/.style={
        at={(-0.05, 1.34)},
        anchor=north west,
        },
        every axis x label/.style={
        at={(0.94, 0.01)},
        anchor=north west,
        },
        xmin=0, xmax=10,
         ymax=950,
        xtick={1,  2, 3, 4, 5, 6, 7, 8, 9 },
        xticklabels = {10, 20, 30, 40, 50, 60, 70, 80, 90},
        scale ticks above exponent=1,
        % ytick={10000, 50000, 100000},
        % y tick label style={/pgf/number format/sci, scaled ticks=base 10:0},
        legend style={
        fill=none,
        draw=none,
        at={(0.0,1.0)},anchor=north west},
        legend columns=2
        ]
    
    \addplot[smooth,mark=*,red] plot coordinates {
        (1, 188.4104104)
        (2, 258.53003)
        (3, 349.2202202)
        (4, 402.8363363)
        (5, 489.8253253)
        (6, 545.2662663)
        (7, 594.5925926)
        (8, 652.5700701)
        (9, 699.8683684)
    };
    \addlegendentry{{\tt {BMT}-{O}}}

    \addplot[smooth,color=blue,mark=star]
        plot coordinates {
        (1, 189.1886887)
        (2, 298.8528529)
        (3, 358.549049)
        (4, 400.961962)
        (5, 452.8548549)
        (6, 499.6061061)
        (7, 554.1646647)
        (8, 600.6011011)
        (9, 635.0735736)
        };
    \addlegendentry{{\tt {BMT}-{FR}}}

    \addplot[smooth,color=black!30!green,mark=o]
        plot coordinates {
        (1, 188.7752753)
        (2, 256.3828829)
        (3, 343.972973)
        (4, 402.2602603)
        (5, 489.0285285)
        (6, 548.0650651)
        (7, 550.9434434)
        (8, 555.7532533)
        (9, 582.8653654)
        
        % (1, 186.7387387)
        % (2, 256.5645646)
        % (3, 347.2082082)
        % (4, 401.3508509)
        % (5, 486.3583584)
        % (6, 479.2742743)
        % (7, 500.8528529)
        % (8, 522.6901902)
        % (9, 546.9089089)
        };
    \addlegendentry{{\tt {BMT}-{PR}}}
    
    \end{axis}

\end{tikzpicture}

%% file: VLDB_Learned_SFC/figures/exp_query_shift_time.tex
    \begin{tikzpicture}[scale=0.7]

    \begin{axis}[
    width=8.5cm,
    height=4cm,
    xlabel= (\%),
    ylabel= Query Latency ($\mu$s), %ymode=log,
    every axis y label/.style={
    at={(-0.05, 1.34)},
    anchor=north west,
    },
    every axis x label/.style={
        at={(0.94, 0.01)},
        anchor=north west,
        },
    xmin=0, xmax=10,
    ymax=15000,
    xtick={1, 2, 3, 4, 5, 6, 7, 8, 9 },
    xticklabels = {10, 20, 30, 40, 50, 60, 70, 80, 90},
    scale ticks above exponent=1,
    % y tick label style={/pgf/number format/sci, scaled ticks=base 10:0},
    legend style={at={(1.0,1.0)},anchor=north east}
    ]
    
    \addplot[smooth,mark=*,red] plot coordinates {
        (1, 3.613677349*1000)
        (2, 4.841171227*1000)
        (3, 6.426287604*1000)
        (4, 7.419407905*1000)
        (5, 9.025894844*1000)
        (6, 9.807888452*1000)
        (7, 10.94739776*1000)
        (8, 11.94587961*1000)
        (9, 12.78195606*1000)
    };
    % \addlegendentry{Original}

    \addplot[smooth,color=blue,mark=star]
        plot coordinates {
        (1, 3.650101097*1000)
        (2, 5.375807946*1000)
        (3, 6.505438635*1000)
        (4, 7.45023705*1000)
        (5, 8.634307721*1000)
        (6, 9.658123638*1000)
        (7, 10.42643777*1000)
        (8, 11.29466325*1000)
        (9, 12.06145141*1000)
        };
    % \addlegendentry{Total Retrain}

    \addplot[smooth,color=black!30!green,mark=o]
        plot coordinates {
        
        (1, 3.645397879*1000)
        (2, 4.844778532*1000)
        (3, 6.39479714*1000)
        (4, 7.380820729*1000)
        (5, 8.90762837*1000)
        (6, 9.712697627*1000)
        (7, 10.01618228*1000)
        (8, 10.34357669*1000)
        (9, 10.8887265*1000)
        
        % (1, 3.783132102*1000)
        % (2, 4.893084904*1000)
        % (3, 6.347979511*1000)
        % (4, 7.42601346*1000)
        % (5, 8.944854722*1000)
        % (6, 8.908125373*1000)
        % (7, 9.355419391*1000)
        % (8, 9.948697057*1000)
        % (9, 10.52163134*1000)
        };
    % \addlegendentry{Partial Retrain}
    
    \end{axis}

\end{tikzpicture}

%% file: VLDB_Learned_SFC/figures/exp_query_shift_train.tex
    \begin{tikzpicture}[scale=0.7]

    \begin{axis}[
    width=8.5cm,
    height=4cm,
    xlabel= (\%),
    ylabel= Training Time (s), %ymode=log,
    every axis y label/.style={
    at={(-0.05, 1.34)},
    anchor=north west,
    },
    every axis x label/.style={
        at={(0.94, 0.01)},
        anchor=north west,
        },
    xmin=0, xmax=10,
    ymin=0,ymax=9000,
    xtick={1, 2, 3, 4, 5, 6, 7, 8, 9 },
    xticklabels = {10, 20, 30, 40, 50, 60, 70, 80, 90},
    scale ticks above exponent=1,
    % y tick label style={/pgf/number format/sci, scaled ticks=base 10:0},
    legend style={at={(0.0,0.0)},anchor=south west}
    ]
    
    % \addplot[smooth,mark=*,red] plot coordinates {
    %     (1, 48.073888888888895)
    %     (2, 2249.9694444444444)
    %     (3, 4498.448333333334)
    %     (4, 8995.433333333332)
    %     (5, 22485.096666666668)
    %     (6, 44964.79277777777)
    %     (7, 67441.42333333332)
    %     (8, 67441.42333333332)
    %     (9, 67441.42333333332)
    % };
    % \addlegendentry{Original}

    \addplot[smooth,color=blue,mark=star]
        plot coordinates {
        (1, 8294.704488)
        (2, 7617.344106)
        (3, 7295.506479)
        (4, 8053.39411)
        (5, 7376.119945)
        (6, 8485.897015)
        (7, 7982.563132)
        (8, 6564.635646)
        (9, 7782.362566)
        };
    % \addlegendentry{Total Retrain}

    \addplot[smooth,color=black!30!green,mark=o]
        plot coordinates {

        (1, 1498.194158)
        (2, 1177.01218)
        (3, 1530.843161)
        (4, 1566.534648)
        (5, 1571.09213)
        (6, 236.3131471)
        (7, 3030.854743)
        (8, 267.8328705)
        (9, 255.9700444)
        
        % (1, 4355.421735)
        % (2, 4320.12562)
        % (3, 3585.860108)
        % (4, 2807.930986)
        % (5, 4118.945395)
        % (6, 4086.830963)
        % (7, 3969.139946)
        % (8, 3788.845397)
        % (9, 3697.6167)
        };
    % \addlegendentry{Partial Retrain}
    
    \end{axis}

\end{tikzpicture}

%% file: VLDB_Learned_SFC/figures/retrain_vary_constraint.tex
    \begin{tikzpicture}[scale=0.7]

    \begin{axis}[
    width=8.5cm,
    height=4cm,
    xlabel= Retrain Constraint Ratio,
    ylabel= I/O Cost, %ymode=log,
    every axis y label/.style={
    at={(-0.05, 1.34)},
    anchor=north west,
    },
    % every axis x label/.style={
        % at={(0.94, 0.01)},
        % anchor=north west,
        % },
    xmin=0, xmax=11,
    ymax=750,
    xtick={1, 2, 3, 4, 5, 6, 7, 8, 9, 10},
    xticklabels = {0.1, 0.2, 0.3, 0.4, 0.5, 0.6, 0.7, 0.8, 0.9, 1.0},
    scale ticks above exponent=1,
    % y tick label style={/pgf/number format/sci, scaled ticks=base 10:0},
    legend style={
    fill=none,
    draw=none,
    scale=0.7,
    font=\small,
    at={(-0.01,1.0)},anchor=north west},
    legend columns=3
    ]
    
    \addplot[smooth,mark=*,red] plot coordinates {
        
        (1, 642.8348348)
        (2, 642.8348348)
        (3, 642.8348348)
        (4, 642.8348348)
        (5, 642.8348348)
        (6, 642.8348348)
        (7, 642.8348348)
        (8, 642.8348348)
        (9, 642.8348348)
        (10, 642.8348348)
        
    };
    \addlegendentry{{\tt {BMT}-{O}}}

    \addplot[smooth,color=blue,mark=star]
        plot coordinates {
        (1, 521.9184184)
        (2, 521.9184184)
        (3, 521.9184184)
        (4, 521.9184184)
        (5, 521.9184184)
        (6, 521.9184184)
        (7, 521.9184184)
        (8, 521.9184184)
        (9, 521.9184184)
        (10, 521.9184184)
        };
    \addlegendentry{{\tt {BMT}-{FR}}}

    % \addplot[smooth,color=black!30!green,mark=o]
    \addplot[color=black!30!green,mark=o]
        plot coordinates {
        (1, 642.6136136)
        (2, 641.9874875)
        (3, 572.7247247)
        (4, 572.5910911)
        (5, 517.0215215)
        (6, 515.1336336)
        (7, 515.9279279)
        (8, 518.2467467)
        (9, 525.6191191)
        (10, 521.9184184)
        };
    \addlegendentry{{\tt {BMT}-{PR}}}
    
    \end{axis}

\end{tikzpicture}

%% file: VLDB_Learned_SFC/figures/retrain_vary_threshold.tex
    \begin{tikzpicture}[scale=0.7]

    \begin{axis}[
    width=8.5cm,
    height=4cm,
    xlabel= Shift Threshold ($\tt 0.XX$),
    ylabel= I/O Cost, %ymode=log,
    every axis y label/.style={
    at={(-0.05, 1.34)},
    anchor=north west,
    },
    % every axis x label/.style={
    %     at={(0.94, 0.01)},
    %     anchor=north west,
    %     },
    xmin=0, xmax=10,
    ymin = 450, ymax=710,
    xtick={1, 2, 3, 4, 5, 6, 7, 8, 9 },
    ytick={100, 200, 300, 400, 500, 600, 700, 800, 900 },
    xticklabels = {.1, .15, .2, .25, .3, .35, .4, .45, .5},
    scale ticks above exponent=1,
    % y tick label style={/pgf/number format/sci, scaled ticks=base 10:0},
    legend style={
    fill=none,
    draw=none,
    at={(0.0,1.0)},anchor=north west},
    legend columns=2
    ]
    
    \addplot[smooth,mark=*,red] plot coordinates {
        
        (1, 642.8348348)
        (2, 642.8348348)
        (3, 642.8348348)
        (4, 642.8348348)
        (5, 642.8348348)
        (6, 642.8348348)
        (7, 642.8348348)
        (8, 642.8348348)
        (9, 642.8348348)
        
    };
    % \addlegendentry{Original}

    \addplot[smooth,color=blue,mark=star]
        plot coordinates {
        (1, 521.9184184)
        (2, 521.9184184)
        (3, 521.9184184)
        (4, 521.9184184)
        (5, 521.9184184)
        (6, 521.9184184)
        (7, 521.9184184)
        (8, 521.9184184)
        (9, 521.9184184)
        };
    % \addlegendentry{Total Retrain}

    \addplot[color=black!30!green,mark=o]
        plot coordinates {
        (1, 517.2052052)
        (2, 516.8493493)
        (3, 517.046046)
        (4, 516.8668669)
        (5, 516.9414414)
        (6, 517.1506507)
        (7, 645.6396396)
        (8, 651.1566567)
        (9, 643.6121121)
        };
    % \addlegendentry{Partial Retrain}
    
    \end{axis}

\end{tikzpicture}

%% file: sections/related_work2.tex
\section{Related Work}
\label{sec:related_works}

%\subsection{Space-Filling Curve}

% \smallskip
\noindent\textbf{Space-Filling Curves (SFCs).}
%The core of Space-Filling Curves (SFCs) is to map multi-dimensional data points into scalar values, and then the data points can be organized in a linear order with the mapped values for indexing. 
Many SFCs have been developed.
The C-curve~\citep{jagadish1990linear} organizes the data points %dimension-by-dimension.
dimension at a time.
%, e.g., it sorts the data points by the first dimension, and then sorts by the next dimension if two points have the same first dimension. }}
%
The Z and Hilbert curves~\cite{orenstem1984class, orenstein1986spatial, orenstein1989redundancy, jagadish1990linear, mokbel2003analysis, moon2001analysis, kipf2020adaptive} are widely used in index design.
% since they order data without assuming the priority of dimensions. 
%Specifically, Z-curve compute SFC based on a bit interleaving rule as described. 
% sorts data following a zig-zag moving.
% , as shown in Figure~\ref{fig:popular_sfcs}(a). 
%
%Hilbert curve designs an ordering algorithm to make the neighboring data points in the original space close to each other in the mapped space. \revise{Hilbert curve's mapping algorithm recursively splits the space quarterly and orders the subspaces with heuristic rules.}
% , and then the subspaces are split and ordered recursively.
%accordingly.
%
%However, those curves have fixed mapping ,
%and are non-adjustable,
% , replacing Universal B Tree from its abbreviation where the rule is fixed and non-adjustable,
%e.g., Z-curve will only map data points following a zig-zag way. Therefore, they cannot be fine-tuned. 
%
Despite the success of these SFCs, they do not consider data and query workload distributions.  
QUILTS~\cite{nishimura2017quilts} is proposed to consider data  and query distributions in designing the mapping function of SFCs. 
%QUILTS uses a heuristic method to rank and select the optimal SFC from a set of candidate SFCs based on a database instance and a given query workload. 
%
%However, Quilts fails to effectively generate candidate SFCs.
% to compute a mapped value and take more mapping rules into account.
% Quilts provide state-of-the-art performance in existing space filling curves. 
%
All 
%the existing 
these SFCs, including 
%those considered by 
QUILTS, adopt a single mapping scheme
% , e.g., Z-curve maps data points following a zig-zag way, and these mapping scheme 
that may not always be suitable for the whole data space and query workload (Section~\ref{sec:intro}). 
This paper proposes the first piecewise SFC that uses different mapping functions for {\CHENG the different data subspaces}. It considers both data  and query distributions. Furthermore, we propose a reinforcement learning-based method to learn SFCs to directly optimize performance. 
%rather than relying on manually designed heuristic rules to choose an SFC. 
 \bluecolor{Following the BMTree~\cite{li2023towards}, there is work~\cite{gao2023lmsfc, liu2023efficient} that leverages learning to construct SFCs. The proposed piecewise SFC design potentially extends the design space of SFCs. Moreover, these studies do not consider fast updating of SFCs.  This paper proposes partially regenerating the BMTree, which reduces the update cost and only requires part of the data to update the SFC values.}
%we develop the first learned SFC method with deep reinforcement learning. 
%\revise{Concurrent to our work, Pai et al.~\citep{pai2022towards} develop the instance-optimal Z-index based on the Z-curve that adapts to the data and workload. 
%which partitions the space and assigns two possible Z-curves to the subspaces.
%However, it does not form an SFC as we do,
%in result but an index structure, which 
%and it cannot be flexibly combined with other indexes such as B+tree as we do.
%(2) it considers Z-curves and supports 2-dimensional data points only.
%}

%Compared with the existing curves, our method provides superior performance handling given database and workload without expertise and experience, especially under  non-uniform distribution. A detailed discussion of these works and our solutions can be found in Section~\ref{sec:prior}.
% \ref{sec:limitation}.
%However, current methods suffer from a  rule design $f$, and they cannot handle a non-uniform  situation. Their limitations leaf us to develop a Learned SFC method. We will discuss the limitations of current works later in detailed.

% \smallskip
\noindent\textbf{SFC-based Index Structures.}
SFCs are used for indexing multi-dimensional data,
%with the mapped values,
%(e.g., the B+Tree index~\citep{skopal2006new, lawder2000using}) can be applied on the SFC values to build an index, 
and 
%this 
is widely adopted by DBMSs. 
%such as SQL Server.
%It uses an SFC to map a data point $\mathbf{x}$ to a scalar value $v$ as the key value, and then any existing one-dimensional indexing structure (e.g., the B+Tree index~\citep{skopal2006new, lawder2000using}) can be applied on the SFC value $v$ to build an index. 
% Kipf et al.~\citep{kipf2020adaptive} applies a Hilbert curve to order the subspaces 
% % cells
% and allows 
% % cells of different sizes. 
% subspaces with different sizes.
% % The order of the grids still follows the Hilbert curve. 
% There is also a concurrent work~\citep{pai2022towards} which partitions the space and assigns two possible Z-curves to the subspaces. Unlike ours, (1) it is not an SFC but an  index structure, which cannot be flexibly combined with other indexes such as B+tree as ours do;
% % does not form an SFC. Instead, it partitions the space and assigns Z-curves, and proposes a specialized index structure to manage pieces. It would be hard to flexibly combine with other indexes such as B+tree; 
% (2) it considers Z-curves and supports 2-dimensional data points only.
%
Also, SFCs are 
%also 
essential for 
%recent works on 
learned multi-dimensional indexes (e.g.,~\citep{ wang2019learned, qi2020effectively, DBLP:journals/jcst/ZhangJWLWX21}).
% , where SFCs are first used to map multi-dimensional data to ordered one-dimensional values, and then learned index techniques for multi-dimensional data are proposed based on those for one-dimensional data~\citep{kraska2018case, nathan2020learning}. 
%Specifically,  
ZM~\citep{wang2019learned} combines a Z-curve with a learned index, namely RMI~\citep{kraska2018case}. RSMI~\citep{qi2020effectively} applies the Hilbert curve together with a learned index structure for spatial data.
\revise{
%Concurrent to our work, 
Pai et al.~\citep{pai2022towards} present preliminary results on the instance-optimal Z-index based on the Z-curve that adapts to  data and workload. 
% However, it can only map a data point to a ranking order, but not a scalar value that is used to order data points as most of SFCs do, and cannot be flexibly integrated into other indexes such as B+tree. %as we do.
%(2) it considers Z-curves and supports 2-dimensional data points only.
}
% 
% 
% 
% 
%\noindent \change{\textbf{Data Skipping.} 
% \citep{raman2013db2, sun2014fine, yang2020qd} 
%Data skipping technique~\citep{raman2013db2} partitions and organizes data into data pages. Data access algorithms~\citep{sun2014fine, yang2020qd} designed based on data skipping only access pages that are relevant to query, supported by the metadata stored in each page. Sun et al.~\citep{sun2014fine} propose a query-driven partition algorithm to minimize the page access, while Yang et al.~\citep{yang2020qd} propose a reinforcement learning based partition algorithm.
SFC-based indexes can also be applied for
data skipping~\citep{raman2013db2, sun2014fine, yang2020qd} that aim to partition and organize data into data pages so that querying algorithms only access pages that are relevant to a query. 
%Sun et al.~\citep{sun2014fine} propose a query-driven partition algorithm to minimize the page access, while Yang et al.~\citep{yang2020qd} propose a reinforcement learning based partition algorithm.
An SFC-based approach~\citep{nishimura2011md}
% (add the ref: MD-HBase: A Scalable Multi-dimensional Data Infrastructure for Location Aware Services)
maps multidimensional  points to scalar values using an SFC, and  uses the B$^+$-Tree or range-partitioned key-value store (e.g., H-base) for partitioning and organizing data. 

% {\bf Discuss  data skipping application}

%For window query of SFC-based Indexes, the key point is to determine the scanning range in the linearized space where all queried data points are within this range. The index then scans the data records within that range with further filtering and return the result records. Z-curve has a criterion called monotonicity, meaning if one data record has an equal or greater value towards another data record on every dimension, it also has an equal or greater mapped value. This criterion enables the query algorithm to only check the SFC values computed by the minimum point and maximum point of the window query to determine the scanning range. If an SFC does not maintain monotonicity, such as Hilbert, bounding scanning range could be complicated. In the following sections, we will discuss how to design a piece-wise SFC while maintain this criterion. 
% We maintain the monotonicity by following the mapping technology that Z-curve and Quilts use but also leverage the goodness of iteratively design ordering that the Hilbert curve does.

% \smallskip
\noindent\textbf{Analysis of SFCs.}  Many studies, e.g.,~\citep{mokbel2001irregularity, mokbel2003analysis, mokbel2003query,moon2001analysis, xu2014optimality, nishimura2017quilts}  evaluate SFCs. Mokbel et al.~\citep{mokbel2001irregularity, mokbel2003analysis, mokbel2003query} discuss the characteristics of good SFCs. %\citep{mokbel2001irregularity} proposes a metric named irregularity that evaluates how close two neighbor points in the original space will be in the linearized space. \citep{mokbel2003query} concludes that the optimal SFC should preserve locality well and proposes a spectral mapping that is good for locality preserving. \citep{mokbel2003analysis} divides the SFC into different types of segments and analyzes the percentage of each segment among different SFCs.  
Moon et al.~\citep{moon2001analysis} propose 
%the clustering number that represents 
% how many times disk seeking 
{\CHENG the number of disk seeks}
during query processing. Xu et al.~\citep{xu2014optimality} prove that the Hilbert curve is a preferable SFC from that respect.
%with a low clustering number. 
Nishimura et al. \citep{nishimura2017quilts} propose a {\em cohesion} cost that evaluates how good SFCs cluster data. 
\bluecolor{Recently, Liu et al.~\cite{liu2023efficient} propose a cost model that evaluates query performance of SFCs. It uses the BMTree and speedups reward computing.}

\noindent\textbf{Reinforcement Learning (RL) in Indexing.}
Our method of generating SFCs is based on RL techniques~\citep{sutton2018reinforcement, browne2012survey}.  
% DBLP:journals/dase/WuLZZC22
% DBLP:journals/chinaf/HuangQZTLC23, 
%If a task can be modeled as a decision making procedure, RL is a powerful paradigm for learning to make good sequential decisions. 
There are several recent studies, e.g.,~\citep{yang2020qd, liang2019neural, gu2021rlr} on applying  RL  to generate tree structures. Yang et al.~\citep{yang2020qd} construct the Qd-tree for partitioning data into blocks on storage with Proximal Policy Optimization  networks (PPO)~\citep{schulman2017proximal}. Gu et al.~\citep{gu2021rlr} 
%propose to 
utilize RL to construct the R-tree for answering spatial queries~\citep{ orenstein1986spatial},
% DBLP:journals/dase/PandeyRKK21,
and Neurocuts~\citep{liang2019neural} constructs a decision tree 
%based on the 
using a 
RL agent.
These RL designs are not suitable for 
%our task of 
learning piecewise SFCs. Our design of RL models is based on MCTS and
%including state, action, and rewards, 
is different from the designs in these studies.
%, and our learning task is different. 

% \bluecolor{

% \noindent \textbf{Distribution drift in AI4DB.}
% paper1 paper2 ...
% }

%The most related work is Neurocuts which deals with the partition problem. Due to the particularity of its partition problem, Neurocuts models the decision making procedure as, instead of a Markov decision process (MDP), a Branch MDP~\citep{hahn2021model} in which one state action pair $(s_t, a_t)$ could be transformed into multiple states, or a state set $\{s^1_{t+1},\ldots, s^k_{t+1}\}$. 
%One issue of these methods is that performance of generated structure remains unknown before fully constructed.
%We introduce a method that generates more than one reward before the whole structure is complete building, which alleviates this problem.
% , which is usually the weakness when applying RL for structure building and could not observe the performance of structure before it's fully constructed.

% In addition, using RL method for structure learning often could only get one performance result when the structure is fully constructed, which appears in both Qd-tree and Neurocuts. They solve this problem by compute multiple rewards based on one performance result.

%% file: sections/conclusion.tex
\section{Conclusion}
\label{sec:conclusion}
In this paper, we study the Space-Filling Curve Design problem, and propose constructing piecewise SFCs that adopt different mapping schemes for different data subspaces.
Specifically, we propose the BMTree for maintaining multiple bit merging patterns in which every path corresponds to a BMP. 
We propose to construct the BMTree in a data-driven manner via reinforcement learning. \bluecolor{Further, we develop a partial retraining procedure 
%which 
that 
supports retraining 
%part 
parts 
of the BMTree while retaining the rest unchanged instead of training from scratch.}
We conduct extensive experiments on both synthetic and real datasets with 
%different 
various
query workloads. Experiments show that the piecewise SFCs are consistently superior over existing SFCs, especially when data and/or queries have a certain degree of skewness. 
% We plan to explore more reinforcement learning techniques to boost other query processing problems in the future.
%{\CHENG We plan to explore learning based techniques to boost the query processing based on SFC-powered indexes in the future.}